\documentclass[letterpaper]{article}

\usepackage[english]{babel}

\usepackage[utf8]{inputenc}
\usepackage{amsmath}
\usepackage{placeins}
\usepackage{graphicx}
\usepackage{setspace}
\usepackage[colorlinks=true, allcolors=blue]{hyperref}
\usepackage[square,numbers]{natbib}
\usepackage{lineno}
\usepackage{ulem}

\title{Modeling Global Surface Dust Deposition Using Physics-Informed Neural Networks\footnote{This manuscript is published in \textit{Communications Earth \& Environment}, volume~5, article~778 (2024) in final form at \url{https://doi.org/10.1038/s43247-024-01942-2}.}}
\author{
	Constanza A. Molina Catricheo\thanks{School of Engineering, Pontificia Universidad Católica de Chile, Santiago, Chile.} \thanks{Institute for Geoinformatics, Universität Münster, Germany.}
	\and Fabrice Lambert\thanks{Geography Institute, Pontificia Universidad Católica de Chile, Santiago, Chile} \thanks{Center for Climate and Resilience Research, Chile}
	\and Julien Salomon\thanks{INRIA Paris, ANGE Project-Team, 75589 Paris Cedex 12, France} \thanks{CNRS, Laboratoire Jacques-Louis Lions, Sorbonne Université, 75005 Paris, France}
	\and Elwin van 't Wout\thanks{Institute for Mathematical and Computational Engineering, School of Engineering and Faculty of Mathematics, Pontificia Universidad Católica de Chile, Santiago, Chile.\newline Corresponding author: E. van 't Wout (e.wout@uc.cl)}
}
\date{\today}

\begin{document}

\maketitle

\begin{abstract}
	Paleoclimatic measurements serve to understand Earth System processes and evaluate climate model performances. However, their spatial coverage is generally sparse and unevenly distributed across the globe. Statistical interpolation methods are the prevalent techniques to grid such data, but these purely data-driven approaches sometimes produce results that are incoherent with our knowledge of the physical world. Physics-Informed Neural Networks (PINNs) follow an innovative approach to data analysis and physical modeling through machine learning, as they incorporate physical principles into the data-driven learning process. Here, we develop PINNs to reconstruct global maps of atmospheric dust surface deposition fluxes from measurement data in paleoclimatic archives for the Holocene and Last Glacial Maximum periods. We design an advection-diffusion equation to consider dominant wind directions at various latitudes, which prevents dust particles from flowing upwind. Our PINN improves on standard kriging interpolation by allowing variable asymmetry around data points. The reconstructions display realistic dust plumes from continental sources towards ocean basins following prevailing winds.
\end{abstract}

\section{Introduction}
\label{section: Introduction}

The deposition of dust on the Earth’s surface provides the terrestrial biosphere and surface ocean with micronutrients, affecting biogeochemical cycles~\cite{mahowald2011aerosol}. Additionally, dust particles can reflect or absorb incoming solar radiation, leading to changes in surface temperature and atmospheric circulation patterns~\cite{kok2023mineral}. Variability in the atmosphere's dust concentration may have affected past climate change, such as glacial-interglacial cycles~\cite{shaffer2018and}.

Previous efforts to simulate dust deposition under past climatic conditions were conducted using various global climate models~\cite{albani2014improved, ohgaito2018effect, sueyoshi2013set, yukimoto2012new, hopcroft2019role}. The complex nature of mineral dust aerosol emission makes it difficult to represent vertical dust fluxes accurately in weather and climate models~\cite{kok2014improved}. Hence, there are significant differences in atmospheric dust load and spatial distribution among various dust simulations, even under modern conditions~\cite{lambert2015dust, kok2023mineral}.

The Dust Indicators and Records of Terrestrial and Marine Paleoenvironments (DIRTMAP) database was designed to serve as a global validation dataset of simulations of the paleo-dust cycle with Earth System Models (ESM)~\cite{kohfeld2001dirtmap}. The dataset includes information on aeolian dust in ice cores, marine sediment cores, and loess-paleosol sections for average Holocene and Last Glacial Maximum (LGM) climatic conditions. These periods are characterized by low and high deposition rates, respectively. Only a few hundred sites around the world met the quality requirements to be included in the DIRTMAP database. Updated versions of this dataset~\cite{maher2010global, cosentino2023paleodust} are still the reference against which paleoclimatic dust simulations are compared. However, these data are unevenly distributed worldwide and unsuitable as boundary conditions or input fields for climate models. In practice, the kriging algorithm is the most appropriate geostatistical technique to interpolate the data across the globe~\cite{lambert2015dust}. These global interpolations are in gridded lat-lon format and can be used as input for climate simulations~\cite{kageyama2017pmip4, lambert2021regional, saini2021southern}. However, the kriging algorithm distributes the influence of a measured data point symmetrically along the anisotropy axis, thus spreading the influence of observations both upwind and downwind. This isotropy is not desirable at sites that are close to source dust regions. As an example, dust emitted in Patagonia in South America should be transported towards the South Atlantic in the direction of the Southern Westerly Winds (SWW) and not against these winds towards the South Pacific.

Deep neural networks could theoretically capture the behavior of global dust deposition if they are trained on enormous datasets. However, copious amounts of data are usually unavailable in the paleoclimatic context. For example, the DIRTMAP database contains a few hundred sites for both Holocene and LGM conditions, which is far from enough to train a neural network correctly. The idea of a Physics-Informed Neural Network (PINN) is to add information about the physical processes to the training algorithm so that predictions are restricted to physically realistic outcomes~\cite{raissi2019physics}. This physics-data fusion is achieved by adding a model loss term to the neural network that measures the consistency of the estimation with regard to a partial differential equation (PDE). In this study, we develop a PINN that optimizes the gridded field of global dust depositions according to two objectives: fitting the data and satisfying the PDE model.

PINNs were recently introduced as a powerful technique in situations where limited data are available and the physical processes are partially known~\cite{karniadakis2021physics}. The versatility of PINNs allows for different types of uncertainties, such as measurement errors in the data and unknown parameters in the PDE model~\cite{sahli2020physics}. Specifically, unknown parameters in the PDE or boundary conditions can be added to the neural network and optimized during the training phase, a procedure also known as an inverse problem or model discovery~\cite{jagtap2020conservative, yang2019adversarial}. These capabilities of the PINN are crucial to our goal of reconstructing global dust deposition since the empirical database has significant measurement errors, and several variables of the PDE are unknown in paleoclimatic settings.

While PINNs have attracted much attention from researchers, there are still many limitations in this field of scientific machine learning~\cite{cuomo2022scientific}. One prominent challenge is the application of PINNs to real-world data from field observations. Most research involves solving PDEs with validation against analytical solutions~\cite{he2021physics} and numerical solvers such as finite element methods~\cite{okazaki2022physics}. Often, the ground truth comes from synthetic data generated by high-fidelity simulations~\cite{penwarden2022multifidelity}. Among the few examples of PINNs for empirical data, we mention earthquake hypocentres~\cite{smith2022hyposvi}, fluid mechanics~\cite{cai2021flow}, cardiovascular flow~\cite{kissas2020machine}, and the covid-19 epidemic~\cite{linka2022bayesian}. While algorithmic innovations develop rapidly, there is still a large untapped potential for artificial intelligence and machine learning in the geosciences~\cite{tuia2021toward} and climate models~\cite{rolnick2022tackling}. For example, innovative approaches such as the PINN may improve global reconstructions of paleoclimatic variables significantly as compared to geostatistical approaches, and with much lower computational costs than coupled climate simulations.

In this study, we design a PINN to reconstruct a global dust deposition map from empirical paleoclimatic datasets with dust transport guided by physical processes such as advection and diffusion. The results are as good as kriging interpolations in regions with high data availability and, most importantly, significantly improve the physical realism of the calculations in data-sparse regions. Plumes of high dust concentration are accurately reconstructed along the prevailing westerlies at mid-latitudes and trade winds near the equator. These achievements confirm that PINNs can live up to their high expectations by combining data analysis and physical modeling within a single framework, even in realistic and challenging settings in geosciences such as reconstructing dust deposition fields from paleoclimatic records.

\section{Results}
\label{section: Results}

We designed a PINN for global dust deposition rates and reconstructed a global map from the paleoclimate dust flux measurements collected in the DIRTMAP database for the Holocene and LGM periods~\cite{lambert2015dust}.

\subsection{Global dust deposition reconstruction}
\label{section: pinn result}

\begin{figure}[!ht] 
	\centering
	(a)
	\includegraphics[width=0.44\textwidth]{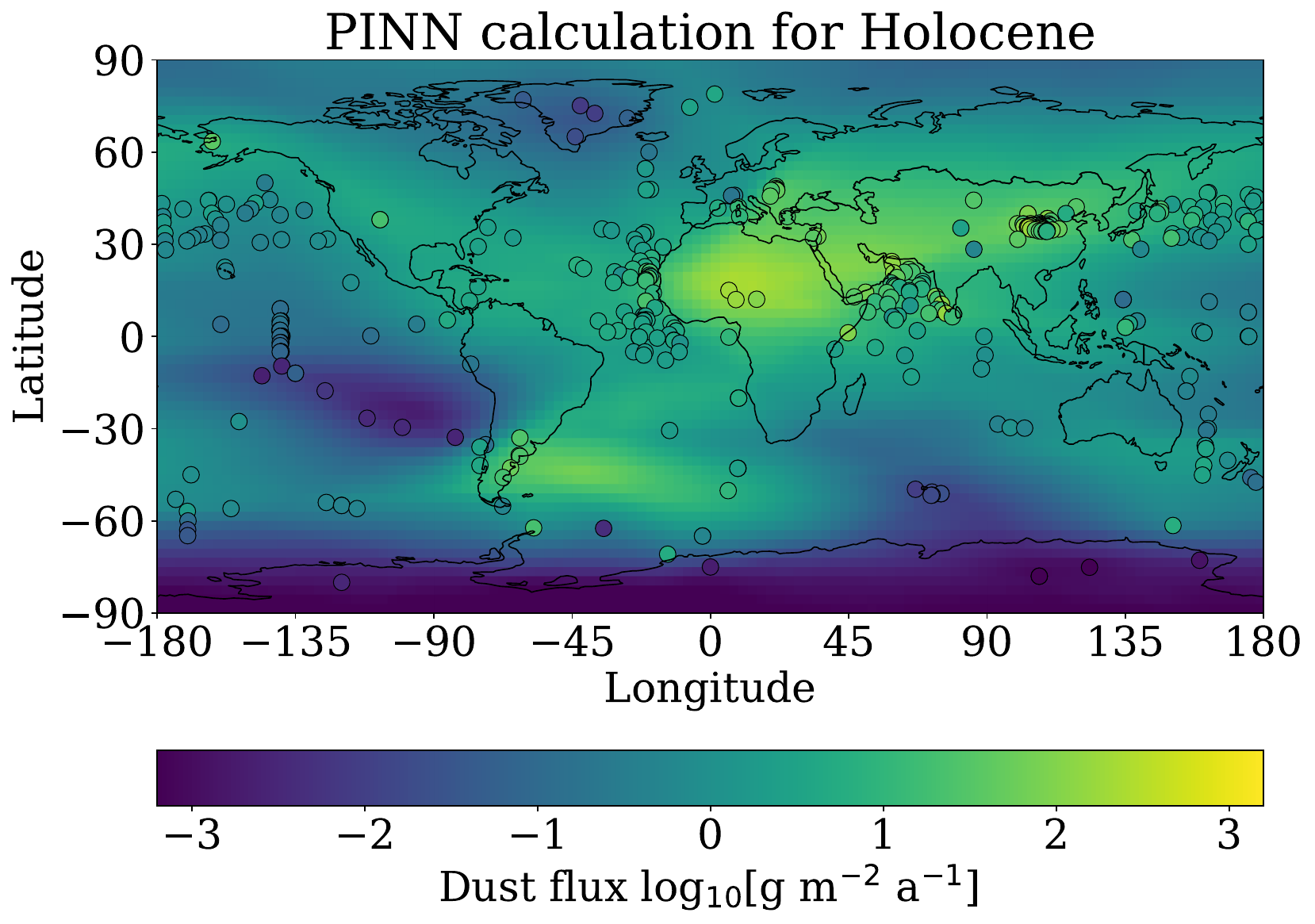}
	(b)
	\includegraphics[width=0.44\textwidth]{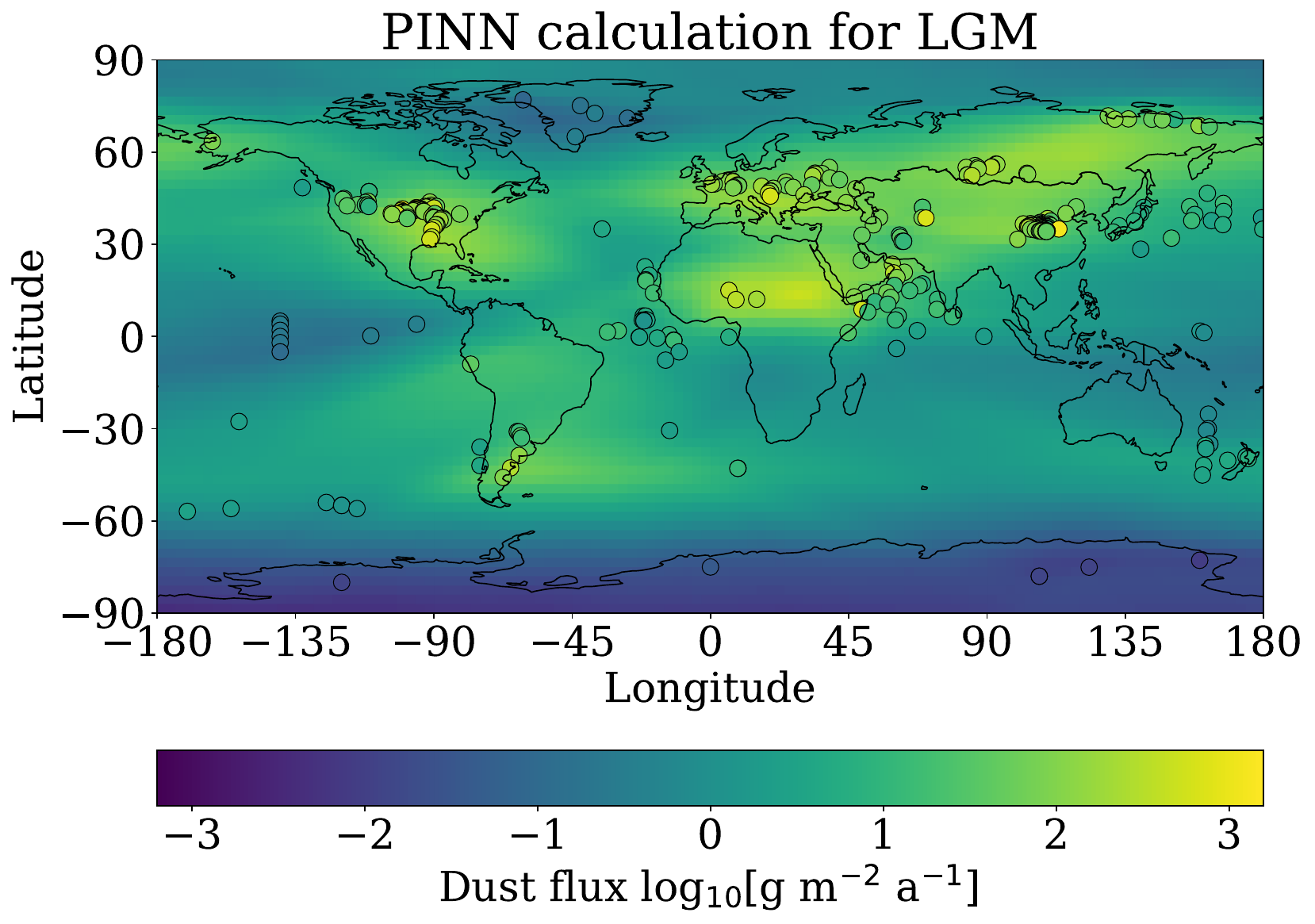}
	\\ (c)
	\includegraphics[width=0.44\textwidth]{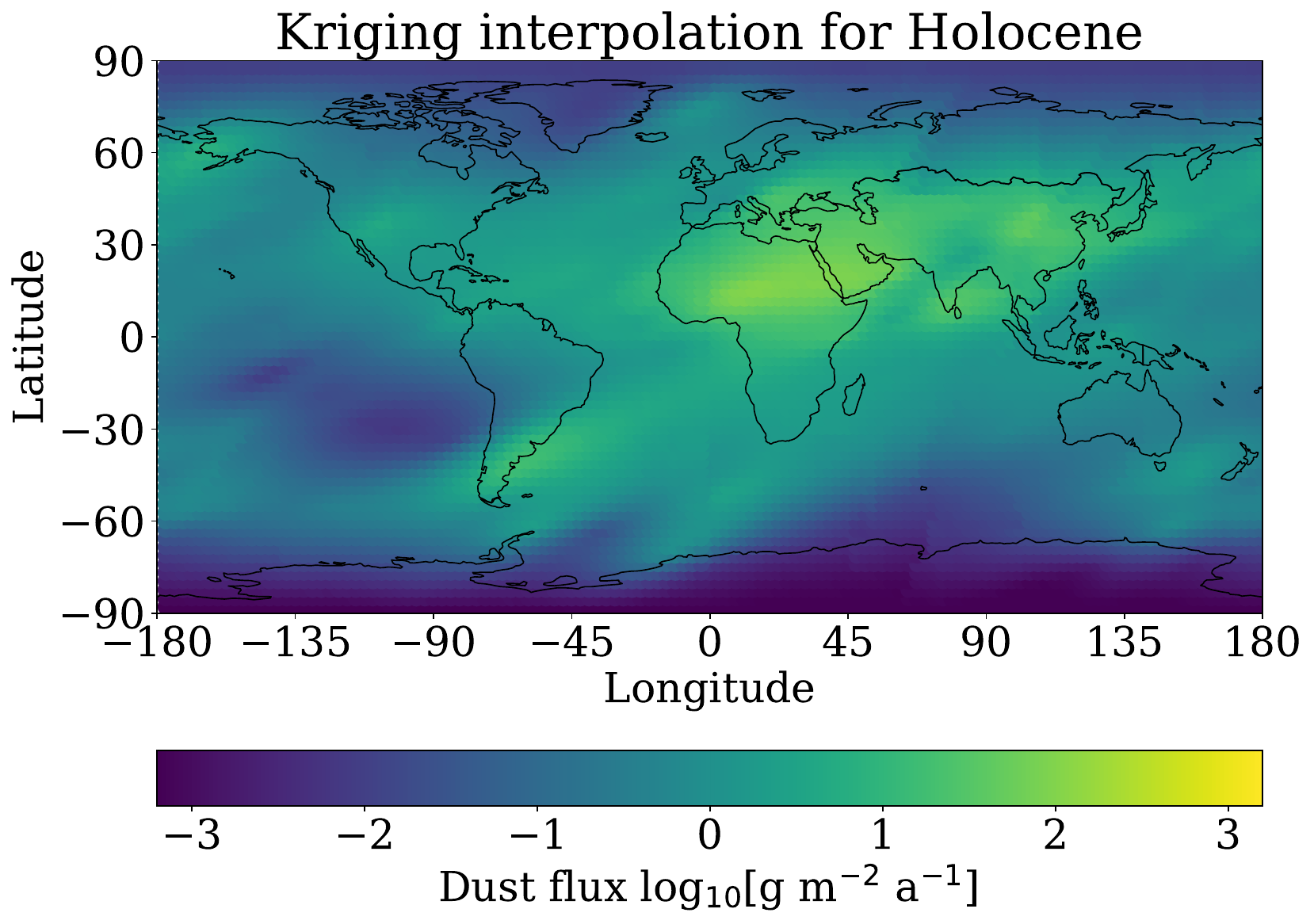}
	(d)
	\includegraphics[width=0.44\textwidth]{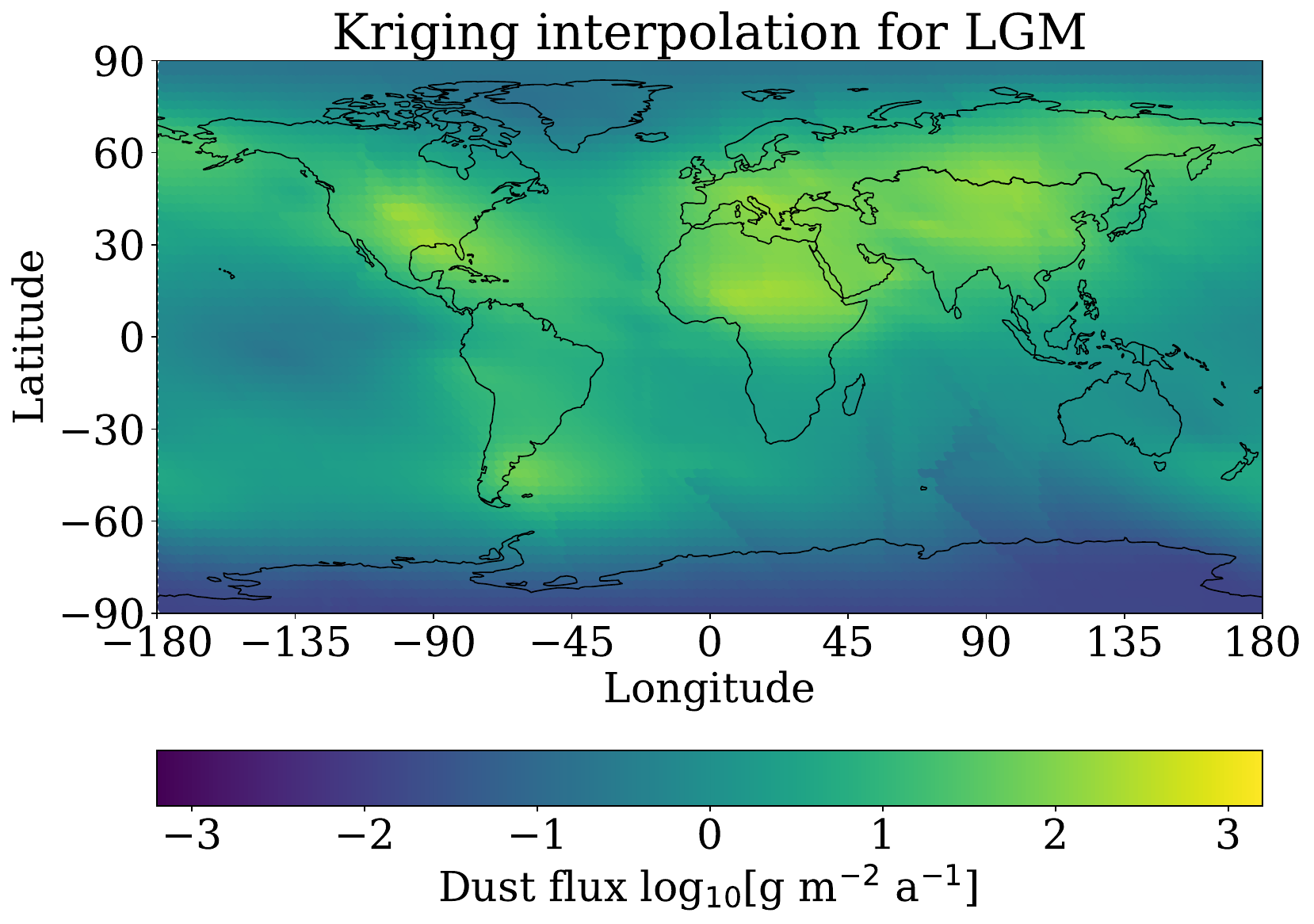}
	\caption{The global dust flux for the Holocene (panels a and c) and the LGM (panels b and d). The markers in panels (a) and (b) indicate the empirical data. The global fields in panels (a) and (b) are calculated with the PINN, while panels (c) and (d) show kriging interpolation.}
	\label{fig: deposition maps}
\end{figure}

We reconstructed the global dust deposition rates with our novel PINN algorithm and a dedicated kriging interpolation from~\cite{lambert2015dust}. The results in Figure~\ref{fig: deposition maps} display the reconstructions along with the 397 measurement points for the Holocene and the 317 measurements under LGM conditions. The PINN calculates the qualitative patterns of dust flow correctly, with high dust deposition values in the expected regions, such as the Sahara, East Asian deserts, North America, and Patagonia. The polar areas and oceans have low deposition rates as they are further away from major dust sources. We note that both the kriging interpolation and PINN calculation show minor dust levels in Australia and New Zealand for both Holocene and LGM conditions. This is because both are based on the same empirical dataset from 2015 that does not include continental information about this region~\cite{lambert2015dust}. Furthermore, the higher concentrations during the LGM compared to Holocene conditions are also visible in the reconstructions. Hence, the results confirm that the expected first-order features of global dust depositions are modeled correctly.

\subsection{Upwind dust flow in data-sparse regions}

As is common in paleoclimatology, global datasets are biased towards regions conducive to field campaigns. Hence, large parts of the world have no measurements or are sparsely covered by few data points. This lack of empirical data yields large uncertainties in geostatistical approaches like kriging interpolation~\cite{lambert2015dust}. Indeed, Figures~\ref{fig: deposition maps}(c) and (d) display numerical artifacts such as stripes and discontinuities in kriging's reconstructions in data-sparse regions such as the oceans in the southern hemisphere. In contrast, the PINN's calculated maps are smooth around the globe and accurately model the sphericity of the Earth.

Most importantly, the PINN models the physics of upwind dust flow accurately. The kriging interpolation is isotropic and displays symmetric peaks of dust concentration around the major sources. In contrast, the PINN includes a physical model for advection and diffusion and reconstructs the expected dust plumes along the dominant wind direction. For example, dust from the Patagonian deserts is transported along the SWW towards the Atlantic Ocean.

\subsection{Reconstruction in regions with uncertain data}

Any global reconstruction algorithm has to fit the empirical data well in regions with sufficient measurements. However, the overfitting of data clusters has to be avoided since data come with significant measurement errors and depend on local geographic conditions that are out of the scope of reconstructions on a global scale. For example, loess data in China may vary significantly between nearby measurement sites, with differences in deposition rate that may reach up to 80\% in some cases~\cite{cosentino2023paleodust}. Our goal is to reconstruct a global dust deposition map with a resolution of a few degrees in latitude and longitude, a scale at which diffusion of dust yields smooth fields. The physics-based loss term of the PINN effectively regularizes the reconstructions, also near the clusters of uncertain data points.

\begin{figure}[!ht] 
 \centering
	(a)
	\includegraphics[width=0.43\textwidth]{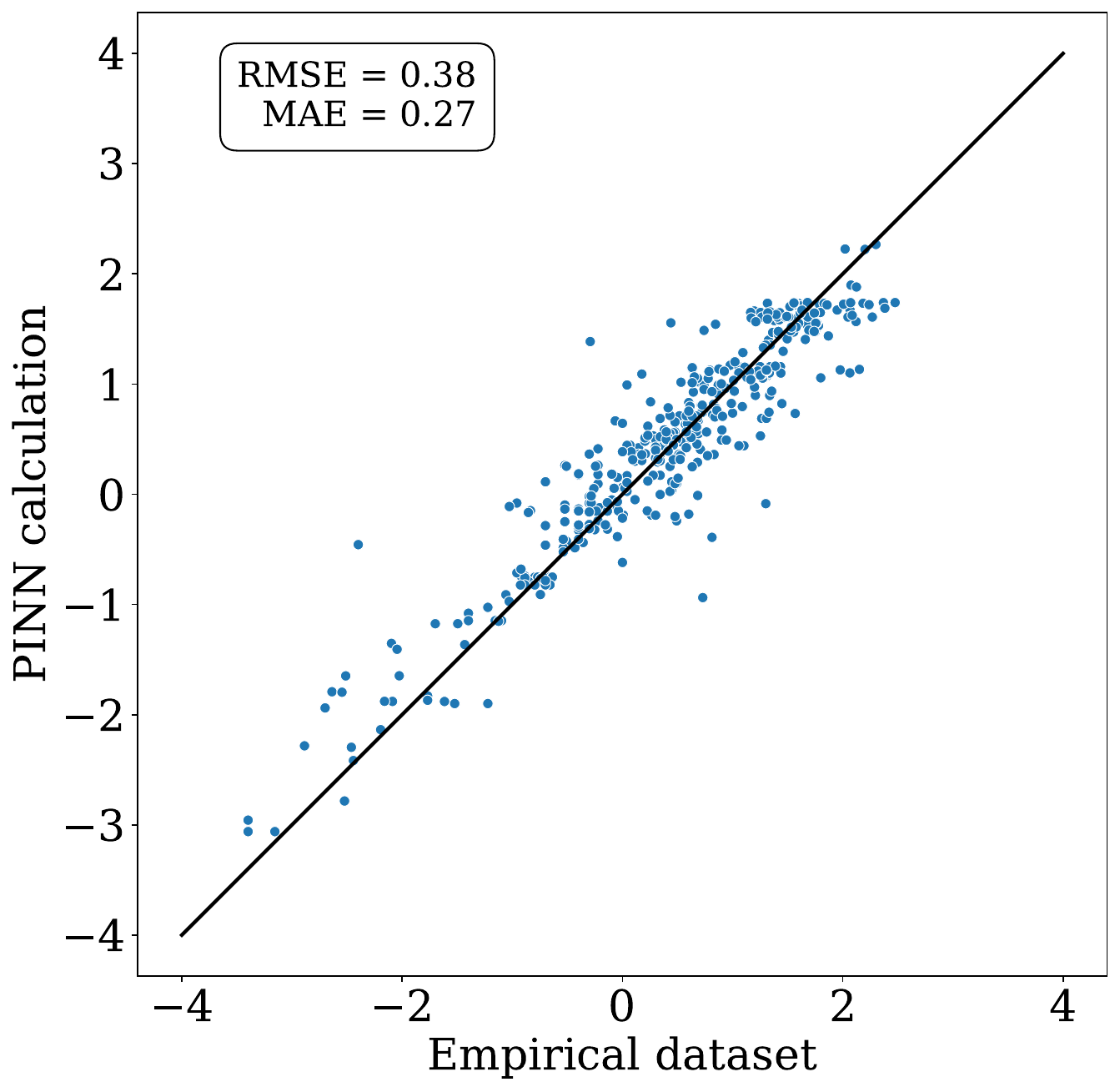}
	(b)
	\includegraphics[width=0.43\textwidth]{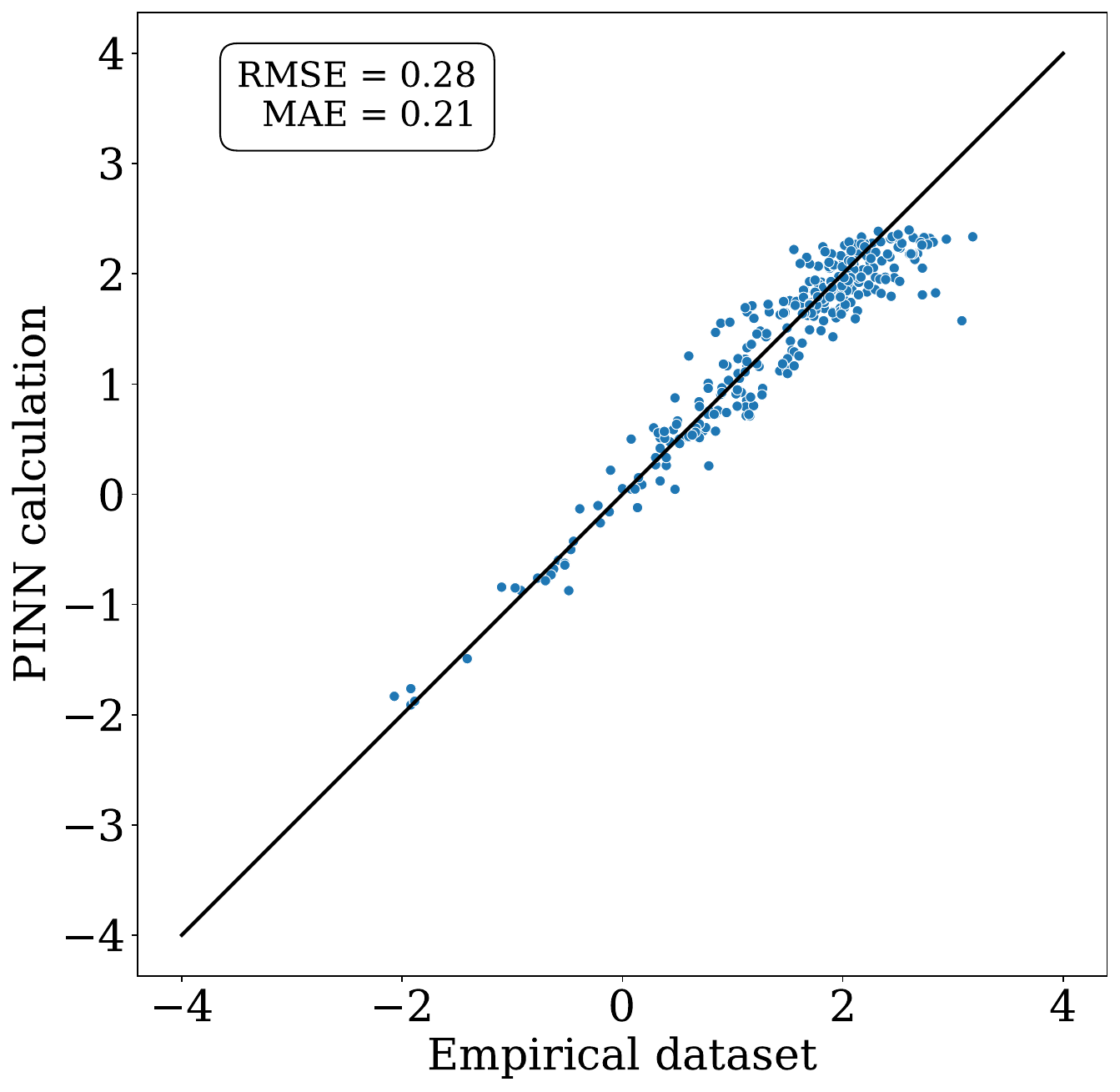}
	\\ (c)
	\includegraphics[width=0.43\textwidth]{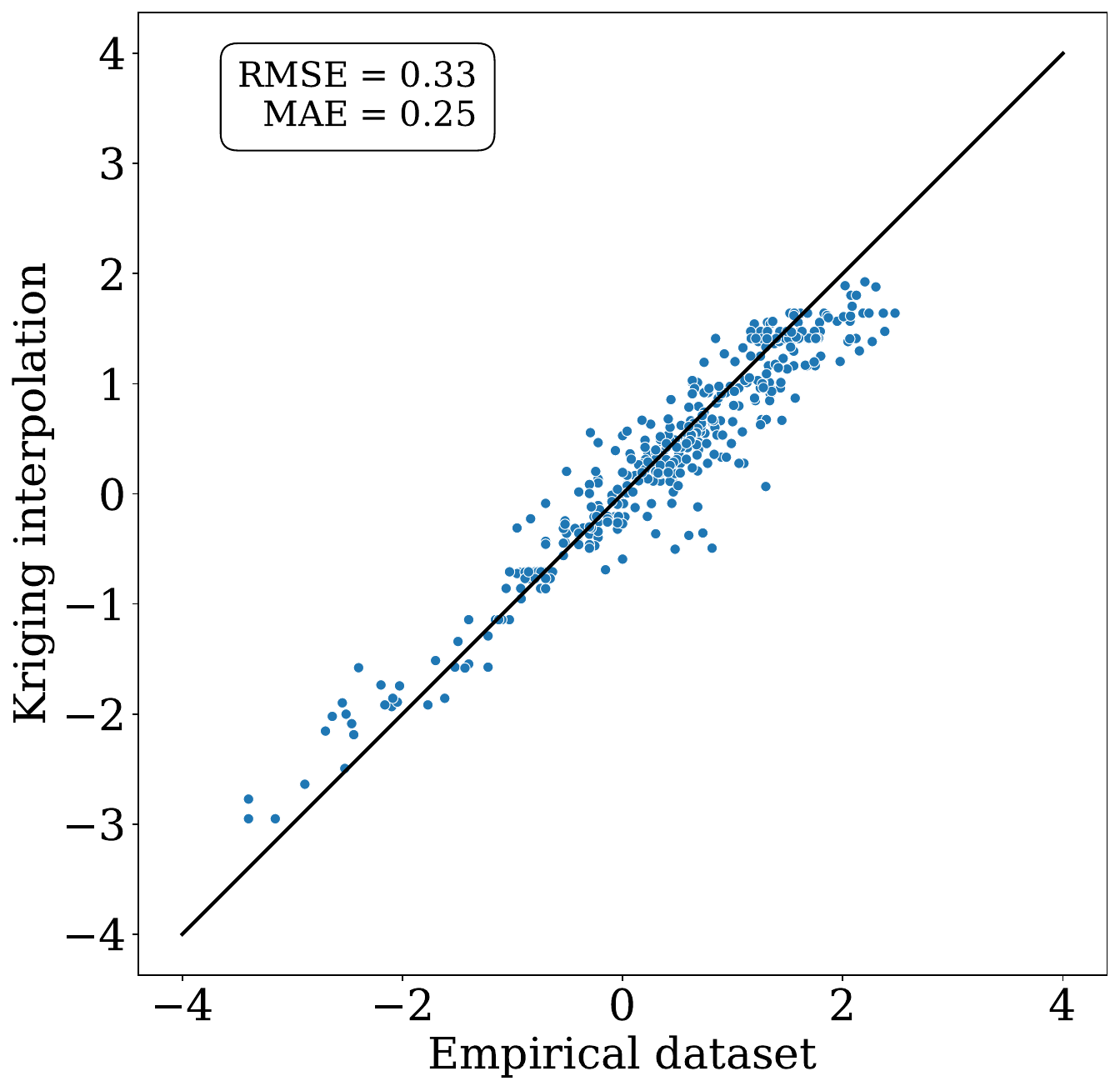} 
	(d)
	\includegraphics[width=0.43\textwidth]{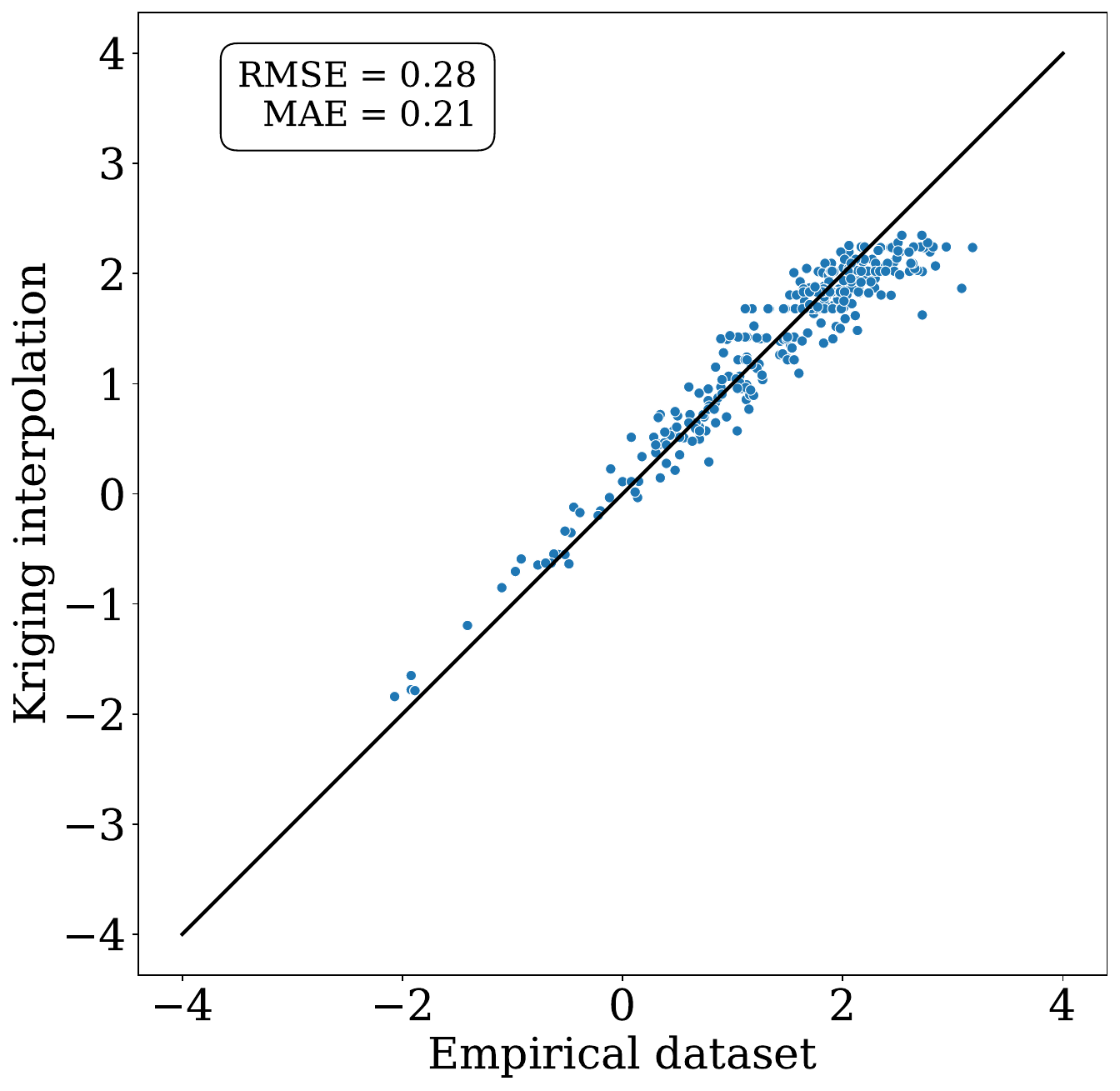}
	\\ (e)
	\includegraphics[width=0.43\textwidth]{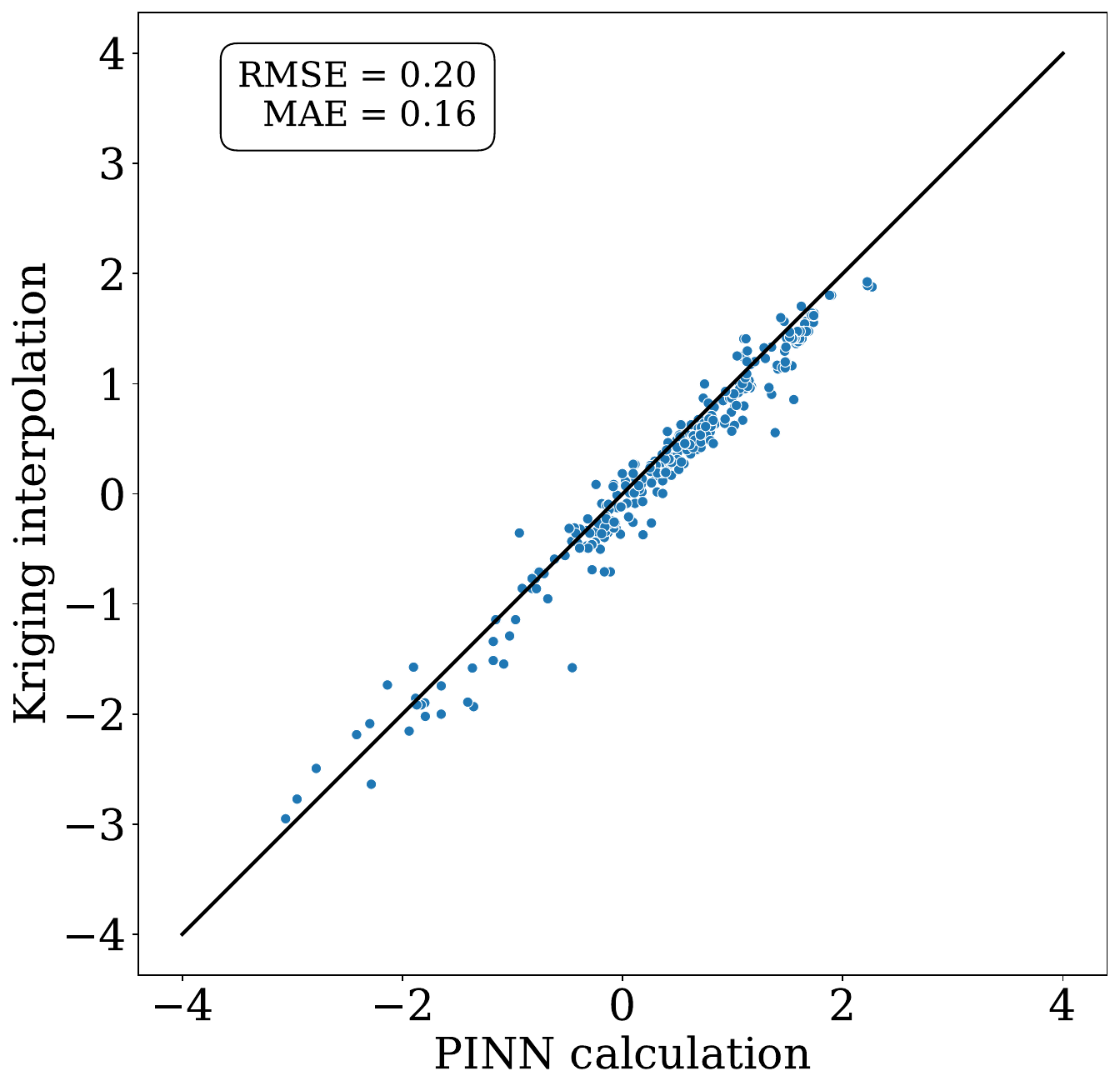}
	(f)
	\includegraphics[width=0.43\textwidth]{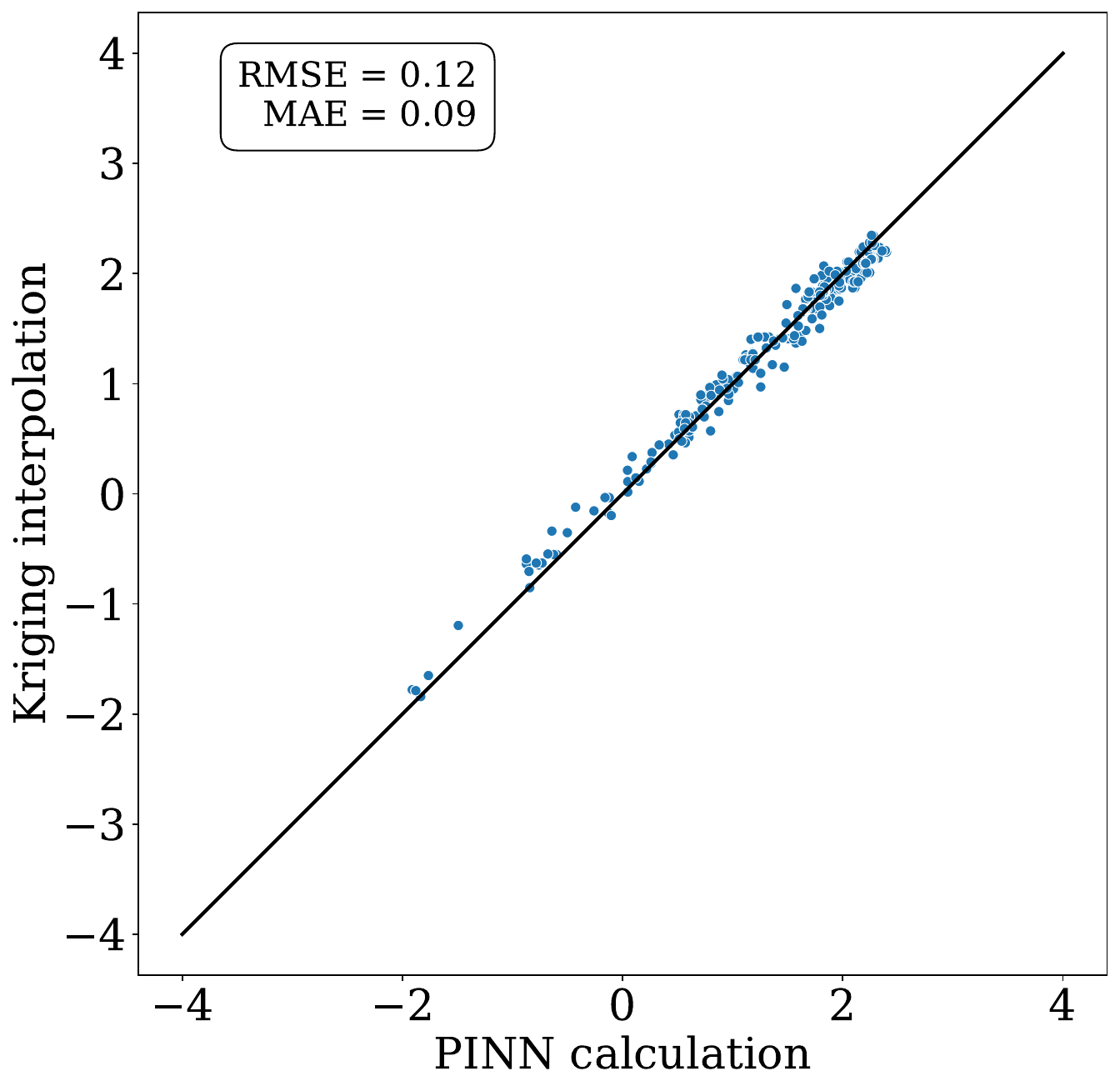}
	\caption{A comparison between the empirical dataset, PINN calculation, and kriging interpolation. The values of dust deposition rate are in log(g/m$^2$/a) as in Figure~\ref{fig: deposition maps}. The results of the PINN and kriging are evaluated at the locations of the empirical dataset. Panels (a) and (b) compare the empirical dataset with the PINN calculation. Panels (c) and (d) compare the empirical dataset with kriging interpolation at the empirical locations. Panels (e) and (f) compare the PINN calculation with kriging interpolation. Panels (a), (c), and (e) correspond to the Holocene, and panels (b), (d), and (f) use LGM conditions. }
	\label{fig: scatterplots}
\end{figure}

\FloatBarrier

Figure~\ref{fig: scatterplots} compares the empirical dataset, our PINN calculation, and the kriging interpolation of the dust fluxes. In general, the PINN and kriging fit well with the data in the measurement locations, with low values of the Root Mean Squared Error (RMSE) and Mean Absolute Error (MAE). Furthermore, the estimations of the dust fluxes in the data points are consistent between the two computational approaches. We emphasize that lower statistical values of the data fit do not necessarily mean that the reconstruction is improved. That is, paleoclimatic data have significant measurement errors and are biased toward locations where field experiments are possible, e.g., ice caps, loess fields, and high-sedimentation oceanic regions. The almost identical statistical error measures of the PINN and kriging to the empirical data confirm that the PINN performs as well as kriging in data-rich regions of the global reconstruction.

\section{Discussion}
\label{section: Discussion}

The PINN is an innovative approach cleverly combining data analysis with physical models. It promises to become a competitive alternative to traditional geostatistical methods by enforcing physical principles in data reconstructions with reasonable calculation times on common computer facilities. At the same time, current research into scientific machine learning targets open questions about the PINN's mathematical foundations and its merits in real-world settings. Our study confirms that PINNs can work with empirical datasets and improve upon dedicated statistical approaches such as kriging.

\subsection{Comparison between PINN and kriging}

Kriging interpolation techniques are the community standard for gridded reconstructions of paleoclimatic data of dust deposition on a global scale for the Holocene and LGM periods~\cite{lambert2015dust, cosentino2023paleodust, cosentino2023global}. However, this geostatistical algorithm is isotropic and yields elevated uncertainties in data-sparse regions. This study shows that PINNs yield global reconstructions that, on the one hand, have a similar goodness of fit to empirical data in regions where measurements are available and, on the other hand, avoid unphysical results by enforcing dust to flow along dominant wind directions. Hence, PINNs improve upon kriging.

The global reconstruction maps in Figure~\ref{fig: deposition maps} confirm that both the PINN and kriging correctly model the first-order features of global dust deposition, with high rates in arid zones and lower values over the oceans. Furthermore, the statistical fits of the PINN and kriging reconstructions with the empirical data are within the same error ranges; see Figure~\ref{fig: scatterplots}. Notice that the DIRTMAP database for dust deposition comes with significant measurement errors, with relative differences up to 80\% depending on the type of field experiment~\cite{cosentino2023paleodust}. Hence, minor differences between computational results and empirical data are expected when considering models on a global scale. The observed differences between PINN calculation, kriging interpolation, and the empirical data are all well within the range of data uncertainties.

The main reason for using PINNs instead of kriging is that they include physical information in reconstructing global dust deposition maps from sparse data. Interpolation techniques such as kriging are symmetric, thus allowing dust to flow upwind in contradiction with physical laws. This problem is exacerbated in regions with low data availability where statistical approaches lack the information to reproduce dust fields accurately. The improved physical realism of the PINN is most visible in the ocean basins downwind from primary dust sources, such as the South Atlantic, North Atlantic, and North Pacific. Moreover, the diffusion phenomenon incorporated in the PINN generates a smoothing effect of outliers and tight clusters of data points, avoiding overfitting issues common to pure data-driven approaches.

\begin{figure}[!ht] 
	\centering
	(a)
    \includegraphics[width=0.42\textwidth]{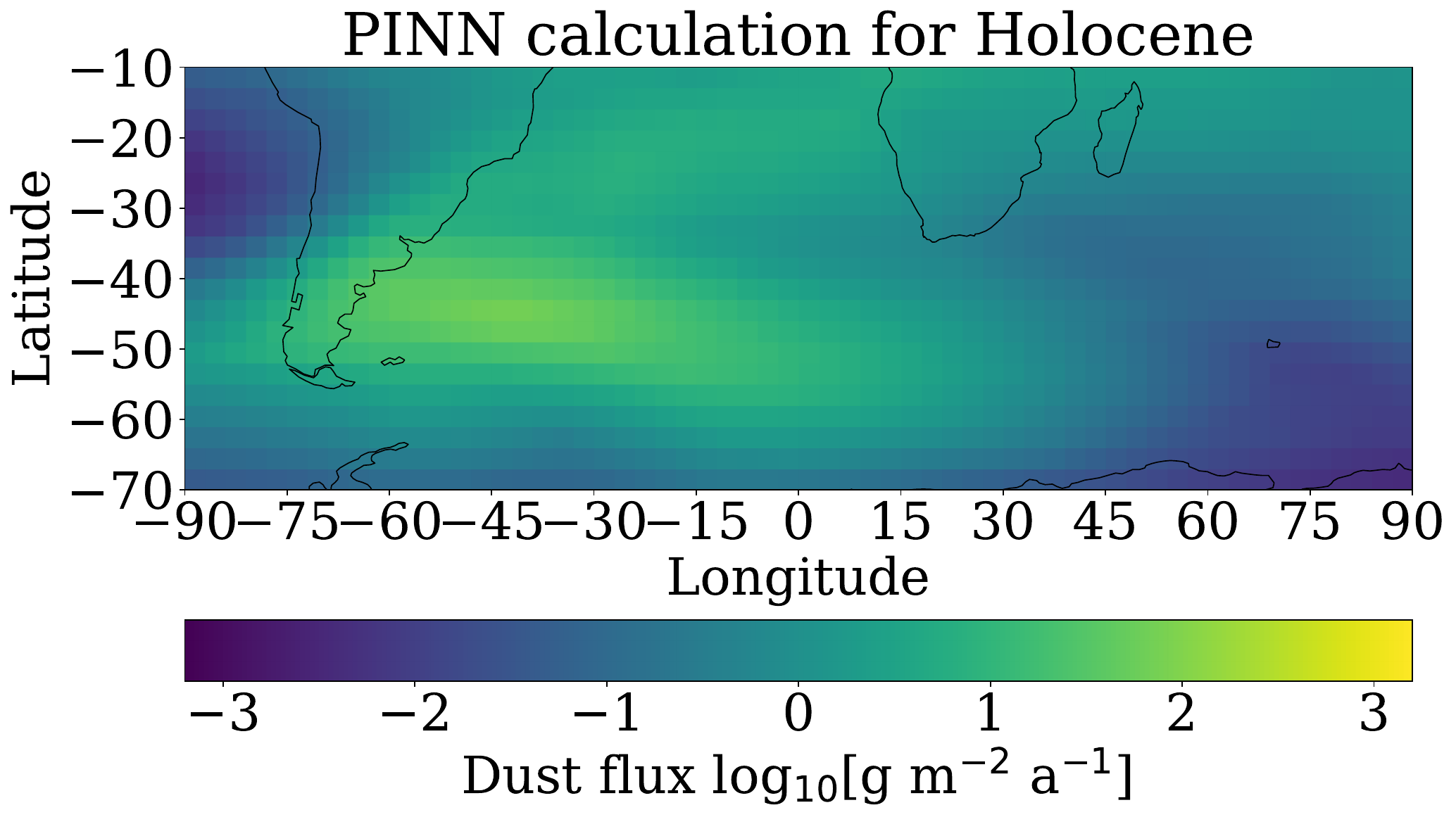}
	(b)
	\includegraphics[width=0.42\textwidth]{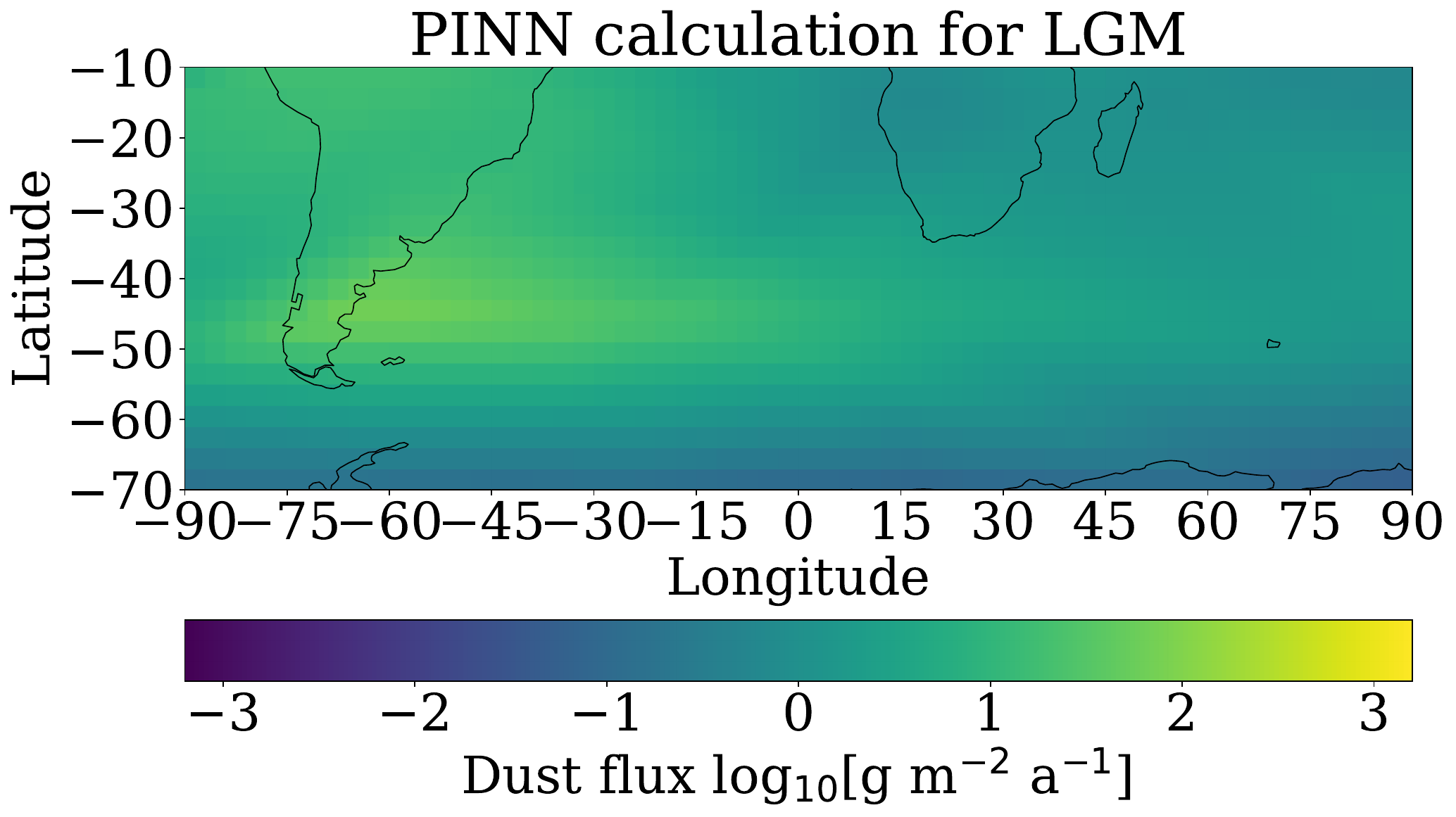}
	\\ (c)
	\includegraphics[width=0.42\textwidth]{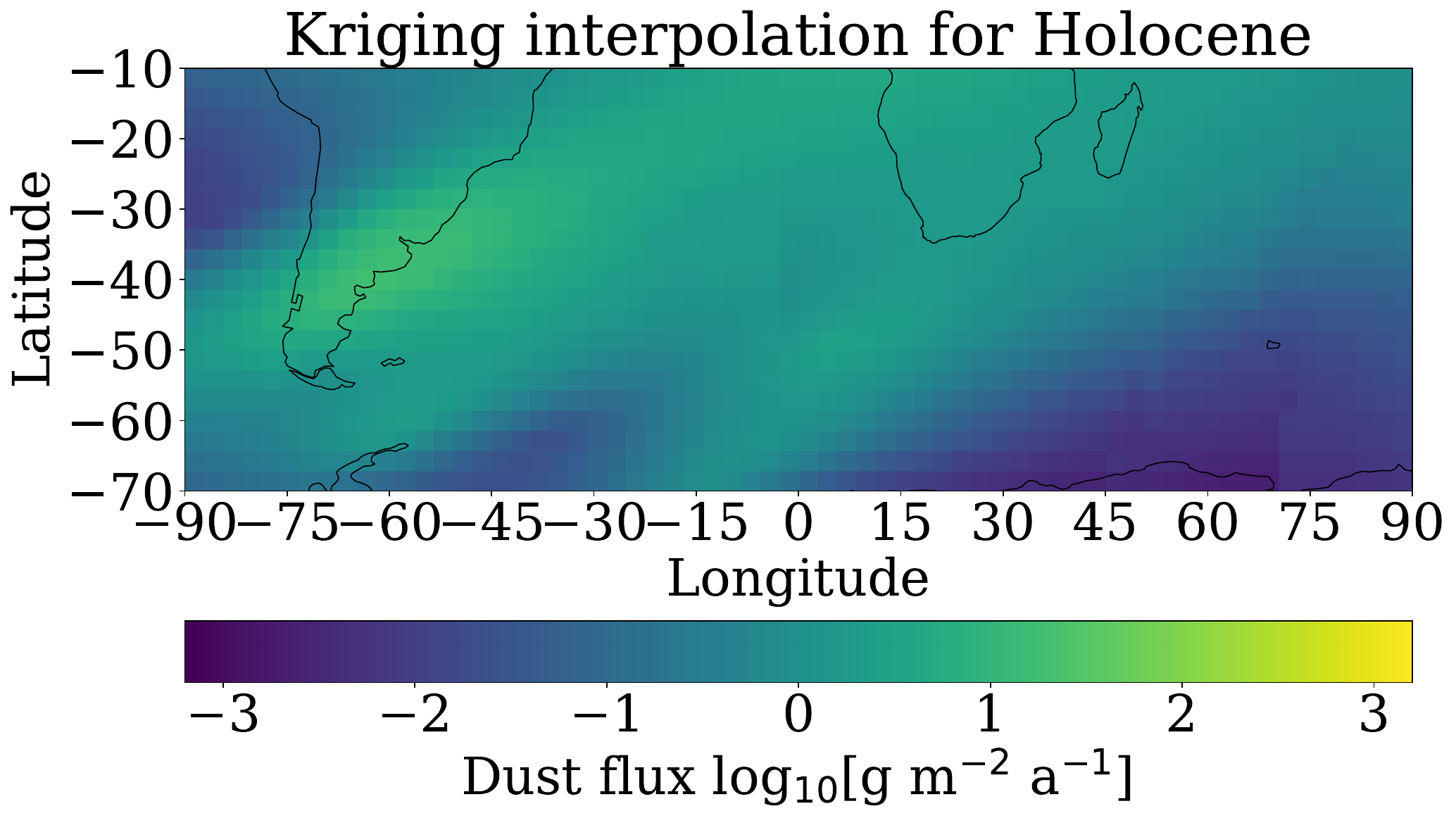}
	(d)
	\includegraphics[width=0.42\textwidth]{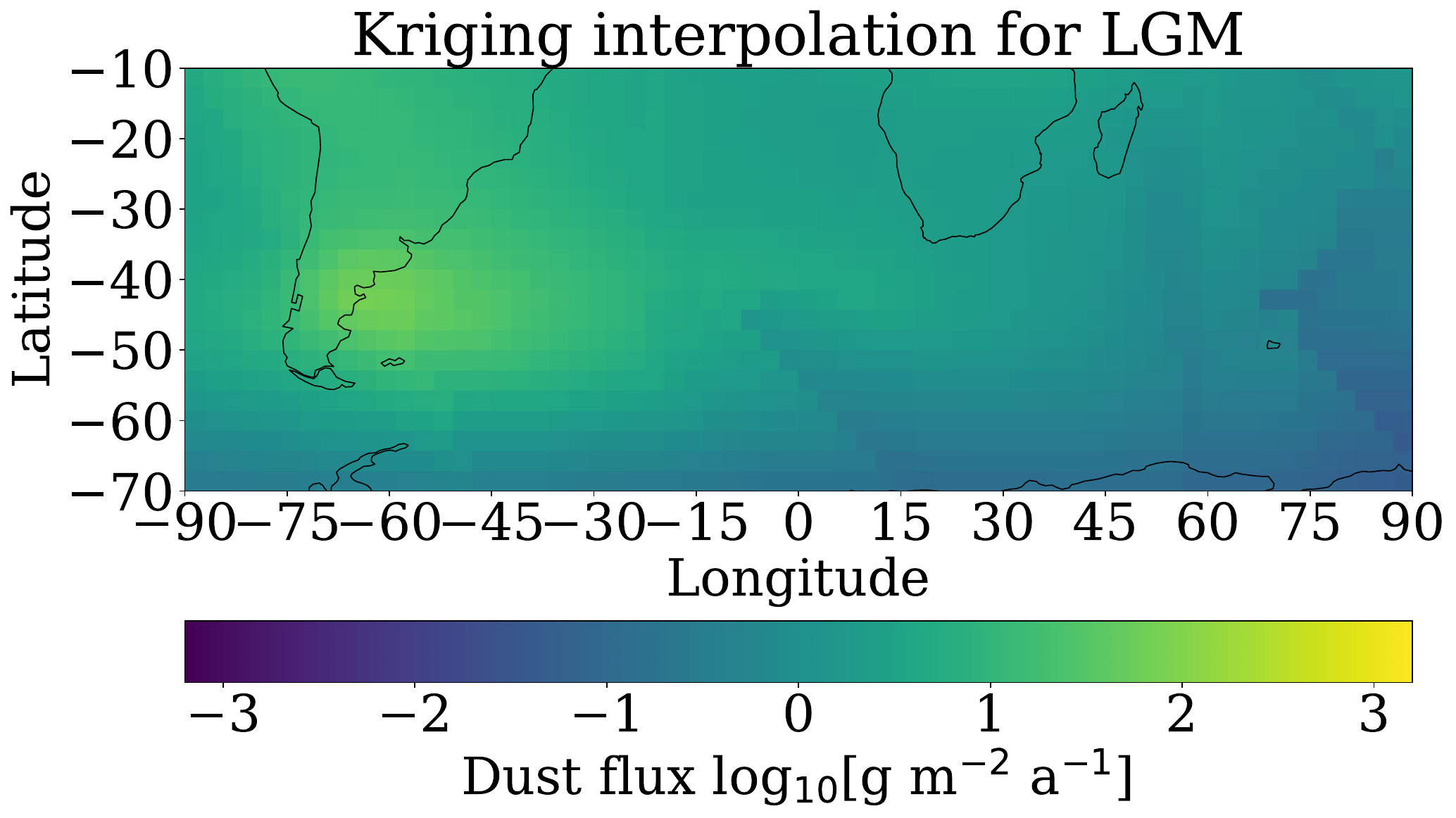}
	\caption{The dust flow from Patagonia towards the southern Atlantic Ocean as calculated by the PINN and kriging algorithms. The reconstructions are the same as in Figure~\ref{fig: deposition maps} but displayed for a smaller region.}
	\label{fig:patagonia}
\end{figure}

As a prime example of improved realism, we highlight that Patagonian dust in South America is transported towards the South Atlantic Ocean by the SWW and not upwind towards the South Pacific Ocean. This expected plume towards the east is clearly visible in the PINN's results; see Figure~\ref{fig:patagonia}. In contrast, the kriging interpolation for the Holocene displays high concentrations towards the northeast, which is inconsistent with the dominant eastward wind direction. In fact, it is an artifact due to the direction of the anisotropy ellipse in kriging, which is calculated globally and produces distinctive southwest-northeast streaks in the interpolation~\cite{lambert2015dust}. The kriging interpolation produces a zonal anisotropy ellipse and a broad and symmetric peak around Patagonia under LGM conditions.

Notably, the dust plume calculated by the PINN is well-focused east of the Patagonian sources and displays a transport pathway reaching the Indian Ocean and even the South Pacific. This advection in the southern hemisphere is particularly strong in the LGM, even stronger than in ESM simulations. While there are several data points in the South Atlantic and the Indian Ocean for the Holocene, there is no empirical data for the LGM in these regions. This lack of measurement sites makes it difficult to assess whether advection is over or under-estimated. However, this marked pathway may be realistic since studies that analyze the chemical composition of dust depositions have detected South American dust transported by the SWW all the way to the Southern Pacific Ocean during the LGM~\cite{struve2020circumpolar}.

Additional examples of dust being correctly transported downwind by the PINN include East Asia, where the Northern Westerly Winds (NWW) transports dust emitted from the central Asian deserts over the North Pacific towards Alaska. Finally, a significant improvement of the PINN over kriging is evident over the north and tropical Atlantic Ocean; the PINN simulates a transport of North American dust towards the North Atlantic following the NWW while also indicating a transport of Saharan dust towards the Caribbean, Central America, and the Amazonas, following the trade winds. In contrast, the kriging interpolation shows a transport of North American dust towards the tropical Atlantic against the prevailing easterlies. Saharan dust transport to the Amazon Forest has been evidenced for modern times~\cite{swap1992saharan, bristow2010fertilizing}.

\subsection{Verification of the PINN methodology}

The PINN is a relatively new algorithm that has spurred a very active research agenda among mathematicians and practitioners of computational simulations. Being a novel technology, we assessed PINN's versatility in handling sparse and irregular datasets and its predictive skill for paleoclimatic processes by four verification strategies.

The first indicator supporting the accuracy of the PINN is the consistency between the calculated and empirical data, as shown in the scatterplots in Figure~\ref{fig: scatterplots}(a) and (b) for the Holocene and LGM, respectively. Second, we analyzed the statistical distribution of the calculated dust deposition rates, which is expected to be approximately log-normally distributed, as is the empirical dataset. The results presented in Section~\ref{sec:si:statistical_distribution} of the Supporting Information confirm that the PINN maintains the log-normal distribution of the data. In the third approach, we applied the PINN to simulated data from an ESM. The PINN's calculations show excellent agreement with the simulated data on a 1-degree global grid; see Section~\ref{sec:si:esm_comparison} of the Supporting Information. Fourth, we analyzed the goodness of fit of the PINN with cross-validation on left-out empirical data and random subsets of simulated data; see Section~\ref{sec:si:goodness_of_fit} of the Supporting Information.

\subsection{Outlook on PINNs for geosciences}

The PINN provides a general framework for combining data analysis with physical models and promises to be beneficial for a broad range of applications in geosciences. Our study on PINNs for global dust depositions serves as a proof-of-concept for reconstructing physical fields from sparse, irregular, and uncertain datasets. We summarize several key advantages of PINNs. First, PINNs work well when limited data is available. In this study, we worked with only 397 and 317 measurements for the Holocene and LGM periods, respectively. Second, the physical model incorporated in the PINN design tends to regularize the solution and thus handles data uncertainties well, even when applied to tight clusters of nearby measurement locations. This smoothing effect is achieved without needing special preprocessing techniques, such as averaging small regions, which is standard in kriging. Third, the PINN handles unknown parameters in the physical model by estimating their value within the training of the neural network. For example, our PINN design included the estimation of the diffusion coefficient and deposition at the polar boundaries, which cannot be taken from direct measurements. The results presented in Section~\ref{sec:si:parameters} of the Supporting Information verify the accuracy of the estimations of unknown parameters in the physical model. Fourth, PINNs work naturally with data in a lat-lon format, and PDEs can easily be written in spherical coordinates to model the Earth's surface. Fifth, training a PINN on datasets of several hundreds of measurements takes only a few minutes on a standard workstation.

One of the main challenges of the PINN approach is creating a robust and accurate neural network. Design choices such as the type of neural network, number of layers, and activation functions strongly impact the convergence during training~\cite{wang2022when}. These issues are due to PINNs solving a non-convex optimization problem with local minima~\cite{krishnapriyan2021characterizing}. Unfortunately, current research into guidelines for setting these technical parameters is still ongoing~\cite{cuomo2022scientific}. Hence, future research should explore other kinds of neural networks, such as graph networks~\cite{gao2022physics} and Fourier neural operators~\cite{jiang2021digital}. Another challenge is specifying the correct PDE for modeling the physical processes of interest. We chose to include convection and diffusion only, but more elaborate models could be incorporated in the future, such as the Navier-Stokes equations~\cite{jin2021nsfnets}. Finally, Bayesian PINNs can be investigated to quantify the prediction uncertainty due to noisy data~\cite{yang2021bayesian}. The promising results of this study could also be extended in future research for emulating, downscaling, and forecasting other weather and climate processes~\cite{kashinath2021physics}.

\FloatBarrier

\section{Methods}
\label{section: Methodology}

\subsection{Data}
\label{section: Data}

The empirical data we use in this study are the surface dust deposition fluxes measured in marine sediment cores, ice cores, and loess deposits published in~\cite{lambert2015dust}, including the DIRTMAP database and some additional sites. The data are averaged dust deposition rates over the Holocene (last 12 thousand years) and LGM periods (between 19 and 26 thousand years ago) and consist of 397 and 317 measurements, respectively.

\begin{figure}[!ht]
	\centering
	(a) \includegraphics[width=0.8\columnwidth]{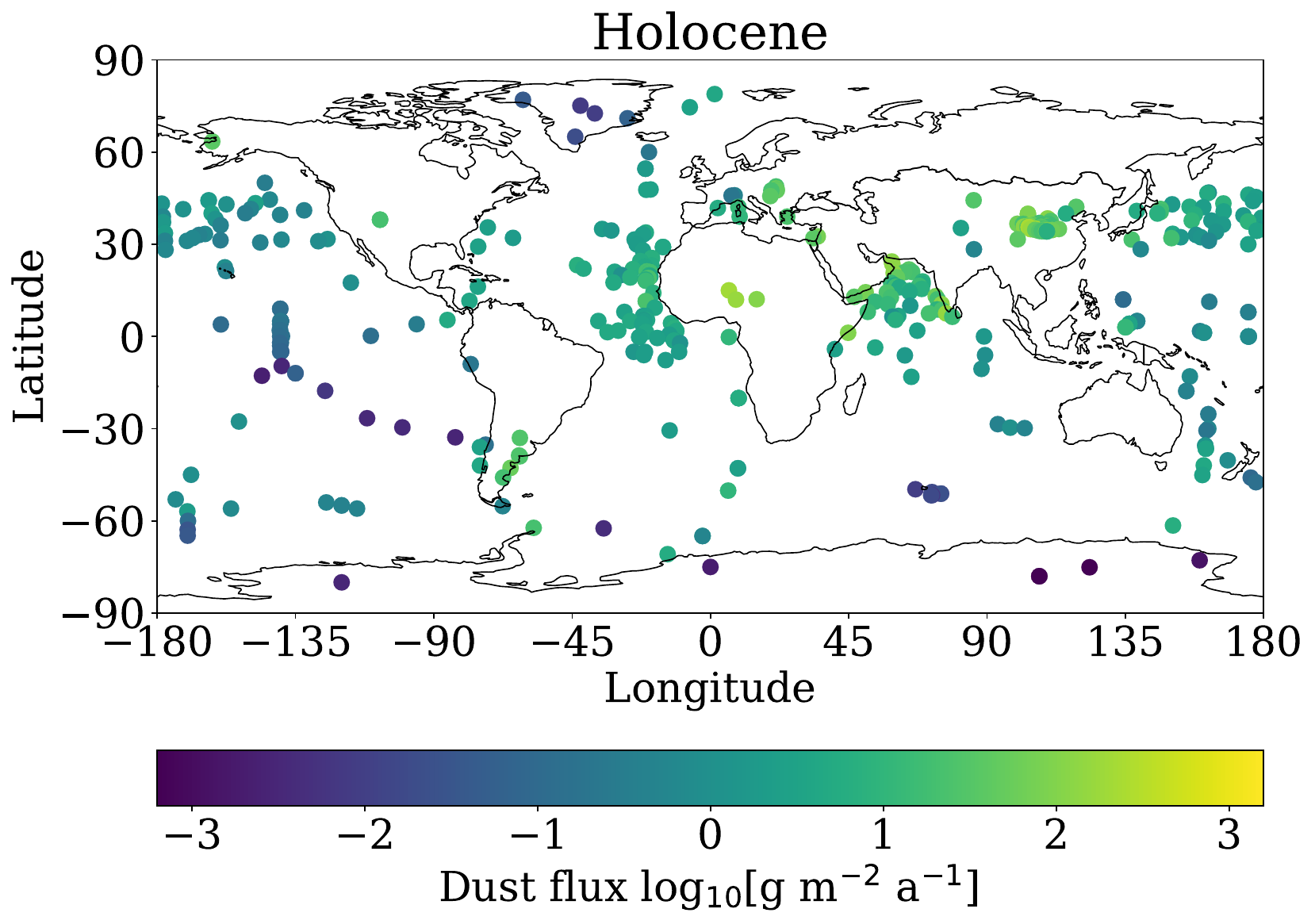}
	\\
	(b) \includegraphics[width=0.8\columnwidth]{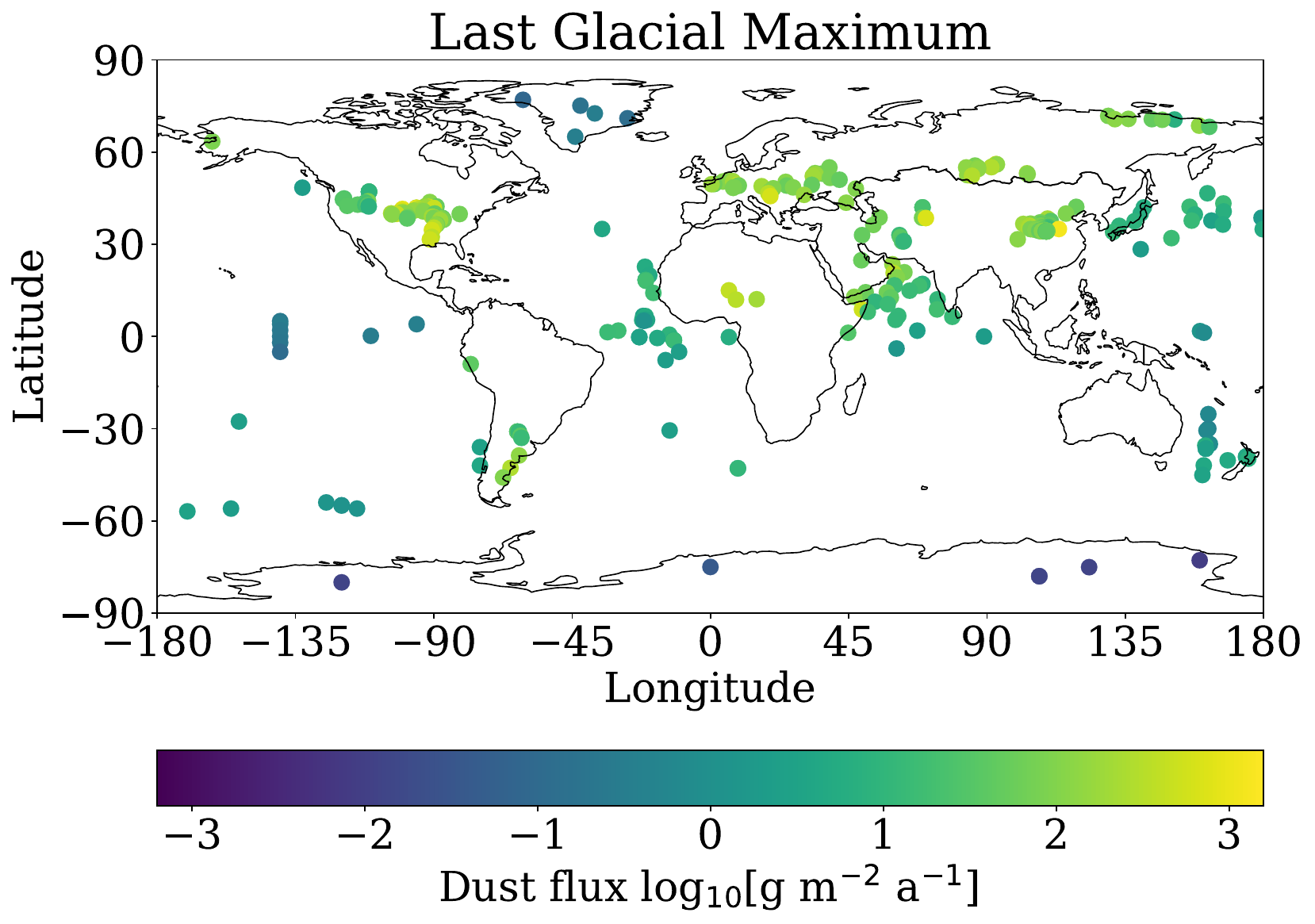}
	\caption{The surface dust flux measurements from paleoclimatic archives~\cite{lambert2015dust} for (a) the Holocene and (b) the LGM.}
	\label{fig: empirical data}
\end{figure}

The global map of the measurement sites in Figure~\ref{fig: empirical data} clearly shows the limited amount of paleoclimate data available for analysis. The data points are also irregularly located around the globe, leaving large parts of the Earth with deficient spatial coverage. Moreover, clusters of data points show significant differences in dust deposition despite being close to each other. These inconsistencies are due to relatively large measurement errors in the paleoclimate archives. Specifically, the errors are in the order of 10-20\% for ice cores, 50-70\% for marine sediments, and 50-80\% for loess~\cite{cosentino2023paleodust}.

Since the dust flux measurements are approximately log-normally distributed~\cite{lambert2015dust}, we work with the logarithmic transformation of the data and use a standard normalization based on the $z$-scores.

\FloatBarrier

\subsection{Dust fluxes model}
\label{section: Dust Fluxes Model}

\begin{figure}[!ht]
	\centering
	\includegraphics[width=0.8\columnwidth]{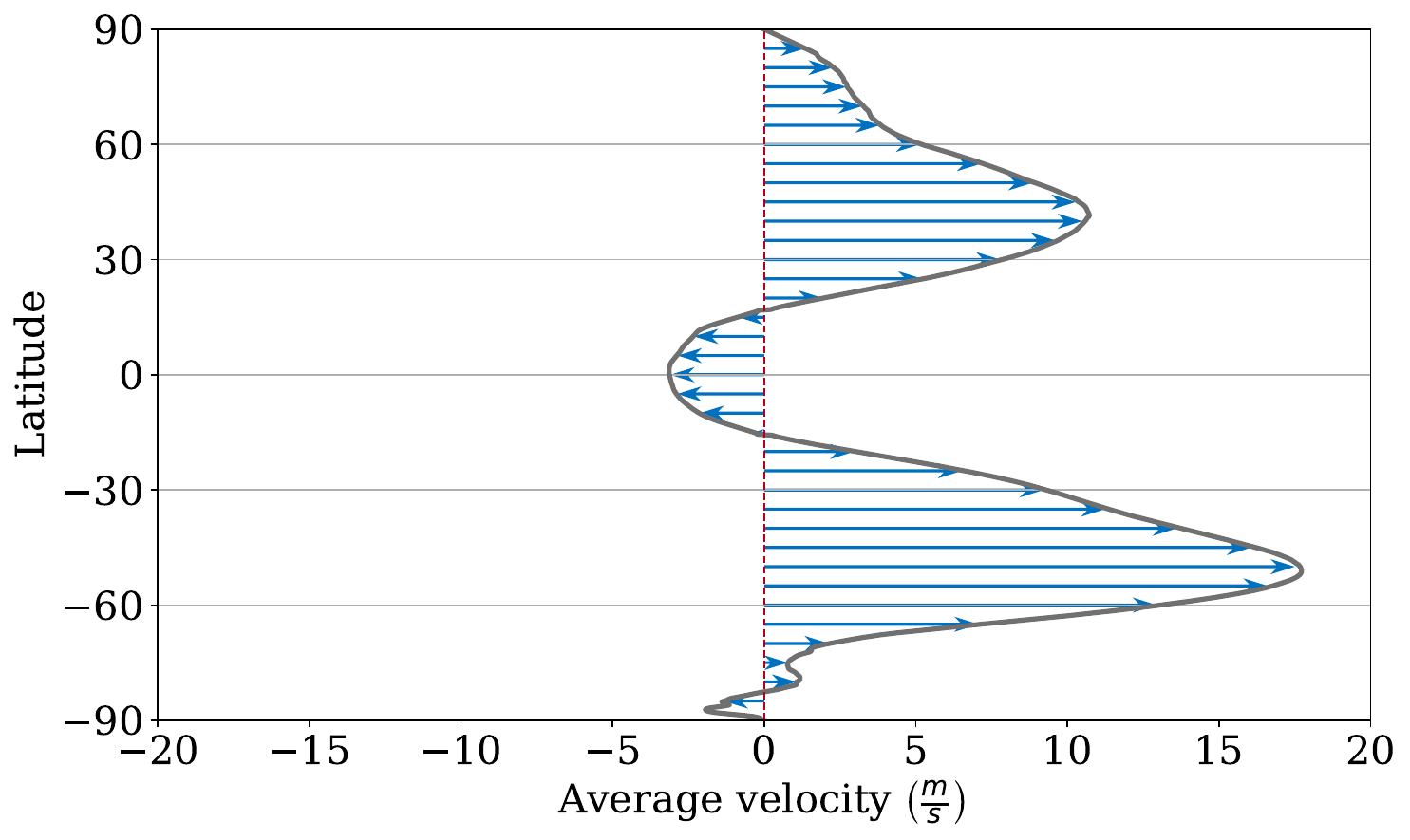}
	\caption{Wind velocity in east-west directions at each latitude from ERA5 reanalysis data~\cite{bell2021era5}. The average is calculated for the period 2011-2020 and integrated from 1000 to 200 hPa.}
	\label{fig: wind}
\end{figure}

The main physical processes we include for atmospheric dust particle transport are the advection along the dominant wind direction and diffusion towards surrounding regions. The most common model is the advection-diffusion equation 
\begin{linenomath}
	\begin{equation}
		\frac{\partial U}{\partial t} + \nabla \cdot (\textbf{v} U) - \nabla \cdot (D \nabla U) = 0,
		\label{advection-diffusion-general}
	\end{equation}
\end{linenomath}
where $U(\mathbf{x}, t)$ denotes the dust fluxes, $\textbf{v}$ is the wind field, and $D$ is the diffusion coefficient \citep{stocker2011introduction}. Since we only consider static dust fluxes averaged over thousands of years, we use the steady-state advection-diffusion equation 
\begin{linenomath}
	\begin{align}
		\nabla \cdot (\mathbf{v} u) - \nabla \cdot (D \nabla u)  &= 0
		\label{advection-diffusion-general-no-time}
	\end{align}
\end{linenomath}
where $u(\mathbf{x})$ denotes the time-averaged dust fluxes. The advection velocity is taken as the average wind speed in zonal direction $\mathbf{v} = (v_1, 0)$ taken from ERA5 reanalysis data (see Figure \ref{fig: wind}), which we used for both Holocene and LGM conditions. Detailed wind information is unavailable for the LGM, as there is no agreement in model simulations, except that the prevailing wind patterns were not significantly different from today~\cite{rojas2013sensitivity, kageyama2021pmip4}. Furthermore, we assume the diffusion parameter $D$ to be constant globally.

We model the Earth's surface as a sphere and write the advection-diffusion equation in spherical coordinates~\cite{fletcher2020semi}. For this purpose, let us denote the longitude by $\lambda \in (-\pi, \pi)$ and the latitude by $\theta \in (-\frac{\pi}{2}, \frac{\pi}{2})$ in radians. Hence, 
\begin{linenomath}
	\begin{align}
		\frac{v_1}{\cos(\theta)} \frac{\partial u}{\partial \lambda} - D \left( \frac{1}{\cos^2(\theta)} \frac{\partial^2 u}{\partial \lambda^2} + \frac{\partial^2 u}{\partial \theta ^2} - \tan(\theta) \frac{\partial u}{\partial \theta} \right) &= 0
		\label{spherical_randians}
	\end{align}
\end{linenomath}
is the model we use in the PINN's design, together with conditions at the boundaries of the spherical coordinate system. In the direction of the longitude, we have the periodic boundary conditions 
\begin{linenomath}
	\begin{align}
		\begin{cases}
			u(-\pi, \theta) = u(\pi, \theta) ,\\
			\frac{\partial u (-\pi, \theta)}{\partial \lambda} = \frac{\partial u  (\pi, \theta)}{\partial \lambda}.
			\label{eq: periodic boundary conditions}    
		\end{cases}
	\end{align}
\end{linenomath}
Concerning latitude, we model continuity in the poles by using
\begin{linenomath}
	\begin{align}
		\begin{cases}
			u(\lambda, -\frac{\pi}{2}) = u_\text{south}, \\
			u(\lambda, \frac{\pi}{2}) = u_\text{north},
		\end{cases}
		\label{eq: poles boundary conditions} 
	\end{align}
\end{linenomath}
where $u_\text{south}$ and $u_\text{north}$ are prescribed values at the south and north poles, respectively.

\subsection{Physics-Informed Neural Networks}
\label{section: Physics-Informed Neural Network}

\begin{figure}[!ht]
	\centering
	\includegraphics[width=0.9\textwidth]{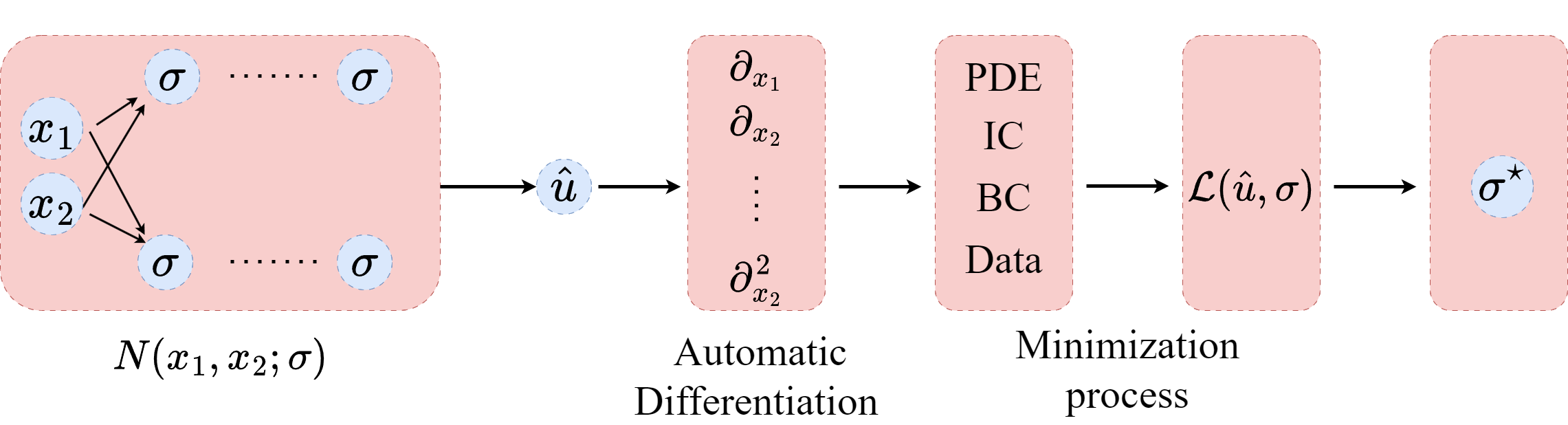}
	\caption{A general PINN training scheme where $x_1$ and $x_2$ are the input variables, $\sigma$ are the parameters of the neural network and $u$ the objective function. Automatic differentiation calculates the derivatives of the objective function, and the data and model fits are combined in a loss function $\mathcal{L}$. The minimization process through backpropagation gives updated parameters $\sigma^{\star}$. }
	\label{fig:PINN}
\end{figure}

A PINN is an algorithm based on supervised machine learning that can intrusively combine empirical data with physical information modeled as a PDE. In our case, the input of the PINN is the spatial location on the Earth's surface, and the output is the dust deposition. During the training process, the PINN optimizes a nonlinear function that fits both data and the PDE model. This optimization process requires specifying a loss function to find optimal parameters for the neural network. Since the PINN uses automatic differentiation to calculate the derivatives of the objective function, we can include our PDE and boundary conditions in the loss function. Figure~\ref{fig:PINN} shows a general workflow of the PINN.

\subsubsection{Loss function design}
\label{section: Loss Function Design}

The objective of a PINN is to fit a function $\hat{u}(\mathbf{x})$ to the PDE model and the empirical data. For this purpose, we define a combined loss function
\begin{linenomath}
	\begin{equation}
		\mathcal{L} = w_1 \mathcal{L}_{\text{data}} + \mathbf{w}_2 \cdot \mathcal{L}_{\text{model}},
		\label{loss_function_global}
	\end{equation}
\end{linenomath}
where $\mathcal{L}_{\text{data}}$ and $\mathcal{L}_{\text{model}}$ are the loss functions corresponding to the data and model, respectively, and $w_1$ and $\mathbf{w}_2$ are the respective weights. The data loss function is given by
\begin{linenomath}
	\begin{equation}
		\mathcal{L}_{\text{data}} = \left| \left| \hat{u}(\mathbf{\tilde{x}};\sigma) - \tilde{u}(\mathbf{\tilde{x}}) \right| \right|
		\label{data_loss_function}
	\end{equation}
\end{linenomath}
where $ \tilde{u}$ denotes the observations at the points $\tilde{x}_j$ for $j=1, \dots, N_{\text{data}}$ with $N_{\text{data}}$ the number of training items, and $\hat{u}$ denotes the PINN's estimation. We always use the standard $L_2$ norm. The model loss function consists of two parts, one for the PDE ($\mathcal{L}_{\text{PDE}}$) and one for the boundary conditions ($\mathcal{L}_{\text{BC}}$), that is, 
\begin{linenomath}
	\begin{equation}
		\mathbf{w}_2 \cdot \mathcal{L}_{\text{model}} = w_{2,1} \mathcal{L}_{\text{PDE}} + \mathbf{w}_{2,2} \cdot \mathcal{L}_{\text{BC}}.
		\label{loss_function_model}
	\end{equation}
\end{linenomath}
The PDE loss function quantifies the fit of the PINN's estimation regarding our model PDE~\eqref{spherical_randians} as
\begin{linenomath}
	\begin{equation}
		\mathcal{L}_{\text{PDE}} = \left| \left| \frac{v_1}{\cos(\theta)} \frac{\partial \hat{u}(\mathbf{\overline{x}})}{\partial \lambda} - D \left( \frac{1}{\cos^2(\theta)} \frac{\partial^2 \hat{u}(\mathbf{\overline{x}})}{\partial \lambda^2} + \frac{\partial^2 \hat{u}(\mathbf{\overline{x}})}{\partial \theta ^2} - \tan(\theta) \frac{\partial \hat{u}(\mathbf{\overline{x}})}{\partial \theta} \right) \right| \right|
		\label{eq: PDE loss function}        
	\end{equation}
\end{linenomath}
where $\overline{x}_j$ for $j=1 \dots, N_{\text{PDE}}$ are the collocation points in which the PINN evaluates the PDE. Finally, the loss function regarding the boundary conditions~\eqref{eq: periodic boundary conditions} and~\eqref{eq: poles boundary conditions} is given by
\begin{linenomath}
	\begin{align}
		\mathbf{w}_{2,2} \cdot \mathcal{L}_{\text{BC}} = \ & w_{2,2,1} \left| \left| \hat{u}\left(-\pi, \overline{\theta}\right) - \hat{u} \left(\pi, \overline{\theta}\right) \right| \right| + w_{2,2,2} \left| \left| \frac{\partial}{\partial \lambda} \hat{u}\left(-\pi, \overline{\theta}\right) - \frac{\partial}{\partial \lambda} \hat{u}\left(\pi, \overline{\theta}\right) \right| \right| + \nonumber \\
		& w_{2,2,3} \left| \left| \hat{u}\left(\overline{\lambda}, - \frac{\pi}{2} + \epsilon\right) - u_\text{south} \right| \right| + w_{2,2,4} \left| \left| \hat{u}\left(\overline{\lambda}, \frac{\pi}{2} - \epsilon\right) - u_\text{north}\right| \right|
	\end{align}
\end{linenomath}
where $\overline{\theta}_j$ and $\overline{\lambda}_j$ for $j=1 \dots, N_{\text{BC}}$ are the collocation points for the boundary conditions. Here, we include the offset $\epsilon$ to avoid numerical issues due to singularities in the spherical coordinate system at the poles. Depending on the specific problem setting, the relative emphasis between the data and the model can be adjusted by setting the weights $w_1$ and $\mathbf{w_2}$.

\subsubsection{Neural Network Design}
\label{section: Neural Network Design}

We normalize all empirical data with their $z$-scores in a logarithmic scale and normalize the computational domain to $[-2,2]$ for the longitude and $[-1, 1]$ for the latitude to improve the stability of the training algorithm and mitigate vanishing gradient pathologies~\cite{kissas2020machine}. The polar regions are excluded from training to avoid singularities in the model. Specifically, we consider latitudes between $-80$ to $+80$ degrees and impose a constant solution at these polar boundaries. This choice is a reasonable modeling approach since the excluded zones are ice-covered regions far away from any potential dust source, and the atmospheric circulation patterns are symmetric around the polar high-pressure systems. Hence, excluding the polar regions will not produce noticeable discrepancies in global dust transport.

We use a fully-connected neural network with five hidden layers, each containing 32 neurons with a Scaled Exponential Linear Unit (SELU) activation function and Glorot normal initializer. The optimizer is Adam, and the learning rate is set to $10^{-5}$. Computational experience of various network designs supports these choices as having sufficient expressiveness while avoiding overfitting of our specific empirical dataset. The PINN is trained on the entire dataset to incorporate all information available from the sparse and stratified data. We programmed the PINN using the DeepXDE library~\cite{lu2021deepxde}. This open-source Python package provides essential methods such as mesh creation, neural network definitions, initial and boundary conditions, and optimization algorithms.

\subsubsection{Model Parameters}
\label{section: Physical Parameters Design}

The PINN evaluates the PDE model loss term~\eqref{eq: PDE loss function} in collocation points that can be specified in arbitrary locations. We choose a regular grid with 5~degrees spacing between the points. This resolution is fine enough to capture the dominant advection and diffusion phenomena of the global dust deposition and sufficiently coarse to avoid overfitting. The boundary conditions are evaluated at 5~degrees resolution also.

Regarding the physical parameters in the PDE model~\eqref{spherical_randians}, the value of $v_1$ is taken from the wind data presented in Figure~\ref{fig: wind} and normalized by its maximum value. However, no reliable information on the diffusion coefficient $D$ is available in the literature on measurements or simulations of global dust fluxes. Hence, we opted to use an inverse problem approach in the PINN to infer this parameter from the empirical data~\cite{tartakovsky2020physics}. Using the same approach, we also estimated the values of $u_\text{north}$ and $u_\text{south}$. These parameters are necessary for the boundary conditions~\eqref{eq: poles boundary conditions}, but no measurement sites are available close to the poles. More specifically, these physical parameters are incorporated as unknown parameters in the neural network, and their values are approximated within the same training process as for the field estimation~\cite{raissi2019physics}. Hence, these physical parameters are optimized to fit the data and the model best.

As part of the design of the PINN, we have to choose the weights of the separate terms in the combined loss function~\eqref{loss_function_global}. These weights give relative importance between the data and the model in the training process. We normalize the weights by choosing a weight equal to one for the north and south pole boundary conditions. We use one-half each for the periodic boundary conditions since they are in the same location. Then, we set the data loss weight equal to ten to emphasize the importance of closely fitting the empirical data. Setting the model loss weight is arguably the most subjective decision~\cite{cuomo2022scientific}. Automatic approaches such as adaptive weights and multi-objective optimization have been proposed in the literature~\cite{wang2021understanding, wolff2022mopinns, meer2022optimally}. However, there is no consensus on setting the weights, and the best choice strongly depends on the application of interest. The common practice is manually setting the weights and evaluating the implications with domain knowledge~\cite{rasht2022physics}. In our case, a large model weight yields too smooth fields, and a small model weight overfits the observations. Section~\ref{sec:si:sensitivity} of the Supporting Information provides more details. We found a good balance for a weight of $w_{2,1} = 1$ for the PDE loss term.

\subsection{Kriging Method}

We compare our innovative PINN design with the more conventional kriging method. Kriging is a geostatistical interpolation technique that uses a weighted average of nearby samples~\cite{cressie1990origins}. We use the same kriging interpolation as performed in~\cite{lambert2015dust} for the DIRTMAP database. For brevity, we refer to that publication for the algorithmic details and model choices.

\section*{Data availability}
The paleoclimate measurements of dust deposition rates are available at the World Data Center PANGAEA, dataset~847983, as a supplement to~\cite{lambert2015dust}.

\section*{Code availability}
The software code is publicly available on GitHub: \url{https://github.com/evantwout/PINN-global-dust}.

\bibliographystyle{unsrtnat}
\bibliography{refs.bib}

\begin{thebibliography}{55}
\providecommand{\natexlab}[1]{#1}
\providecommand{\url}[1]{\texttt{#1}}
\expandafter\ifx\csname urlstyle\endcsname\relax
  \providecommand{\doi}[1]{doi: #1}\else
  \providecommand{\doi}{doi: \begingroup \urlstyle{rm}\Url}\fi

\bibitem[Mahowald(2011)]{mahowald2011aerosol}
Natalie Mahowald.
\newblock Aerosol indirect effect on biogeochemical cycles and climate.
\newblock \emph{Science}, 334\penalty0 (6057):\penalty0 794--796, 2011.
\newblock \doi{10.1126/science.1207374}.

\bibitem[Kok et~al.(2023)Kok, Storelvmo, Karydis, Adebiyi, Mahowald, Evan, He,
  and Leung]{kok2023mineral}
Jasper~F Kok, Trude Storelvmo, Vlassis~A Karydis, Adeyemi~A Adebiyi, Natalie~M
  Mahowald, Amato~T Evan, Cenlin He, and Danny~M Leung.
\newblock Mineral dust aerosol impacts on global climate and climate change.
\newblock \emph{Nature Reviews Earth \& Environment}, 4:\penalty0 71--86, 2023.
\newblock \doi{10.1038/s43017-022-00379-5}.

\bibitem[Shaffer and Lambert(2018)]{shaffer2018and}
Gary Shaffer and Fabrice Lambert.
\newblock In and out of glacial extremes by way of dust-climate feedbacks.
\newblock \emph{Proceedings of the National Academy of Sciences}, 115\penalty0
  (9):\penalty0 2026--2031, 2018.
\newblock \doi{10.1073/pnas.1708174115}.

\bibitem[Albani et~al.(2014)Albani, Mahowald, Perry, Scanza, Zender, Heavens,
  Maggi, Kok, and Otto-Bliesner]{albani2014improved}
S~Albani, NM~Mahowald, AT~Perry, RA~Scanza, CS~Zender, NG~Heavens, V~Maggi,
  JF~Kok, and BL~Otto-Bliesner.
\newblock Improved dust representation in the {C}ommunity {A}tmosphere {M}odel.
\newblock \emph{Journal of Advances in Modeling Earth Systems}, 6\penalty0
  (3):\penalty0 541--570, 2014.
\newblock \doi{10.1002/2013MS000279}.

\bibitem[Ohgaito et~al.(2018)Ohgaito, Abe-Ouchi, O'ishi, Takemura, Ito, Hajima,
  Watanabe, and Kawamiya]{ohgaito2018effect}
Rumi Ohgaito, Ayako Abe-Ouchi, Ryouta O'ishi, Toshihiko Takemura, Akinori Ito,
  Tomohiro Hajima, Shingo Watanabe, and Michio Kawamiya.
\newblock Effect of high dust amount on surface temperature during the {L}ast
  {G}lacial {M}aximum: a modelling study using {MIROC-ESM}.
\newblock \emph{Climate of the Past}, 14\penalty0 (11):\penalty0 1565--1581,
  2018.
\newblock \doi{10.5194/cp-14-1565-2018}.

\bibitem[Sueyoshi et~al.(2013)Sueyoshi, Ohgaito, Yamamoto, Chikamoto, Hajima,
  Okajima, Yoshimori, Abe, O'ishi, Saito, Watanabe, Kawamiya, and
  Abe-Ouchi]{sueyoshi2013set}
T.~Sueyoshi, R.~Ohgaito, A.~Yamamoto, M.~O. Chikamoto, T.~Hajima, H.~Okajima,
  M.~Yoshimori, M.~Abe, R.~O'ishi, F.~Saito, S.~Watanabe, M.~Kawamiya, and
  A.~Abe-Ouchi.
\newblock Set-up of the {PMIP3} paleoclimate experiments conducted using an
  {E}arth system model, {MIROC-ESM}.
\newblock \emph{Geoscientific Model Development}, 6\penalty0 (3):\penalty0
  819--836, 2013.
\newblock \doi{10.5194/gmd-6-819-2013}.

\bibitem[Yukimoto et~al.(2012)Yukimoto, Adachi, Hosaka, Sakami, Yoshimura,
  Hirabara, Tanaka, Shindo, Tsujino, Deushi, Mizuta, Yabu, Obata, Nakano,
  Koshiro, Ose, and Kitoh]{yukimoto2012new}
Seiji Yukimoto, Yukimasa Adachi, Masahiro Hosaka, Tomonori Sakami, Hiromasa
  Yoshimura, Mikitoshi Hirabara, Taichu~Y Tanaka, Eiki Shindo, Hiroyuki
  Tsujino, Makoto Deushi, Ryo Mizuta, Shoukichi Yabu, Atsushi Obata, Hideyuki
  Nakano, Tsuyoshi Koshiro, Tomoaki Ose, and Akio Kitoh.
\newblock A new global climate model of the {M}eteorological {R}esearch
  {I}nstitute: {MRI}-{CGCM3}. {M}odel description and basic performance.
\newblock \emph{Journal of the Meteorological Society of Japan}, 90\penalty0
  (A):\penalty0 23--64, 2012.
\newblock \doi{10.2151/jmsj.2012-A02}.

\bibitem[Hopcroft and Valdes(2019)]{hopcroft2019role}
Peter~O Hopcroft and Paul~J Valdes.
\newblock On the role of dust-climate feedbacks during the mid-{H}olocene.
\newblock \emph{Geophysical Research Letters}, 46\penalty0 (3):\penalty0
  1612--1621, 2019.
\newblock \doi{10.1029/2018GL080483}.

\bibitem[Kok et~al.(2014)Kok, Albani, Mahowald, and Ward]{kok2014improved}
JF~Kok, S~Albani, NM~Mahowald, and DS~Ward.
\newblock An improved dust emission model--{P}art 2: Evaluation in the
  {C}ommunity {E}arth {S}ystem {M}odel, with implications for the use of dust
  source functions.
\newblock \emph{Atmospheric Chemistry and Physics}, 14\penalty0 (23):\penalty0
  13043--13061, 2014.
\newblock \doi{10.5194/acp-14-13043-2014}.

\bibitem[Lambert et~al.(2015)Lambert, Tagliabue, Shaffer, Lamy, Winckler,
  Farias, Gallardo, and De~Pol-Holz]{lambert2015dust}
Fabrice Lambert, Alessandro Tagliabue, Gary Shaffer, Frank Lamy, Gisela
  Winckler, Laura Farias, Laura Gallardo, and Ricardo De~Pol-Holz.
\newblock Dust fluxes and iron fertilization in {H}olocene and {L}ast {G}lacial
  {M}aximum climates.
\newblock \emph{Geophysical Research Letters}, 42\penalty0 (14):\penalty0
  6014--6023, 2015.
\newblock \doi{10.1002/2015GL064250}.

\bibitem[Kohfeld and Harrison(2001)]{kohfeld2001dirtmap}
Karen~E Kohfeld and Sandy~P Harrison.
\newblock {DIRTMAP}: the geological record of dust.
\newblock \emph{Earth-Science Reviews}, 54\penalty0 (1-3):\penalty0 81--114,
  2001.
\newblock \doi{10.1016/S0012-8252(01)00042-3}.

\bibitem[Maher et~al.(2010)Maher, Prospero, Mackie, Gaiero, Hesse, and
  Balkanski]{maher2010global}
BA~Maher, JM~Prospero, D~Mackie, D~Gaiero, PP~Hesse, and Yves Balkanski.
\newblock Global connections between aeolian dust, climate and ocean
  biogeochemistry at the present day and at the last glacial maximum.
\newblock \emph{Earth-Science Reviews}, 99\penalty0 (1-2):\penalty0 61--97,
  2010.
\newblock \doi{10.1016/j.earscirev.2009.12.001}.

\bibitem[Cosentino et~al.(2024)Cosentino, Torre, Lambert, Albani,
  De~Vleeschouwer, and Bory]{cosentino2023paleodust}
Nicolás~J. Cosentino, Gabriela Torre, Fabrice Lambert, Samuel Albani,
  François De~Vleeschouwer, and Aloys Bory.
\newblock Paleo$\pm$dust: Quantifying uncertainty in paleo-dust deposition
  across archive types.
\newblock \emph{Earth System Science Data}, 16\penalty0 (2):\penalty0 941--959,
  2024.
\newblock \doi{10.5194/essd-2023-241}.

\bibitem[Kageyama et~al.(2017)Kageyama, Albani, Braconnot, Harrison, Hopcroft,
  Ivanovic, Lambert, Marti, Peltier, Peterschmitt, Roche, Tarasov, Zhang,
  Brady, Haywood, LeGrande, Lunt, Mahowald, Mikolajewicz, Nisancioglu,
  Otto-Bliesner, Renssen, Tomas, Zhang, Abe-Ouchi, Bartlein, Cao, Li, Lohmann,
  Ohgaito, Shi, Volodin, Yoshida, Zhang, and Zheng]{kageyama2017pmip4}
M.~Kageyama, S.~Albani, P.~Braconnot, S.~P. Harrison, P.~O. Hopcroft, R.~F.
  Ivanovic, F.~Lambert, O.~Marti, W.~R. Peltier, J.-Y. Peterschmitt, D.~M.
  Roche, L.~Tarasov, X.~Zhang, E.~C. Brady, A.~M. Haywood, A.~N. LeGrande,
  D.~J. Lunt, N.~M. Mahowald, U.~Mikolajewicz, K.~H. Nisancioglu, B.~L.
  Otto-Bliesner, H.~Renssen, R.~A. Tomas, Q.~Zhang, A.~Abe-Ouchi, P.~J.
  Bartlein, J.~Cao, Q.~Li, G.~Lohmann, R.~Ohgaito, X.~Shi, E.~Volodin,
  K.~Yoshida, X.~Zhang, and W.~Zheng.
\newblock The {PMIP4} contribution to {CMIP6} -- {P}art 4: Scientific
  objectives and experimental design of the {PMIP4}-{CMIP6} {L}ast {G}lacial
  {M}aximum experiments and {PMIP4} sensitivity experiments.
\newblock \emph{Geoscientific Model Development}, 10\penalty0 (11):\penalty0
  4035--4055, 2017.
\newblock \doi{10.5194/gmd-10-4035-2017}.

\bibitem[Lambert et~al.(2021)Lambert, Opazo, Ridgwell, Winckler, Lamy, Shaffer,
  Kohfeld, Ohgaito, Albani, and Abe-Ouchi]{lambert2021regional}
Fabrice Lambert, Natalia Opazo, Andy Ridgwell, Gisela Winckler, Frank Lamy,
  Gary Shaffer, Karen Kohfeld, Rumi Ohgaito, Samuel Albani, and Ayako
  Abe-Ouchi.
\newblock Regional patterns and temporal evolution of ocean iron fertilization
  and {CO2} drawdown during the last glacial termination.
\newblock \emph{Earth and Planetary Science Letters}, 554:\penalty0 116675,
  2021.
\newblock \doi{10.1016/j.epsl.2020.116675}.

\bibitem[Saini et~al.(2021)Saini, Kvale, Chase, Kohfeld, Meissner, and
  Menviel]{saini2021southern}
Himadri Saini, Karin Kvale, Zanna Chase, Karen~E Kohfeld, Katrin~J Meissner,
  and Laurie Menviel.
\newblock Southern ocean ecosystem response to {L}ast {G}lacial {M}aximum
  boundary conditions.
\newblock \emph{Paleoceanography and Paleoclimatology}, 36\penalty0
  (7):\penalty0 e2020PA004075, 2021.
\newblock \doi{10.1029/2020PA004075}.

\bibitem[Raissi et~al.(2019)Raissi, Perdikaris, and
  Karniadakis]{raissi2019physics}
Maziar Raissi, Paris Perdikaris, and George~E Karniadakis.
\newblock Physics-informed neural networks: A deep learning framework for
  solving forward and inverse problems involving nonlinear partial differential
  equations.
\newblock \emph{Journal of Computational Physics}, 378:\penalty0 686--707,
  2019.
\newblock \doi{10.1016/j.jcp.2018.10.045}.

\bibitem[Karniadakis et~al.(2021)Karniadakis, Kevrekidis, Lu, Perdikaris, Wang,
  and Yang]{karniadakis2021physics}
George~Em Karniadakis, Ioannis~G Kevrekidis, Lu~Lu, Paris Perdikaris, Sifan
  Wang, and Liu Yang.
\newblock Physics-informed machine learning.
\newblock \emph{Nature Reviews Physics}, 3\penalty0 (6):\penalty0 422--440,
  2021.
\newblock \doi{10.1038/s42254-021-00314-5}.

\bibitem[Sahli~Costabal et~al.(2020)Sahli~Costabal, Yang, Perdikaris, Hurtado,
  and Kuhl]{sahli2020physics}
Francisco Sahli~Costabal, Yibo Yang, Paris Perdikaris, Daniel~E Hurtado, and
  Ellen Kuhl.
\newblock Physics-informed neural networks for cardiac activation mapping.
\newblock \emph{Frontiers in Physics}, 8:\penalty0 42, 2020.
\newblock \doi{10.3389/fphy.2020.00042}.

\bibitem[Jagtap et~al.(2020)Jagtap, Kharazmi, and
  Karniadakis]{jagtap2020conservative}
Ameya~D Jagtap, Ehsan Kharazmi, and George~Em Karniadakis.
\newblock Conservative physics-informed neural networks on discrete domains for
  conservation laws: Applications to forward and inverse problems.
\newblock \emph{Computer Methods in Applied Mechanics and Engineering},
  365:\penalty0 113028, 2020.
\newblock \doi{10.1016/j.cma.2020.113028}.

\bibitem[Yang and Perdikaris(2019)]{yang2019adversarial}
Yibo Yang and Paris Perdikaris.
\newblock Adversarial uncertainty quantification in physics-informed neural
  networks.
\newblock \emph{Journal of Computational Physics}, 394:\penalty0 136--152,
  2019.
\newblock \doi{10.1016/j.jcp.2019.05.027}.

\bibitem[Cuomo et~al.(2022)Cuomo, Di~Cola, Giampaolo, Rozza, Raissi, and
  Piccialli]{cuomo2022scientific}
Salvatore Cuomo, Vincenzo~Schiano Di~Cola, Fabio Giampaolo, Gianluigi Rozza,
  Maziar Raissi, and Francesco Piccialli.
\newblock Scientific machine learning through physics--informed neural
  networks: where we are and what’s next.
\newblock \emph{Journal of Scientific Computing}, 92\penalty0 (3):\penalty0 88,
  2022.
\newblock \doi{10.1007/s10915-022-01939-z}.

\bibitem[He and Tartakovsky(2021)]{he2021physics}
QiZhi He and Alexandre~M Tartakovsky.
\newblock Physics-informed neural network method for forward and backward
  advection-dispersion equations.
\newblock \emph{Water Resources Research}, 57\penalty0 (7):\penalty0
  e2020WR029479, 2021.
\newblock \doi{10.1029/2020WR029479}.

\bibitem[Okazaki et~al.(2022)Okazaki, Ito, Hirahara, and
  Ueda]{okazaki2022physics}
Tomohisa Okazaki, Takeo Ito, Kazuro Hirahara, and Naonori Ueda.
\newblock Physics-informed deep learning approach for modeling crustal
  deformation.
\newblock \emph{Nature Communications}, 13\penalty0 (1):\penalty0 7092, 2022.
\newblock \doi{10.1038/s41467-022-34922-1}.

\bibitem[Penwarden et~al.(2022)Penwarden, Zhe, Narayan, and
  Kirby]{penwarden2022multifidelity}
Michael Penwarden, Shandian Zhe, Akil Narayan, and Robert~M Kirby.
\newblock Multifidelity modeling for physics-informed neural networks
  ({PINN}s).
\newblock \emph{Journal of Computational Physics}, 451:\penalty0 110844, 2022.
\newblock \doi{10.1016/j.jcp.2021.110844}.

\bibitem[Smith et~al.(2022)Smith, Ross, Azizzadenesheli, and
  Muir]{smith2022hyposvi}
Jonthan~D Smith, Zachary~E Ross, Kamyar Azizzadenesheli, and Jack~B Muir.
\newblock Hypo{SVI}: Hypocentre inversion with stein variational inference and
  physics informed neural networks.
\newblock \emph{Geophysical Journal International}, 228\penalty0 (1):\penalty0
  698--710, 2022.
\newblock \doi{10.1093/gji/ggab309}.

\bibitem[Cai et~al.(2021)Cai, Wang, Fuest, Jeon, Gray, and
  Karniadakis]{cai2021flow}
Shengze Cai, Zhicheng Wang, Frederik Fuest, Young~Jin Jeon, Callum Gray, and
  George~Em Karniadakis.
\newblock Flow over an espresso cup: inferring 3-{D} velocity and pressure
  fields from tomographic background oriented schlieren via physics-informed
  neural networks.
\newblock \emph{Journal of Fluid Mechanics}, 915:\penalty0 A102, 2021.
\newblock \doi{10.1017/jfm.2021.135}.

\bibitem[Kissas et~al.(2020)Kissas, Yang, Hwuang, Witschey, Detre, and
  Perdikaris]{kissas2020machine}
Georgios Kissas, Yibo Yang, Eileen Hwuang, Walter~R Witschey, John~A Detre, and
  Paris Perdikaris.
\newblock Machine learning in cardiovascular flows modeling: Predicting
  arterial blood pressure from non-invasive {4D} flow {MRI} data using
  physics-informed neural networks.
\newblock \emph{Computer Methods in Applied Mechanics and Engineering},
  358:\penalty0 112623, 2020.
\newblock \doi{10.1016/j.cma.2019.112623}.

\bibitem[Linka et~al.(2022)Linka, Sch{\"a}fer, Meng, Zou, Karniadakis, and
  Kuhl]{linka2022bayesian}
Kevin Linka, Amelie Sch{\"a}fer, Xuhui Meng, Zongren Zou, George~Em
  Karniadakis, and Ellen Kuhl.
\newblock Bayesian physics informed neural networks for real-world nonlinear
  dynamical systems.
\newblock \emph{Computer Methods in Applied Mechanics and Engineering},
  402:\penalty0 115346, 2022.
\newblock \doi{10.1016/j.cma.2022.115346}.

\bibitem[Tuia et~al.(2021)Tuia, Roscher, Wegner, Jacobs, Zhu, and
  Camps-Valls]{tuia2021toward}
Devis Tuia, Ribana Roscher, Jan~Dirk Wegner, Nathan Jacobs, Xiaoxiang Zhu, and
  Gustau Camps-Valls.
\newblock Toward a collective agenda on {AI} for earth science data analysis.
\newblock \emph{IEEE Geoscience and Remote Sensing Magazine}, 9\penalty0
  (2):\penalty0 88--104, 2021.
\newblock \doi{10.1109/MGRS.2020.3043504}.

\bibitem[Rolnick et~al.(2022)Rolnick, Donti, Kaack, Kochanski, Lacoste,
  Sankaran, Ross, Milojevic-Dupont, Jaques, Waldman-Brown, Luccioni, Maharaj,
  Sherwin, Mukkavilli, Kording, Gomes, Ng, Hassabis, Platt, Creutzig, Chayes,
  and Bengio]{rolnick2022tackling}
David Rolnick, Priya~L Donti, Lynn~H Kaack, Kelly Kochanski, Alexandre Lacoste,
  Kris Sankaran, Andrew~Slavin Ross, Nikola Milojevic-Dupont, Natasha Jaques,
  Anna Waldman-Brown, Alexandra~Sasha Luccioni, Tegan Maharaj, Evan~D. Sherwin,
  S.~Karthik Mukkavilli, Konrad~P. Kording, Carla~P. Gomes, Andrew~Y. Ng, Demis
  Hassabis, John~C. Platt, Felix Creutzig, Jennifer Chayes, and Yoshua Bengio.
\newblock Tackling climate change with machine learning.
\newblock \emph{ACM Computing Surveys (CSUR)}, 55\penalty0 (2):\penalty0 1--96,
  2022.
\newblock \doi{10.1145/3485128}.

\bibitem[Cosentino et~al.(2023)Cosentino, Opazo, Lambert, Osses, and van~'t
  Wout]{cosentino2023global}
NJ~Cosentino, NE~Opazo, F~Lambert, A~Osses, and E~van~'t Wout.
\newblock Global-krigger: A global kriging interpolation toolbox with
  paleoclimatology examples.
\newblock \emph{Geochemistry, Geophysics, Geosystems}, 24\penalty0
  (6):\penalty0 e2022GC010821, 2023.
\newblock \doi{10.1029/2022GC010821}.

\bibitem[Struve et~al.(2020)Struve, Pahnke, Lamy, Wengler, B{\"o}ning, and
  Winckler]{struve2020circumpolar}
Torben Struve, Katharina Pahnke, Frank Lamy, Marc Wengler, Philipp B{\"o}ning,
  and Gisela Winckler.
\newblock A circumpolar dust conveyor in the glacial southern ocean.
\newblock \emph{Nature communications}, 11\penalty0 (1):\penalty0 5655, 2020.
\newblock \doi{10.1038/s41467-020-18858-y}.

\bibitem[Swap et~al.(1992)Swap, Garstang, Greco, Talbot, and
  K{\aa}llberg]{swap1992saharan}
Robert Swap, Michael Garstang, S~Greco, R~Talbot, and P~K{\aa}llberg.
\newblock Saharan dust in the {A}mazon basin.
\newblock \emph{Tellus B}, 44\penalty0 (2):\penalty0 133--149, 1992.
\newblock \doi{10.3402/tellusb.v44i2.15434}.

\bibitem[Bristow et~al.(2010)Bristow, Hudson-Edwards, and
  Chappell]{bristow2010fertilizing}
Charlie~S Bristow, Karen~A Hudson-Edwards, and Adrian Chappell.
\newblock Fertilizing the {A}mazon and equatorial {A}tlantic with {W}est
  {A}frican dust.
\newblock \emph{Geophysical Research Letters}, 37\penalty0 (14), 2010.
\newblock \doi{10.1029/2010GL043486}.

\bibitem[Wang et~al.(2022)Wang, Yu, and Perdikaris]{wang2022when}
Sifan Wang, Xinling Yu, and Paris Perdikaris.
\newblock When and why {PINNs} fail to train: A neural tangent kernel
  perspective.
\newblock \emph{Journal of Computational Physics}, 449:\penalty0 110768, 2022.
\newblock \doi{10.1016/j.jcp.2021.110768}.

\bibitem[Krishnapriyan et~al.(2021)Krishnapriyan, Gholami, Zhe, Kirby, and
  Mahoney]{krishnapriyan2021characterizing}
Aditi Krishnapriyan, Amir Gholami, Shandian Zhe, Robert Kirby, and Michael~W
  Mahoney.
\newblock Characterizing possible failure modes in physics-informed neural
  networks.
\newblock In \emph{35th Conference on Neural Information Processing Systems
  (NeurIPS 2021)}, 2021.

\bibitem[Gao et~al.(2022)Gao, Zahr, and Wang]{gao2022physics}
Han Gao, Matthew~J Zahr, and Jian-Xun Wang.
\newblock Physics-informed graph neural {G}alerkin networks: A unified
  framework for solving {PDE}-governed forward and inverse problems.
\newblock \emph{Computer Methods in Applied Mechanics and Engineering},
  390:\penalty0 114502, 2022.
\newblock \doi{10.1016/j.cma.2021.114502}.

\bibitem[Jiang et~al.(2021)Jiang, Meinert, Jord{\~a}o, Weisser, Holgate, Lavin,
  L{\"u}tjens, Newman, Wainwright, Walker, and Barnard]{jiang2021digital}
Peishi Jiang, Nis Meinert, Helga Jord{\~a}o, Constantin Weisser, Simon Holgate,
  Alexander Lavin, Bj{\"o}rn L{\"u}tjens, Dava Newman, Haruko Wainwright,
  Catherine Walker, and Patrick Barnard.
\newblock Digital twin earth--coasts: Developing a fast and physics-informed
  surrogate model for coastal floods via neural operators.
\newblock In \emph{Fourth Workshop on Machine Learning and the Physical
  Sciences (NeurIPS 2021)}, 2021.

\bibitem[Jin et~al.(2021)Jin, Cai, Li, and Karniadakis]{jin2021nsfnets}
Xiaowei Jin, Shengze Cai, Hui Li, and George~Em Karniadakis.
\newblock {NSF}nets ({N}avier-{S}tokes flow nets): Physics-informed neural
  networks for the incompressible {N}avier-{S}tokes equations.
\newblock \emph{Journal of Computational Physics}, 426:\penalty0 109951, 2021.
\newblock \doi{10.1016/j.jcp.2020.109951}.

\bibitem[Yang et~al.(2021)Yang, Meng, and Karniadakis]{yang2021bayesian}
Liu Yang, Xuhui Meng, and George~Em Karniadakis.
\newblock B-{PINN}s: Bayesian physics-informed neural networks for forward and
  inverse {PDE} problems with noisy data.
\newblock \emph{Journal of Computational Physics}, 425:\penalty0 109913, 2021.
\newblock \doi{10.1016/j.jcp.2020.109913}.

\bibitem[Kashinath et~al.(2021)Kashinath, Mustafa, Albert, Wu, C.,
  Esmaeilzadeh, Azizzadenesheli, Wang, Chattopadhyay, Singh, Manepalli,
  Chirila, Yu, Walters, White, Xiao, Tchelepi, Marcus, Anandkumar, Hassanzadeh,
  and Prabhat]{kashinath2021physics}
K.~Kashinath, M.~Mustafa, A.~Albert, J-L. Wu, Jiang. C., S.~Esmaeilzadeh,
  K.~Azizzadenesheli, R.~Wang, A.~Chattopadhyay, A.~Singh, A.~Manepalli,
  D.~Chirila, R.~Yu, R.~Walters, B.~White, H.~Xiao, H.~A. Tchelepi, P.~Marcus,
  A.~Anandkumar, P.~Hassanzadeh, and Prabhat.
\newblock Physics-informed machine learning: case studies for weather and
  climate modelling.
\newblock \emph{Philosophical Transactions of the Royal Society A},
  379\penalty0 (2194):\penalty0 20200093, 2021.
\newblock \doi{10.1098/rsta.2020.0093}.

\bibitem[Stocker(2011)]{stocker2011introduction}
Thomas Stocker.
\newblock Introduction to climate modelling.
\newblock In \emph{Advances in Geophysical and Environmental Mechanics and
  Mathematics}. Springer, Berlin, Germany, 2011.
\newblock \doi{10.1007/978-3-642-00773-6}.

\bibitem[Rojas(2013)]{rojas2013sensitivity}
Maisa Rojas.
\newblock Sensitivity of southern hemisphere circulation to {LGM} and
  4$\times${CO2} climates.
\newblock \emph{Geophysical Research Letters}, 40\penalty0 (5):\penalty0
  965--970, 2013.
\newblock \doi{10.1002/grl.50195}.

\bibitem[Kageyama et~al.(2021)Kageyama, Harrison, Kapsch, Lofverstrom, Lora,
  Mikolajewicz, Sherriff-Tadano, Vadsaria, Abe-Ouchi, Bouttes, Chandan,
  Gregoire, Ivanovic, Izumi, LeGrande, Lhardy, Lohmann, Morozova, Ohgaito,
  Paul, Peltier, Poulsen, Quiquet, Roche, Shi, Tierney, Valdes, Volodin, and
  Zhu]{kageyama2021pmip4}
Masa Kageyama, Sandy~P Harrison, Marie-L Kapsch, Marcus Lofverstrom, Juan~M
  Lora, Uwe Mikolajewicz, Sam Sherriff-Tadano, Tristan Vadsaria, Ayako
  Abe-Ouchi, Nathaelle Bouttes, Deepak Chandan, Lauren~J. Gregoire, Ruza~F.
  Ivanovic, Kenji Izumi, Allegra~N. LeGrande, Fanny Lhardy, Gerrit Lohmann,
  Polina~A. Morozova, Rumi Ohgaito, André Paul, W.~Richard Peltier,
  Christopher~J. Poulsen, Aurélien Quiquet, Didier~M. Roche, Xiaoxu Shi,
  Jessica~E. Tierney, Paul~J. Valdes, Evgeny Volodin, and Jiang Zhu.
\newblock The {PMIP4} {L}ast {G}lacial {M}aximum experiments: preliminary
  results and comparison with the {PMIP3} simulations.
\newblock \emph{Climate of the Past}, 17\penalty0 (3):\penalty0 1065--1089,
  2021.
\newblock \doi{10.5194/cp-17-1065-2021}.

\bibitem[Bell et~al.(2021)Bell, Hersbach, Simmons, Berrisford, Dahlgren,
  Horányi, Muñoz-Sabater, Nicolas, Radu, Schepers, Soci, Villaume, Bidlot,
  Haimberger, Woollen, Buontempo, and Thépaut]{bell2021era5}
Bill Bell, Hans Hersbach, Adrian Simmons, Paul Berrisford, Per Dahlgren,
  András Horányi, Joaquín Muñoz-Sabater, Julien Nicolas, Raluca Radu,
  Dinand Schepers, Cornel Soci, Sebastien Villaume, Jean-Raymond Bidlot, Leo
  Haimberger, Jack Woollen, Carlo Buontempo, and Jean-Noël Thépaut.
\newblock The {ERA5} global reanalysis: Preliminary extension to 1950.
\newblock \emph{Quarterly Journal of the Royal Meteorological Society},
  147\penalty0 (741):\penalty0 4186--4227, 2021.
\newblock \doi{10.1002/qj.4174}.

\bibitem[Fletcher(2020)]{fletcher2020semi}
Steven~J. Fletcher.
\newblock Semi-{L}agrangian methods on a sphere.
\newblock In \emph{Semi-{L}agrangian Advection Methods and Their Applications
  in Geoscience}, chapter~12, pages 381--469. Elsevier, 2020.
\newblock \doi{10.1016/B978-0-12-817222-3.00016-5}.

\bibitem[Lu et~al.(2021)Lu, Meng, Mao, and Karniadakis]{lu2021deepxde}
Lu~Lu, Xuhui Meng, Zhiping Mao, and George~Em Karniadakis.
\newblock Deep{XDE}: A deep learning library for solving differential
  equations.
\newblock \emph{SIAM Review}, 63\penalty0 (1):\penalty0 208--228, 2021.
\newblock \doi{10.1137/19M1274067}.

\bibitem[Tartakovsky et~al.(2020)Tartakovsky, Marrero, Perdikaris, Tartakovsky,
  and Barajas-Solano]{tartakovsky2020physics}
Alexandre~M Tartakovsky, C~Ortiz Marrero, Paris Perdikaris, Guzel~D
  Tartakovsky, and David Barajas-Solano.
\newblock Physics-informed deep neural networks for learning parameters and
  constitutive relationships in subsurface flow problems.
\newblock \emph{Water Resources Research}, 56\penalty0 (5):\penalty0
  e2019WR026731, 2020.
\newblock \doi{10.1029/2019WR026731}.

\bibitem[Wang et~al.(2021)Wang, Teng, and Perdikaris]{wang2021understanding}
Sifan Wang, Yujun Teng, and Paris Perdikaris.
\newblock Understanding and mitigating gradient flow pathologies in
  physics-informed neural networks.
\newblock \emph{SIAM Journal on Scientific Computing}, 43\penalty0
  (5):\penalty0 A3055--A3081, 2021.
\newblock \doi{10.1137/20M1318043}.

\bibitem[de~Wolff et~al.(2022)de~Wolff, Lincopi, Mart{\'\i}, and
  Sanchez-Pi]{wolff2022mopinns}
Taco de~Wolff, Hugo~Carrillo Lincopi, Luis Mart{\'\i}, and Nayat Sanchez-Pi.
\newblock {MOPINNs}: an evolutionary multi-objective approach to
  physics-informed neural networks.
\newblock In \emph{Proceedings of the Genetic and Evolutionary Computation
  Conference Companion}, pages 228--231, 2022.
\newblock \doi{10.1145/3520304.3529071}.

\bibitem[van~der Meer et~al.(2022)van~der Meer, Oosterlee, and
  Borovykh]{meer2022optimally}
Remco van~der Meer, Cornelis~W Oosterlee, and Anastasia Borovykh.
\newblock Optimally weighted loss functions for solving {PDEs} with neural
  networks.
\newblock \emph{Journal of Computational and Applied Mathematics},
  405:\penalty0 113887, 2022.
\newblock \doi{10.1016/j.cam.2021.113887}.

\bibitem[Rasht-Behesht et~al.(2022)Rasht-Behesht, Huber, Shukla, and
  Karniadakis]{rasht2022physics}
Majid Rasht-Behesht, Christian Huber, Khemraj Shukla, and George~Em
  Karniadakis.
\newblock Physics-informed neural networks ({PINN}s) for wave propagation and
  full waveform inversions.
\newblock \emph{Journal of Geophysical Research: Solid Earth}, 127\penalty0
  (5):\penalty0 e2021JB023120, 2022.
\newblock \doi{10.1029/2021JB023120}.

\bibitem[Cressie(1990)]{cressie1990origins}
Noel Cressie.
\newblock The origins of kriging.
\newblock \emph{Mathematical Geology}, 22:\penalty0 239--252, 1990.
\newblock \doi{10.1007/BF00889887}.

\bibitem[Albani et~al.(2016)Albani, Mahowald, Murphy, Raiswell, Moore,
  Anderson, McGee, Bradtmiller, Delmonte, Hesse, and
  Mayewski]{albani2016paleodust}
S.~Albani, N.~M. Mahowald, L.~N. Murphy, R.~Raiswell, J.~K. Moore, R.~F.
  Anderson, D.~McGee, L.~I. Bradtmiller, B.~Delmonte, P.~P. Hesse, and P.~A.
  Mayewski.
\newblock Paleodust variability since the {L}ast {G}lacial {M}aximum and
  implications for iron inputs to the ocean.
\newblock \emph{Geophysical Research Letters}, 43\penalty0 (8):\penalty0
  3944--3954, 2016.
\newblock \doi{10.1002/2016GL067911}.

\end{thebibliography}

\appendix

\section{Supporting information}

\subsection{Statistical distributions of global reconstructions}
\label{sec:si:statistical_distribution}

\begin{figure}[!ht] 
	\centering
	(a)
    \includegraphics[width=0.44\textwidth]{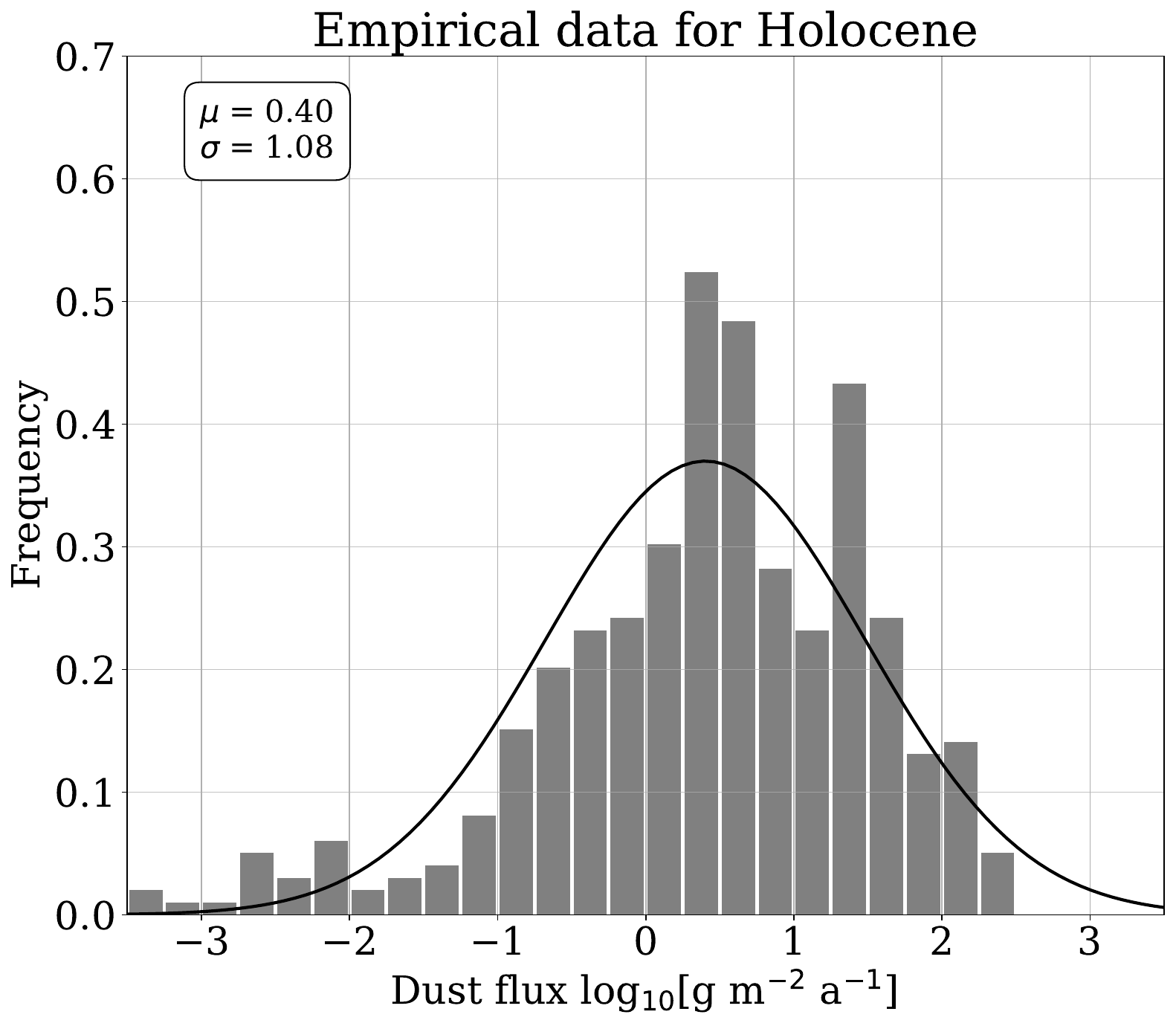}
	(b)
	\includegraphics[width=0.44\textwidth]{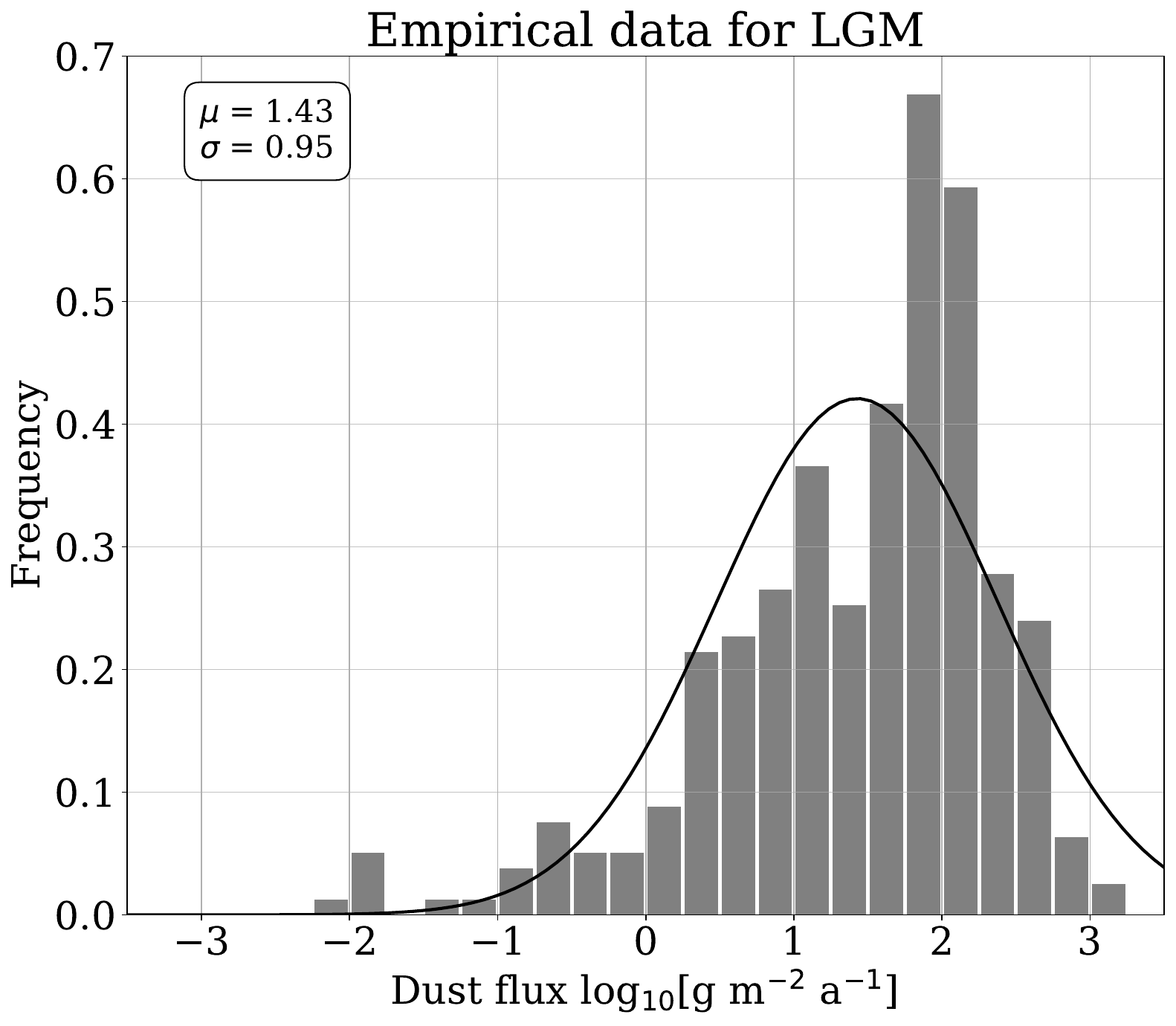}
	\\ (c)
	\includegraphics[width=0.44\textwidth]{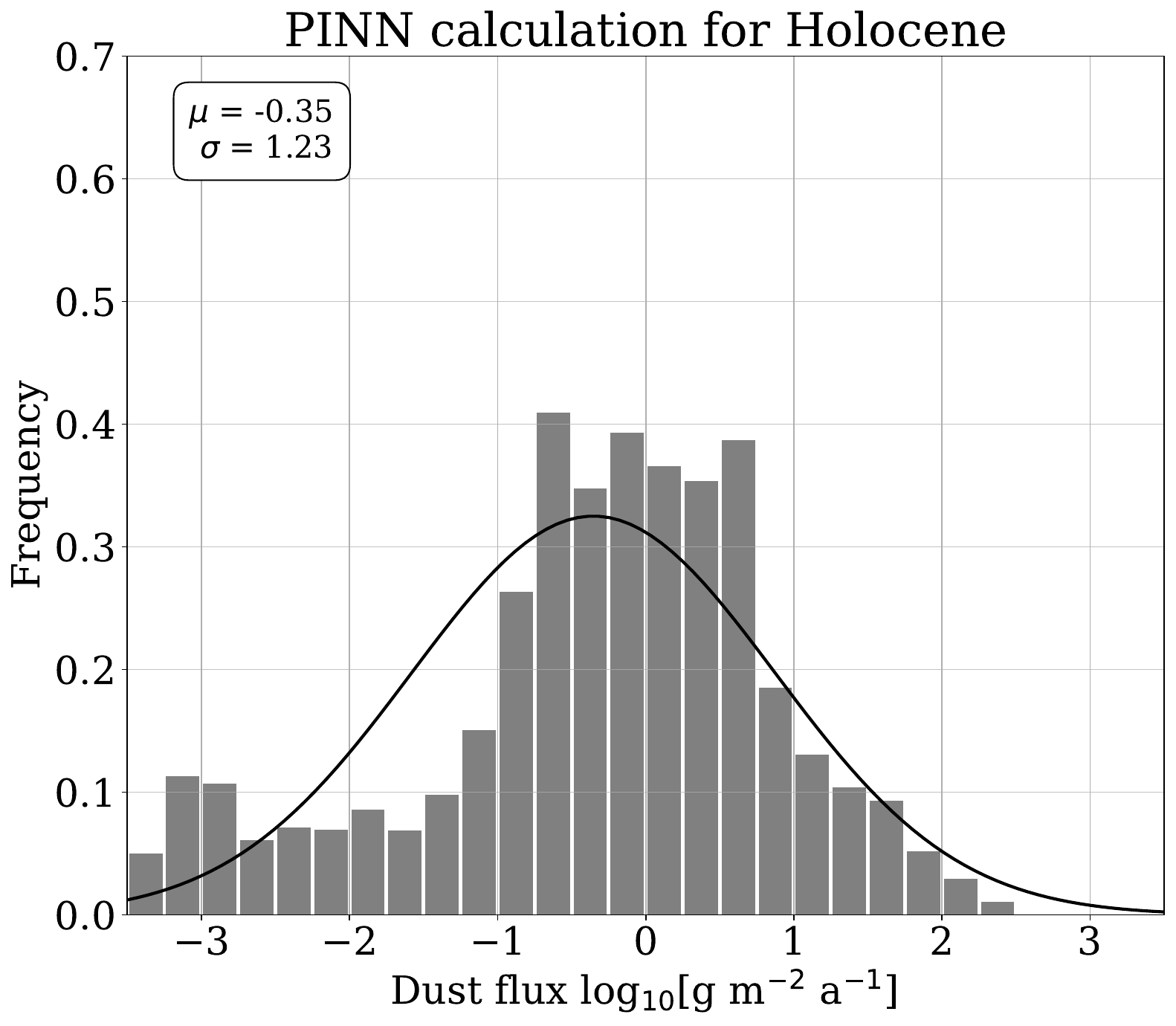}
	(d)
	\includegraphics[width=0.44\textwidth]{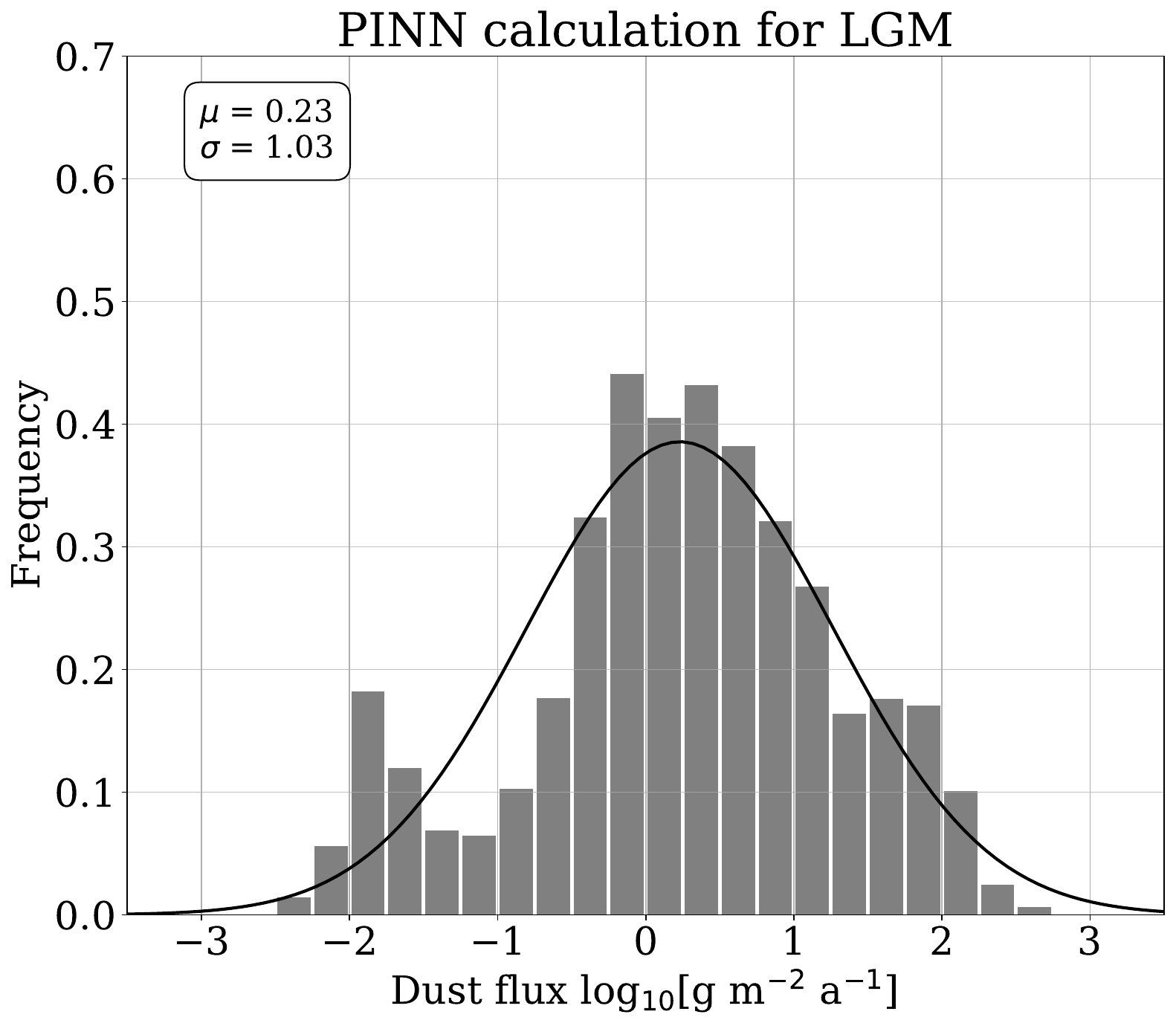}
	\\ (e)
	\includegraphics[width=0.44\textwidth]{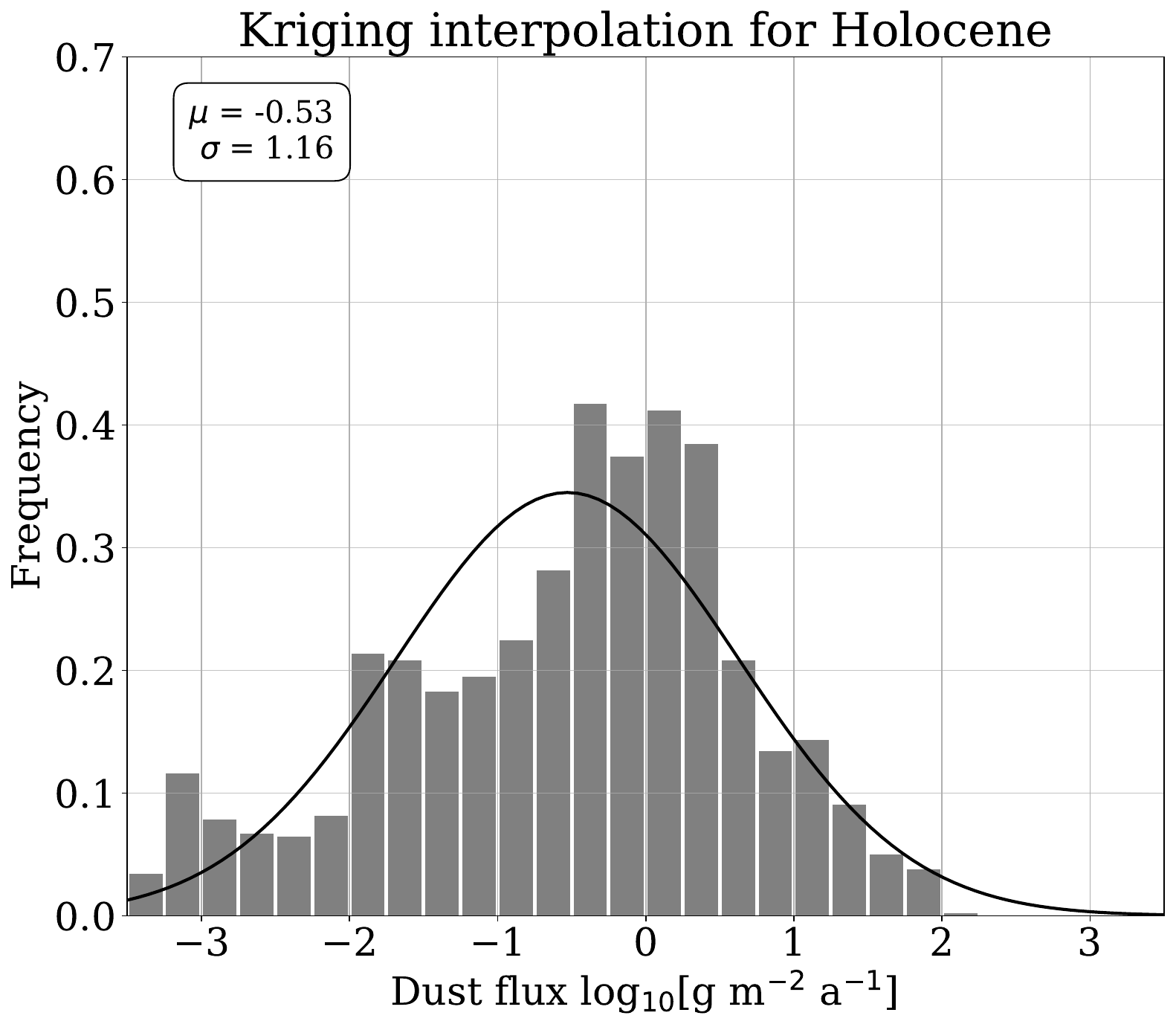}
	(f)
	\includegraphics[width=0.44\textwidth]{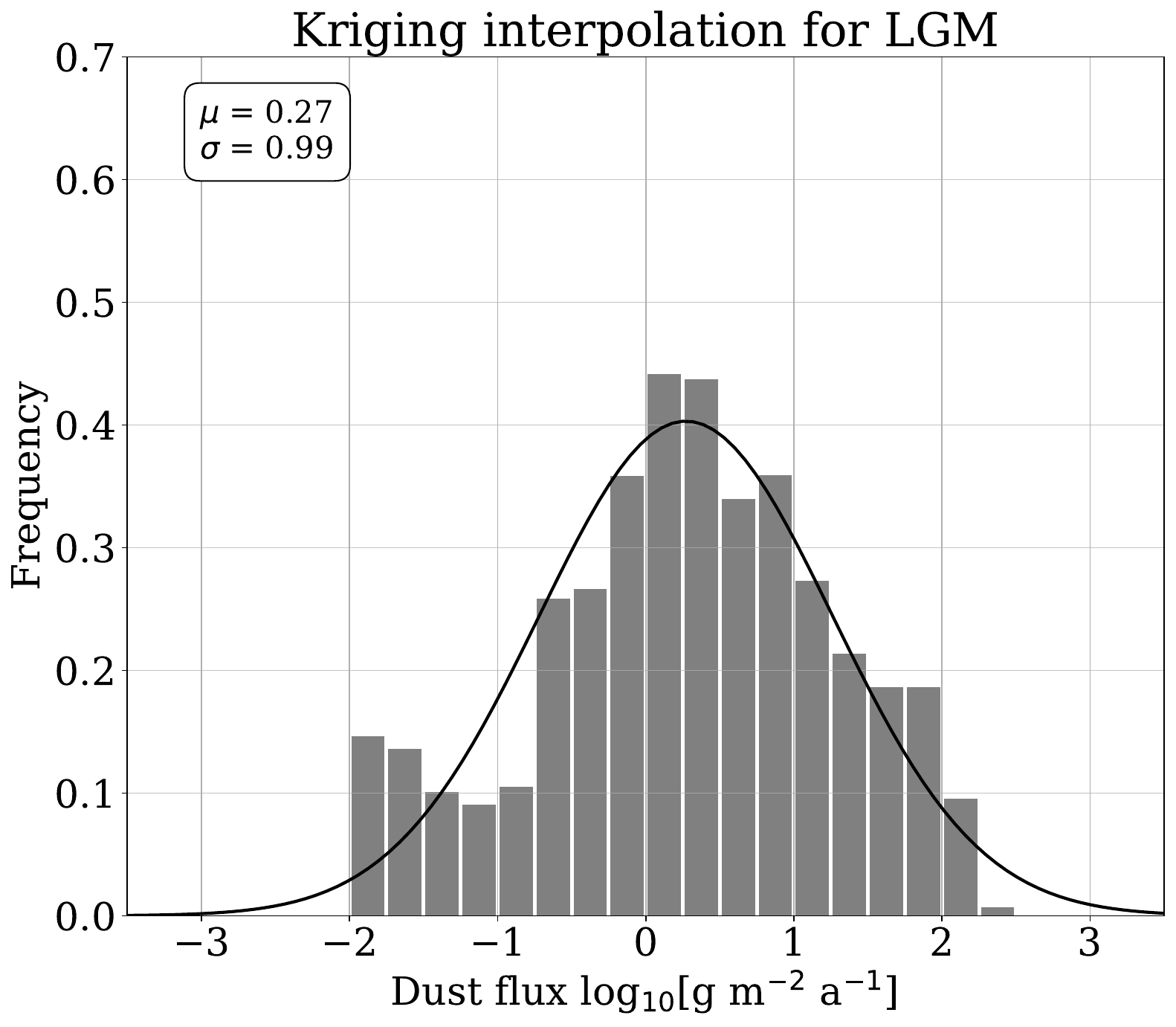}
	\caption{A log-normal fit of the data with mean~$\mu$ and standard deviation~$\sigma$, for: (a) and (b) the empirical data; (c) and (d) the PINN calculation on the global grid; and (e) and (f) the kriging interpolation on the global grid. Panels (a), (c), and (e) are for the Holocene, and (b), (d), and (f) for the LGM.}
	\label{fig: histograms}
\end{figure}

The dust deposition rates in the empirical dataset are approximately log-normal distributed; see Figures~\ref{fig: histograms}(a) and (b). The estimated values on the rectangular grid by the PINN and kriging also obey a log-normal distribution, as depicted in Figures~\ref{fig: histograms}(c) to (f). This consistency supports the accuracy of both computational methods. The sample means of the PINN calculation and kriging interpolation are consistent between them but tend to be lower than the empirical data. We emphasize that the empirical data are biased toward locations where dust flux measurements are possible (ice caps, loess fields, high-sedimentation oceanic regions, etc.). Hence, they are not a representative sample of the global population. This observation explains the lower sample means for the reconstructed fields: the empirical dataset has relatively few points in the oceans and polar regions, where low deposition rates are expected.

\FloatBarrier

\subsection{Comparison with global earth system models}
\label{sec:si:esm_comparison}

\begin{figure}[!ht]
	\centering
	(a)
	\includegraphics[width=0.44\columnwidth]{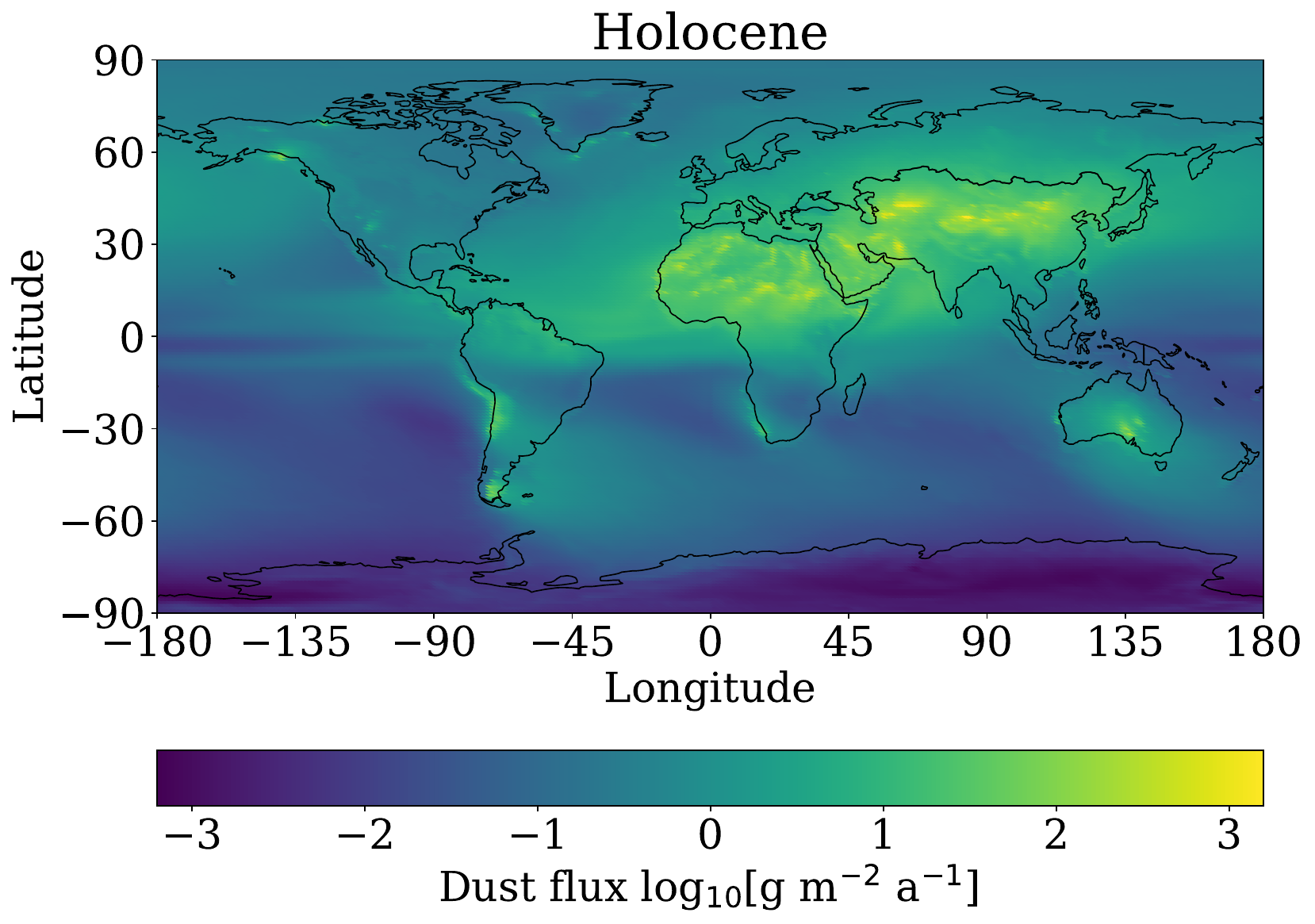}
	(b)
	\includegraphics[width=0.44\columnwidth]{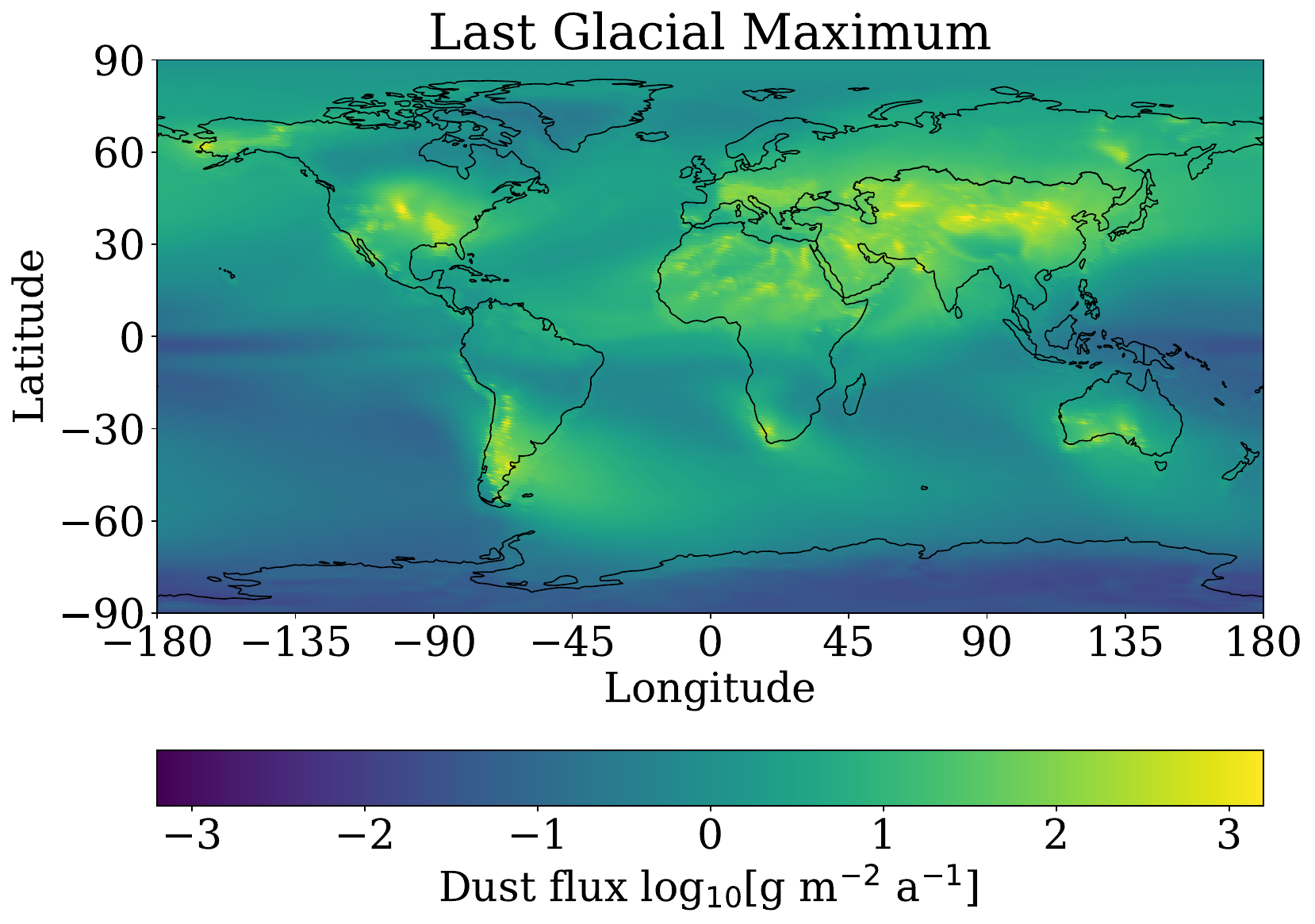}
	\caption{The dust deposition rate simulated by the ESM under (a) Holocene and (b) LGM conditions~\cite{albani2014improved}.}
	\label{fig: ESM data}
\end{figure}

Global ESM use coupled physical simulations of multiple climate variables. The dust deposition is often one of their outputs and is provided on a global grid. There is no intrinsic motivation to apply the PINN to simulated data that is already available on a high-resolution grid. Still, it provides a valuable opportunity for an additional sanity check of our methodology. Importantly, the high-resolution simulated data allows us to estimate the unknown physical parameters in the PINN's advection-diffusion equation and compare these outcomes with the PINN's model trained on the empirical data; see Section~\ref{sec:si:parameters}. The simulated data consists of dust fluxes from the Community Earth System Model 1.0.5~\cite{albani2016paleodust} for pre-industrial Holocene and LGM conditions at a 1-degree resolution, see Figure~\ref{fig: ESM data}.

The PINN considers latitudes from -80 to +80 degrees during training to avoid pole singularities in the spherical coordinates of the advection-diffusion equation. All 49,536 data points in this domain are used to train the PINN. We use the same PINN design and weights in the combined loss function as in the simulations on the empirical dataset. The advection is again taken from the ERA5 reanalysis data shown in Figure~\ref{fig: wind}. Differently, we let the PINN optimize the other physical parameters (diffusion coefficient and values at the north and south poles) on this new dataset.

\begin{figure}[!ht] 
	\centering
	(a)
	\includegraphics[width=0.5\textwidth]{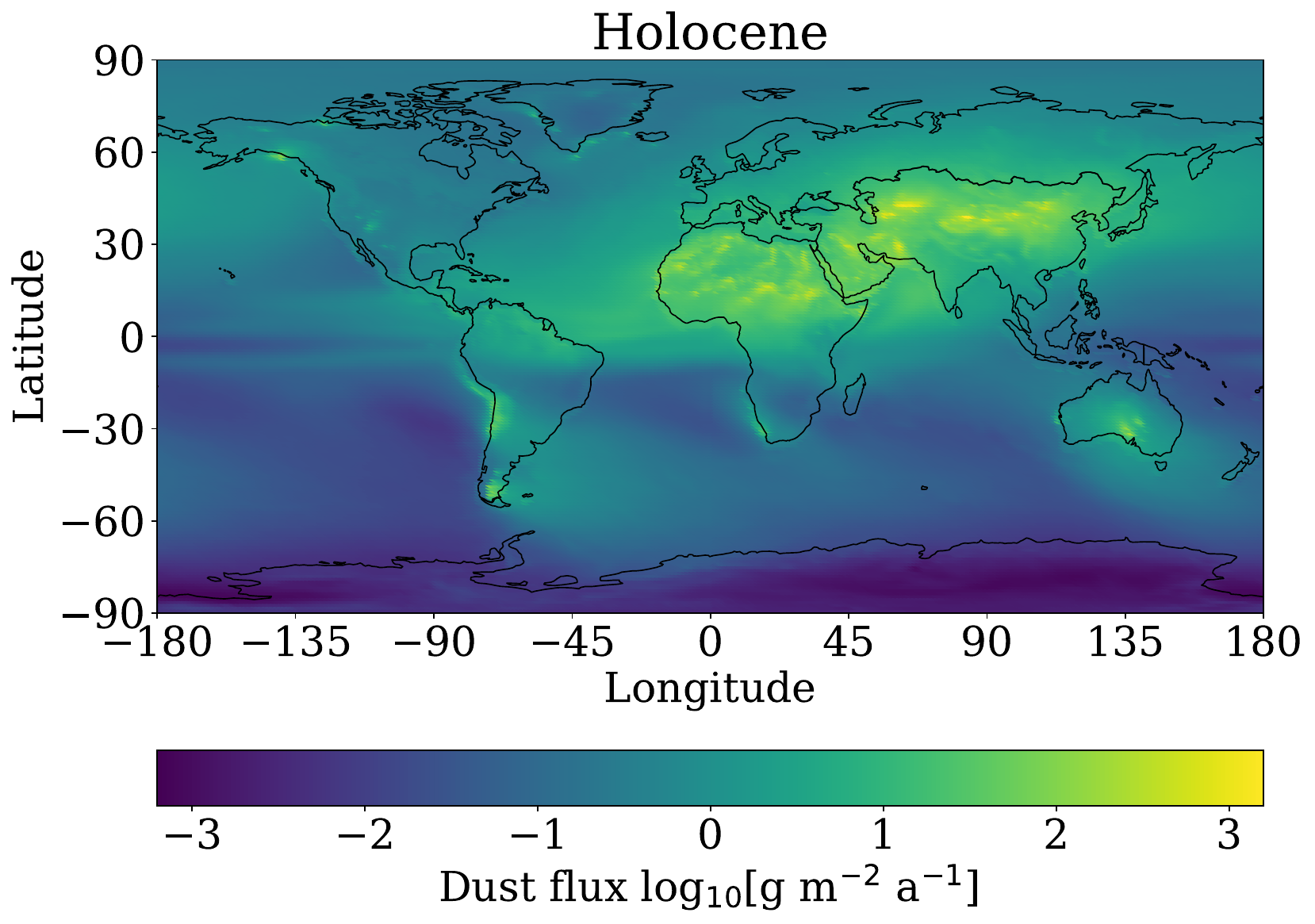}
	(b)
	\includegraphics[width=0.38\textwidth]{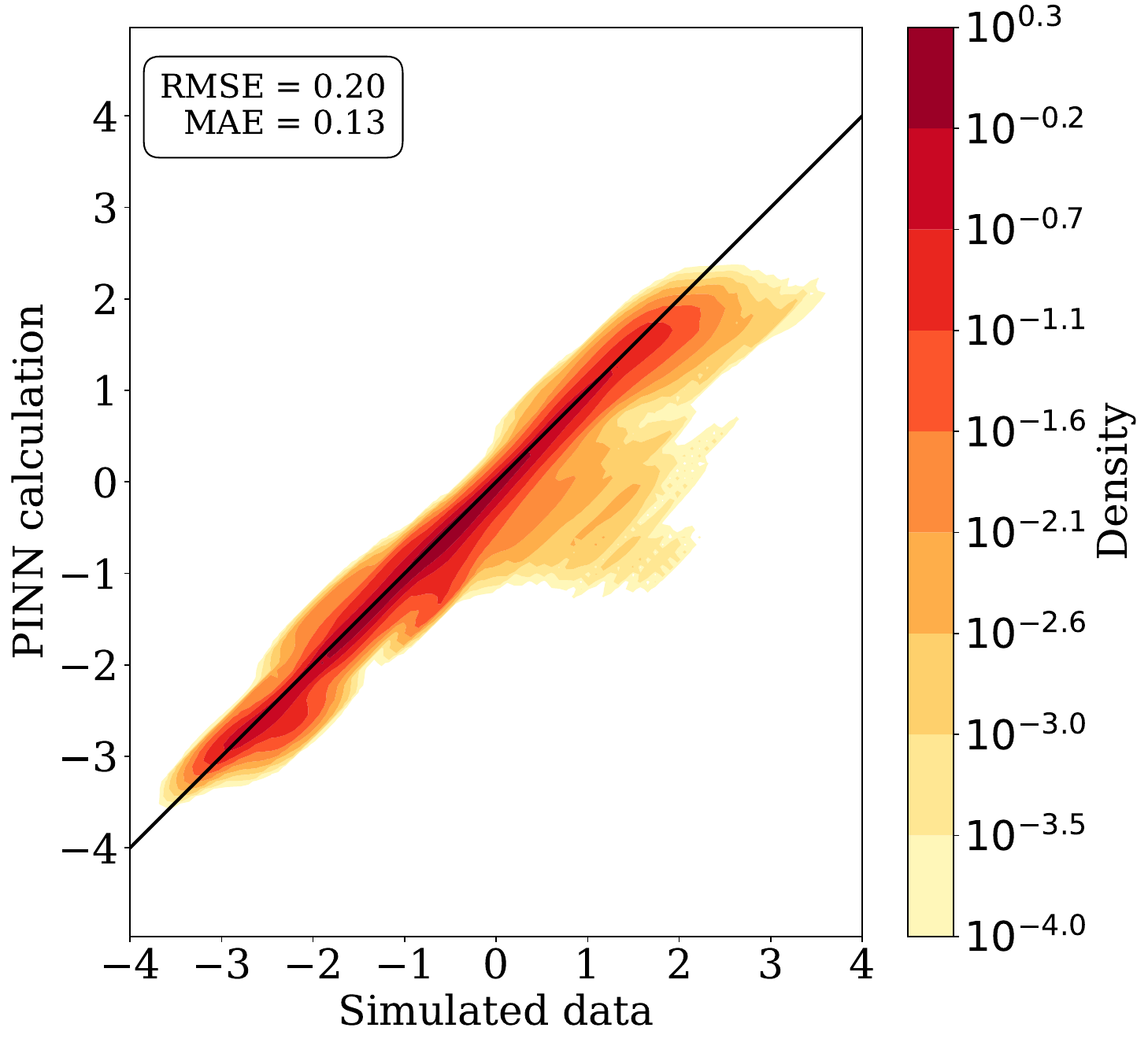}
	\\ (c)
	\includegraphics[width=0.5\textwidth]{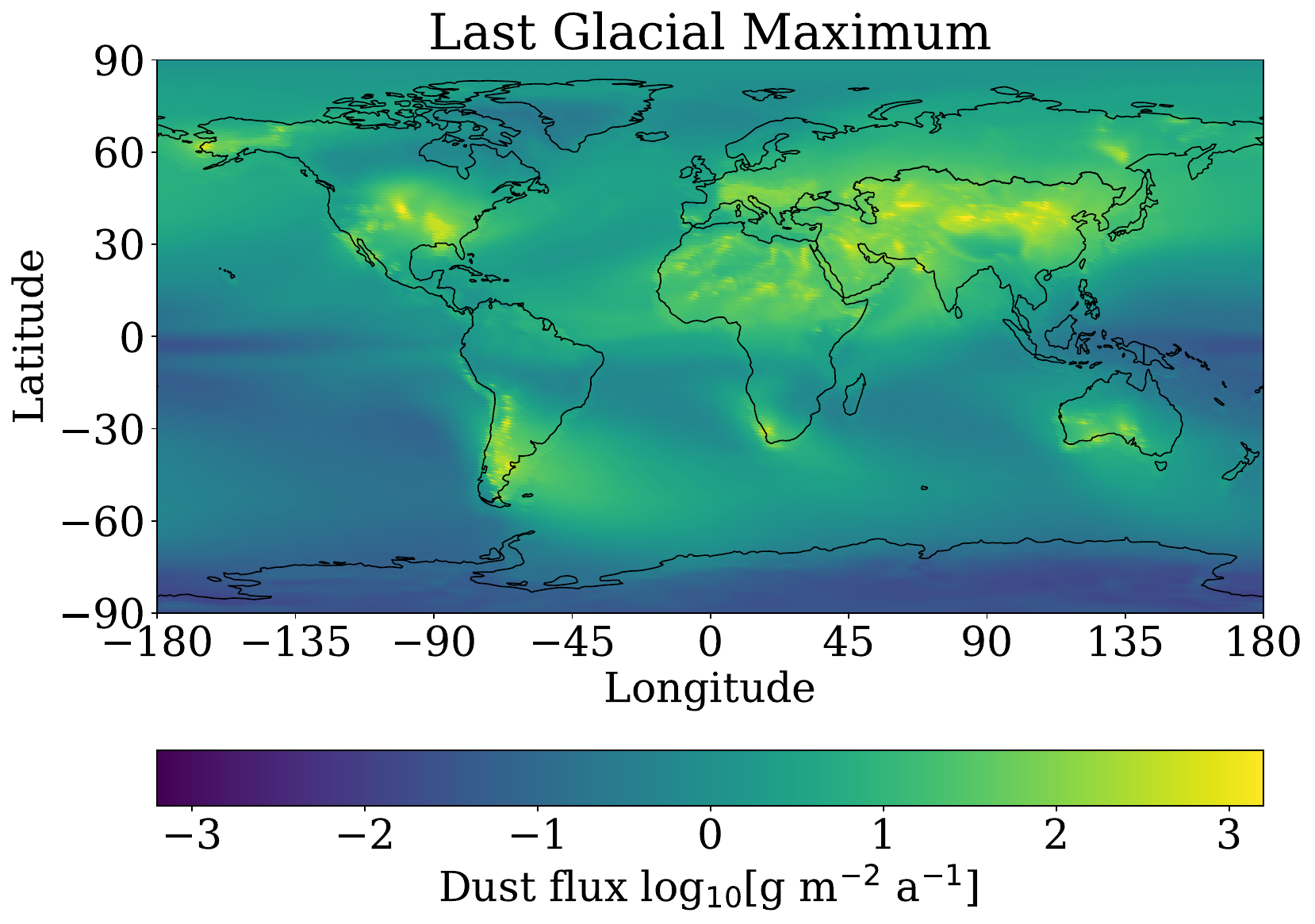}
	(d)
	\includegraphics[width=0.38\textwidth]{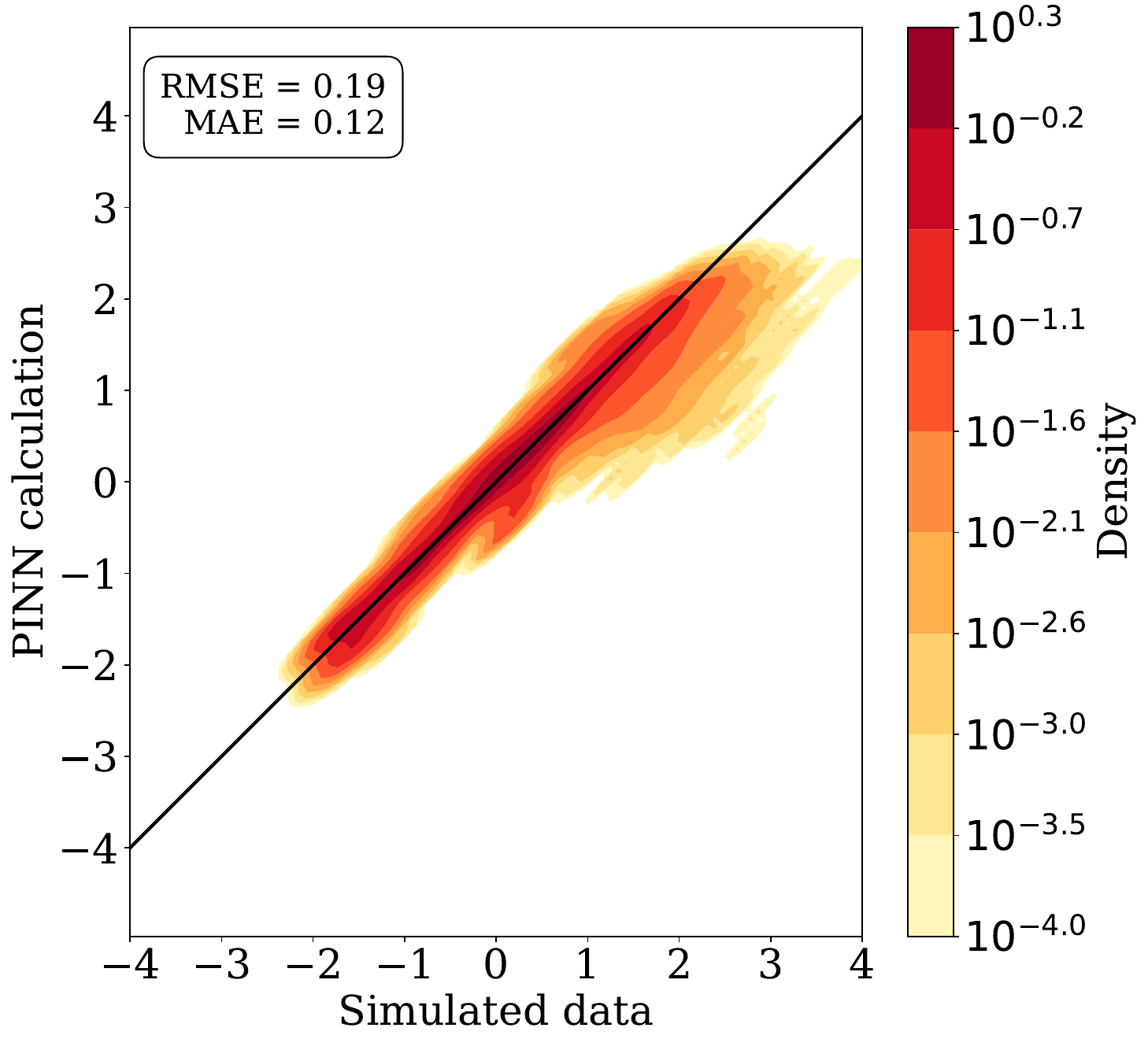}
	\caption{The PINN's calculation of global dust deposition from the ESM data. Panels (a) and (c) show the PINN's calculated field from the simulated dataset for the Holocene and LGM, respectively. Panels (b) and (d) are density plots of the simulated data versus the PINN's calculation in each data point for the Holocene and LGM, respectively.}
	\label{fig: Results of simulated data}
\end{figure}

Figure~\ref{fig: Results of simulated data} compares the PINN's calculation with the simulated dataset for the Holocene and LGM. The overall patterns between the PINN's analysis and the ESM are the same, with minor differences due to the PINN smoothing out the local peaks in deposition rates from the simulated data. This is visible by the lower maximum of the PINN compared to the ESM data and the slightly skewed values in the upper right corner of the density plots. However, these skewed results are a tiny proportion of the data points and are only noticeable with a logarithmic scaling of the density. The vast majority of the data points show a sharp and symmetric fit, indicating that the PINN accurately reproduced the ESM dust field.

\FloatBarrier

\subsection{Goodness of fit}
\label{sec:si:goodness_of_fit}

\begin{figure}[!ht]
	\centering
	(a)
	\includegraphics[width=0.8\columnwidth]{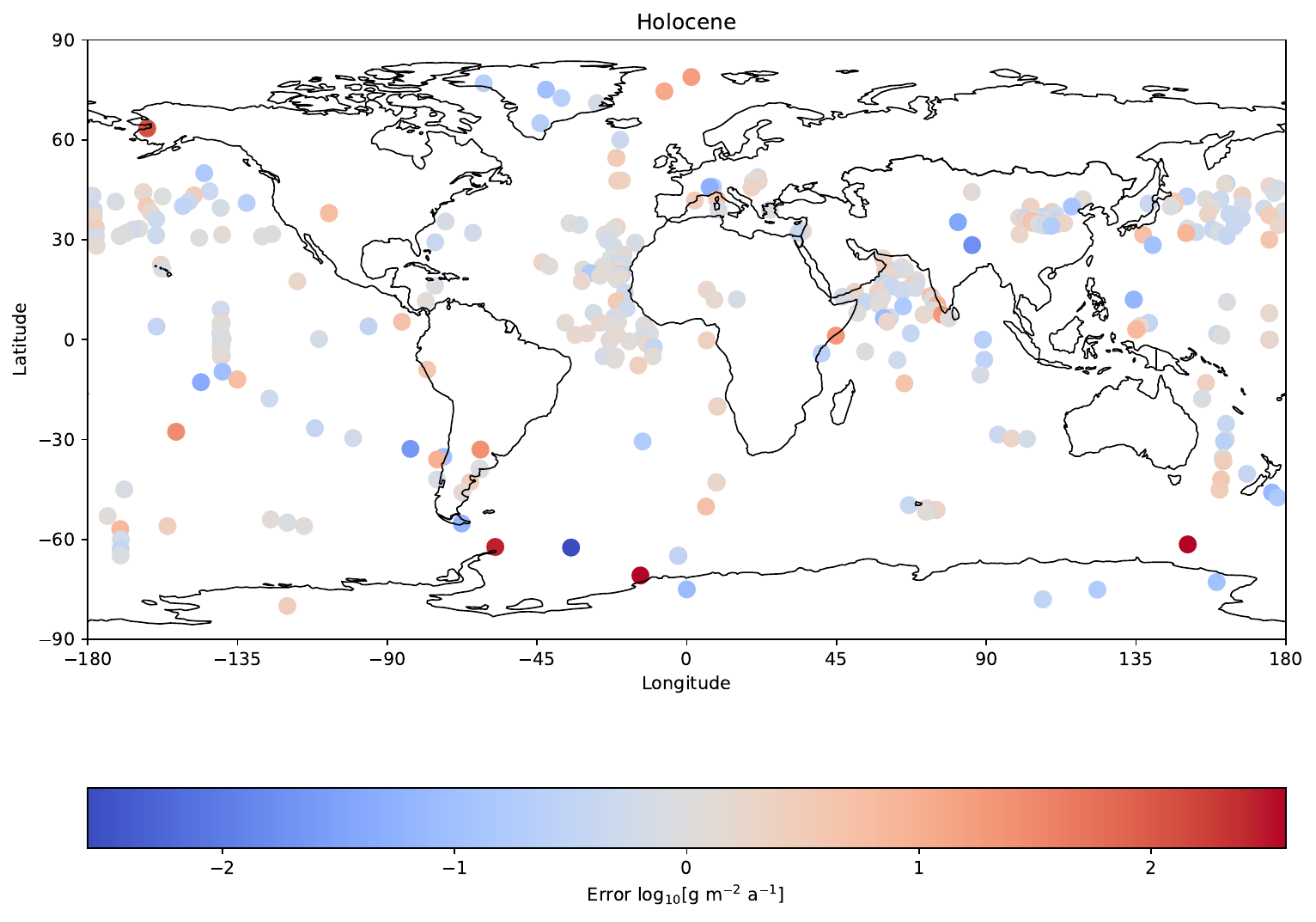}\\
	(b)
	\includegraphics[width=0.8\columnwidth]{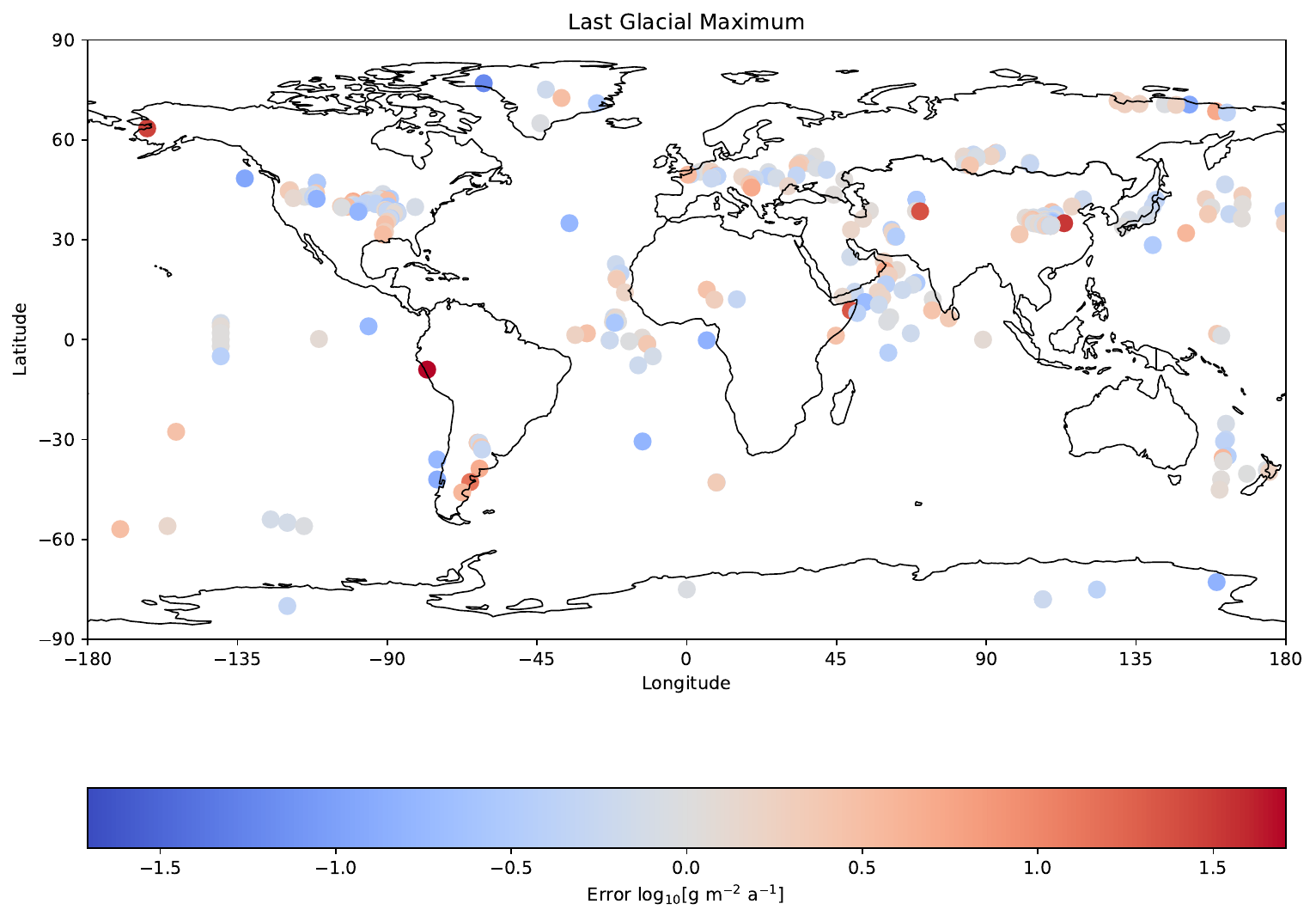}
	\caption{The error of the PINN's leave-one-out cross-validation under (a) Holocene and (b) LGM conditions.}
	\label{fig: cross validation}
\end{figure}

To support the effectiveness of PINNs in reconstructing global dust deposition maps from sparse and uncertain empirical data, let us analyze their goodness of fit by applying cross-validation on left-out empirical data. Given the relatively small dataset sizes, i.e., $n=397$ and $n=317$ for the Holocene and LGM periods, respectively, we perform a jackknife or leave-one-out cross-validation. Specifically, we train the PINN on $n-1$ samples and predict the dust deposition in the left-out location. We use the same PINN design as in the main text and repeat this validation procedure for all $n$ possible combinations. The $n$ estimations are then compared with their respective left-out data values. Figure~\ref{fig: cross validation} displays this test error, i.e., the difference between the observation and the prediction in each data point.

We highlight that the PINN effectively reconstructed the left-out observations, as visible by the small values on the global map. The largest errors are primarily in isolated points or locations with high measurement uncertainty. Furthermore, the differences are symmetric, meaning the PINN does not induce a bias. More precisely, the mean values of the dust deposition reconstructions are 0.402 and 1.427 [log$_{10}$(g\,m$^{-2}$\,a$^{-1}$)] for the Holocene and LGM periods, respectively. These are almost identical to the sample mean of the empirical data; see Figure~\ref{fig: histograms}. Moreover, the mean absolute errors are 0.360 and 0.283 [log$_{10}$(g\,m$^{-2}$\,a$^{-1}$)] for the Holocene and LGM periods, respectively. These error values are much smaller than the standard deviation of the empirical dataset and thus support the PINN's excellent performance.

\begin{figure}[!ht]
	\centering
	(a)
	\includegraphics[width=0.44\columnwidth]{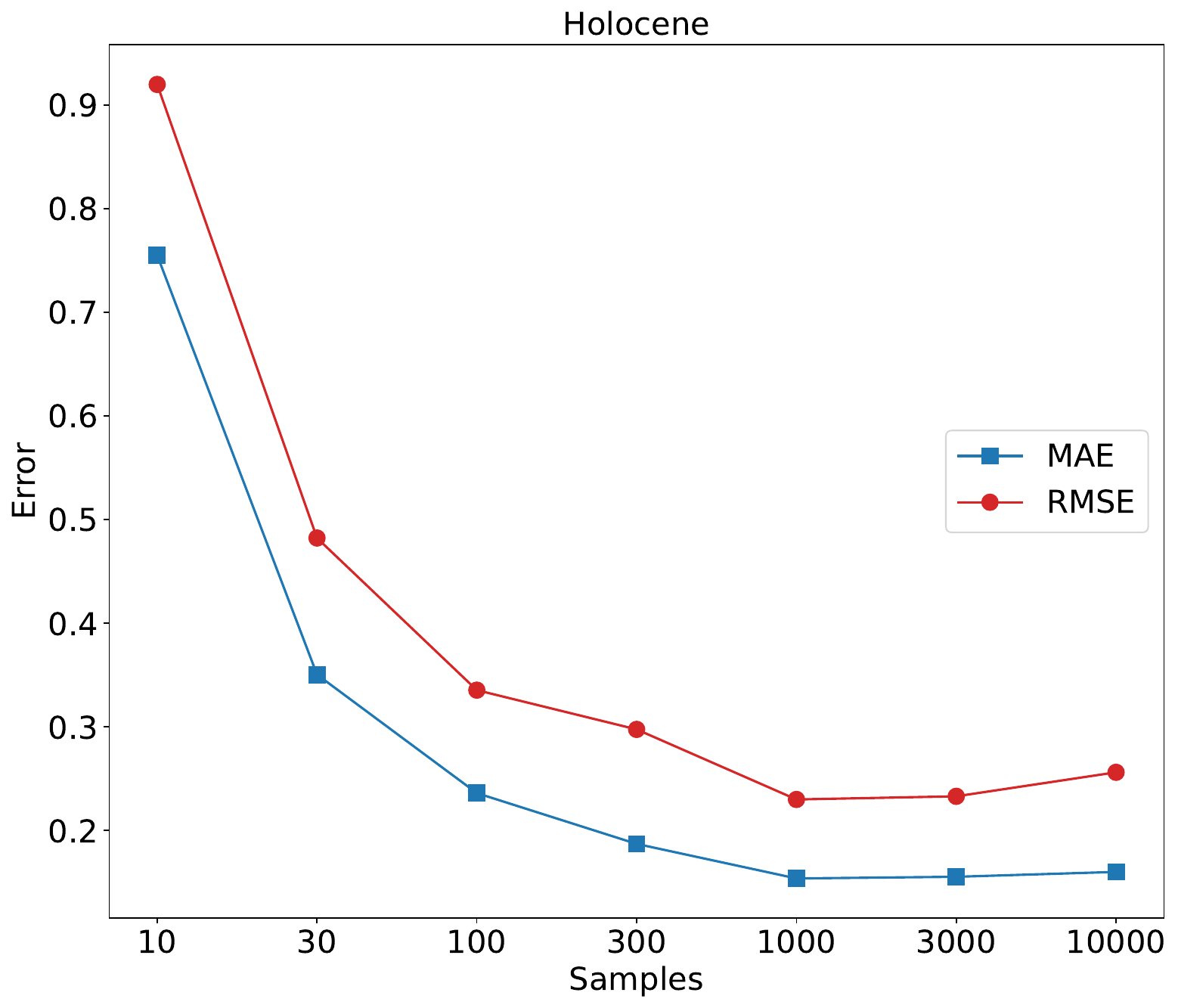}
	(b)
	\includegraphics[width=0.44\columnwidth]{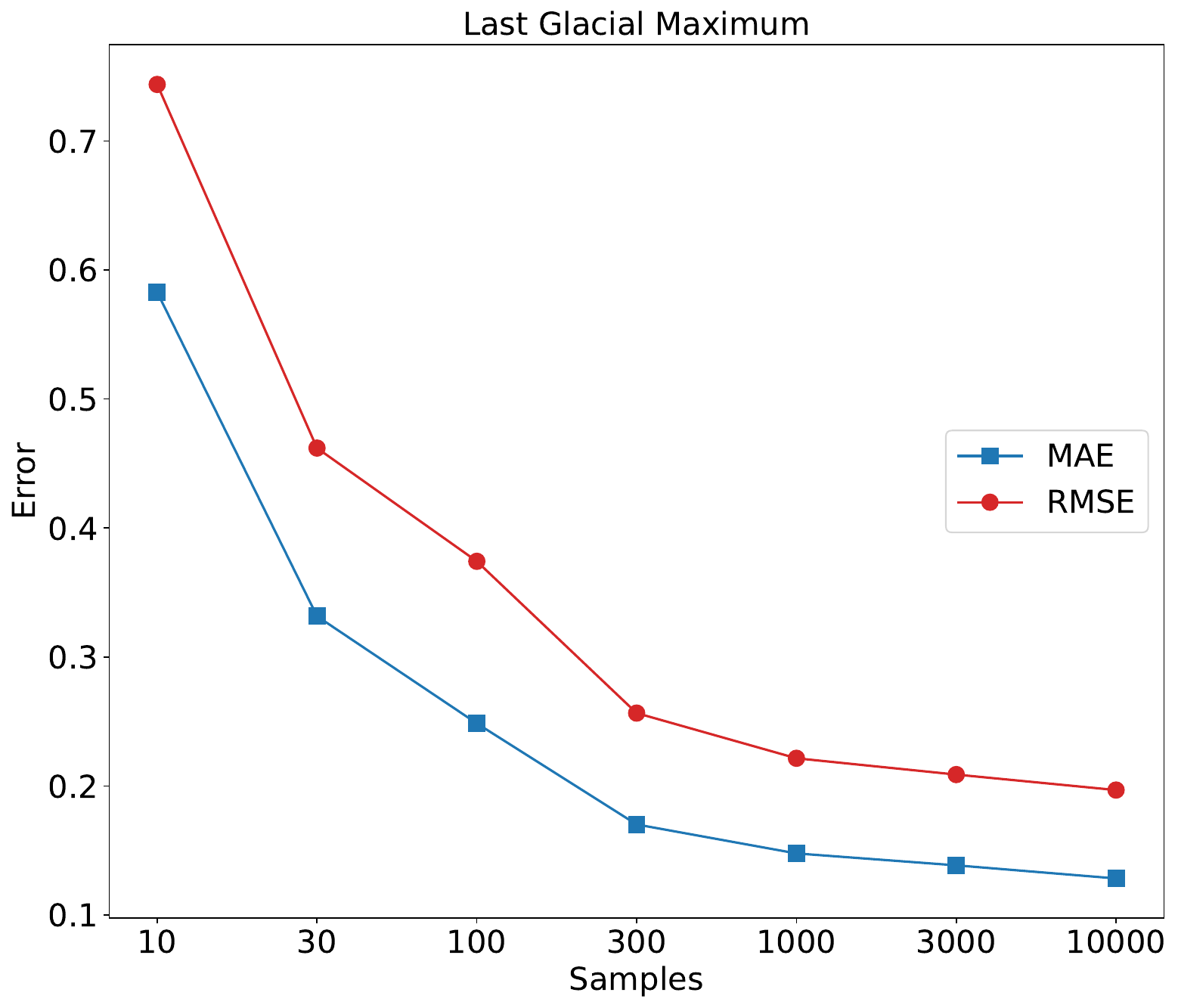}
	\caption{The error of the PINN's validation on subsamples of the ESM dust deposition data, in log$_{10}$(g\,m$^{-2}$\,a$^{-1}$), under (a) Holocene and (b) LGM conditions.}
	\label{fig: sample size}
\end{figure}

Paleoclimatic datasets on dust depositions depend on extensive measurement campaigns in locations conducive to field experiments and, therefore, have limited sample sizes. The empirical dataset used in this study involves all relevant data available to date, which is only 397 and 317 measurements for the Holocene and LGM periods, respectively. To test the influence of sample size on the PINN's accuracy, let us use the ESM datasets presented in Section~\ref{sec:si:esm_comparison}, which have 49,536 data points on a global grid. We train the PINN on random subsamples and reconstruct dust deposition on the entire grid. The cross-validation error is the difference between the ESM data and the PINN's estimations on all 49,536 data points. Figure~\ref{fig: sample size} confirms the PINN's quick convergence with increasing sample size. Most importantly, small error values have already been reached with only 300 samples. The results also highlight that larger datasets may indeed reduce reconstruction errors further, thus making the case for more empirical paleoclimatic studies.

\FloatBarrier

\subsection{Optimized physical parameters}
\label{sec:si:parameters}

The PDE loss term in the PINN includes four physical parameters: the advection and diffusion coefficients in the PDE and the dust deposition at the north and south pole in the boundary conditions. We used wind data from ERA5 reanalysis (see Figure~\ref{fig: wind}) for the advection. The other physical parameters are included as additional unknowns in the PINN and optimized by the neural network itself, as explained in Section~\ref{section: Physical Parameters Design} in the main text. Here, we check the optimal values of the physical parameters found by the PINN and compare them to estimations from the empirical dataset and ESM simulations. Specifically, we take an average of the closest data points near the poles to analyze the latitudinal boundary conditions. In the case of the empirical dataset, we estimate the dust deposition rate at the north and south poles by averaging all observations further north than $+60^\circ$ and further south than $-63^\circ$ latitude, respectively. In the case of the simulated data, we average all data points within 9 degrees latitude from each pole. The diffusion coefficient cannot be estimated straightforwardly from the data since it aggregates all diffusion processes of dust flow in a single parameter.

\begin{table}[!ht]
	\caption{The values of the dust deposition rate at the north and south pole ($u_{\text{north}}$, $u_{\text{south}}$) and the diffusion coefficient $D$ were calculated by: 1) estimations from the empirical dataset, 2) optimized values by the PINN applied to the empirical dataset, 3) estimations from the simulated ESM data, and 4) optimized values by the PINN applied to the simulated ESM data. All for Holocene and LGM conditions.}
	\label{table: PINN's performance}
	\begin{center}
		\begin{tabular}{|p{3cm}|p{17mm}|p{17mm}|p{17mm}|p{17mm}|}
			\hline
			\textbf{Parameter} & \textbf{Empirical data} & \textbf{PINN on empirical data} & \textbf{Simulated ESM data} & \textbf{PINN on simulated data} \\[0.8ex]
			\hline \hline
			\multicolumn{5}{|c|}{\textbf{Holocene}}\\
			\hline \hline
			north pole ($u_{\text{north}}$) & -1.042 & -1.0317 & 0.064 &  0.0810\\
			south pole ($u_{\text{south}}$) & -2.07  & -3.0344 &  -1.70 & -1.8128\\
			diffusion ($D$) & N/A & 0.1366 & N/A &  0.1335\\
			\hline \hline
			\multicolumn{5}{|c|}{\textbf{Last Glacial Maximum}}\\
			\hline \hline
			north pole ($u_{\text{north}}$) & -0.59 & -1.6517 & 0.1267  & 0.1353\\
			south pole ($u_{\text{south}}$) & -3.45 & -3.2270 & -1.785 & -1.7441\\
			diffusion ($D$) & N/A & 0.1289 & N/A & 0.1602\\
			\hline
		\end{tabular}
	\end{center}
\end{table}

The results in Table~\ref{table: PINN's performance} confirm that the PINN gives reasonable estimates of the dust deposition close to the poles. The two most considerable differences are in the value of the south pole in the Holocene and the north pole under LGM conditions. In the first case, the measurements near Antarctica have several points with relatively high values in the Holocene; see Figure~\ref{fig: empirical data}(a). In the second case, there are several measurements in eastern Siberia with relatively high values (see Figure~\ref{fig: empirical data}(b)) that explain the higher value for the estimation from the empirical data compared to the PINN calculation in the north pole. Finally, the optimized values of the diffusion coefficient are similar for all cases. In general, the PINN successfully found reasonable values for the unknown physical parameters in the PDE model for the pole's deposition rate and global diffusion of mineral dust.

\subsection{Sensitivity of the model weight}
\label{sec:si:sensitivity}

\begin{figure}[!ht] 
	\centering
	(a)
	\includegraphics[width=0.44\textwidth]{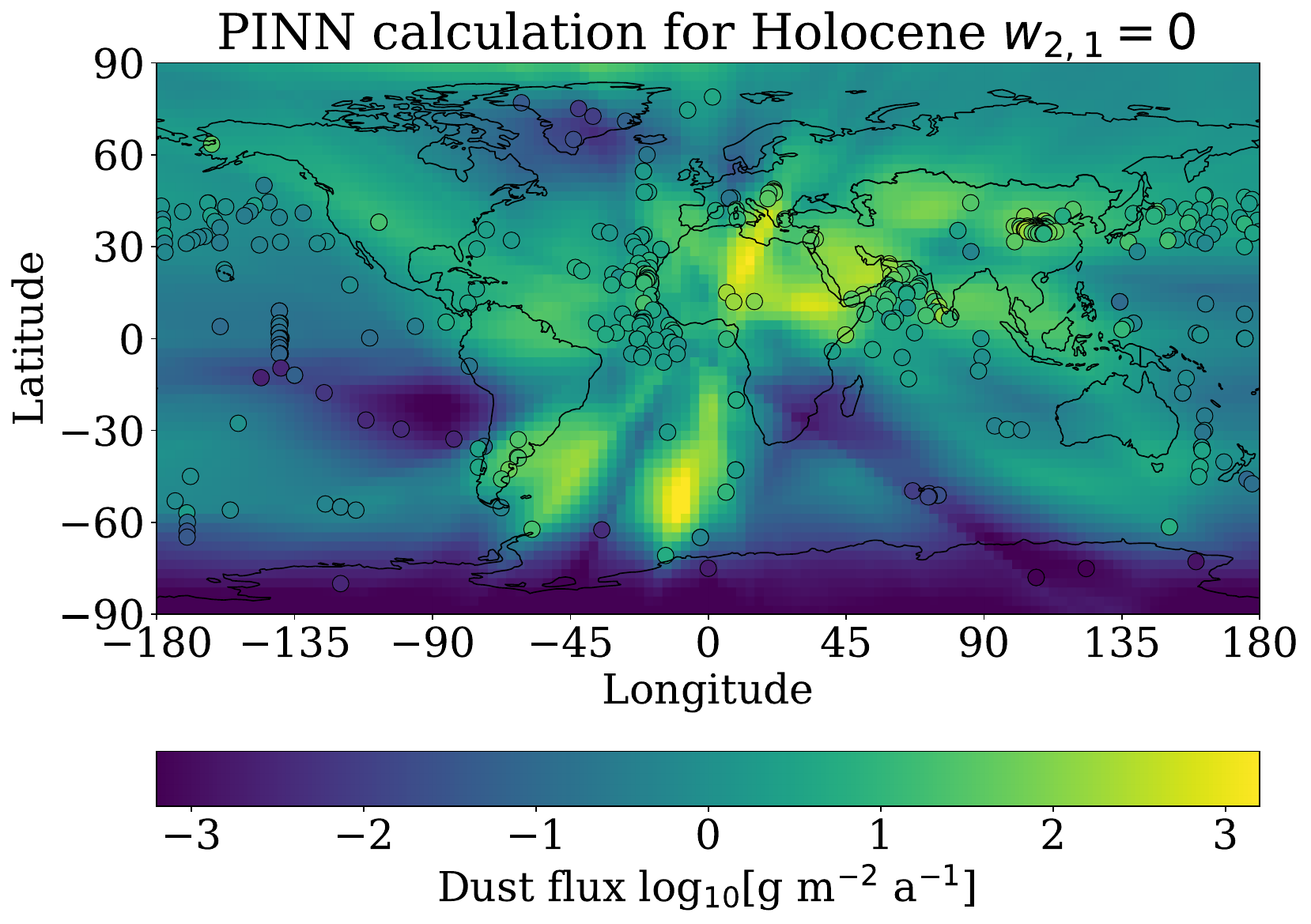}
	(b)
	\includegraphics[width=0.44\textwidth]{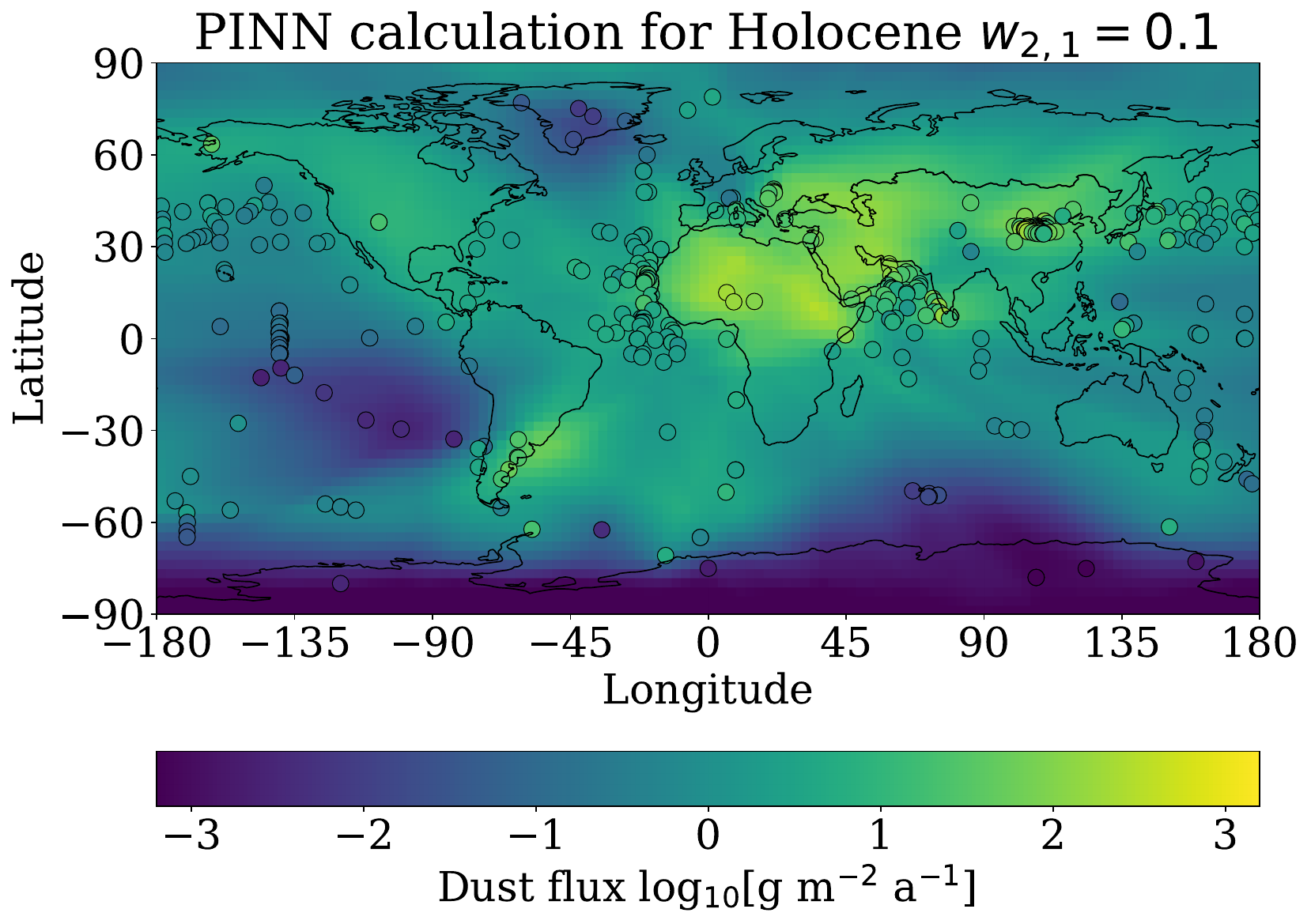}
	\\ (c)
	\includegraphics[width=0.44\textwidth]{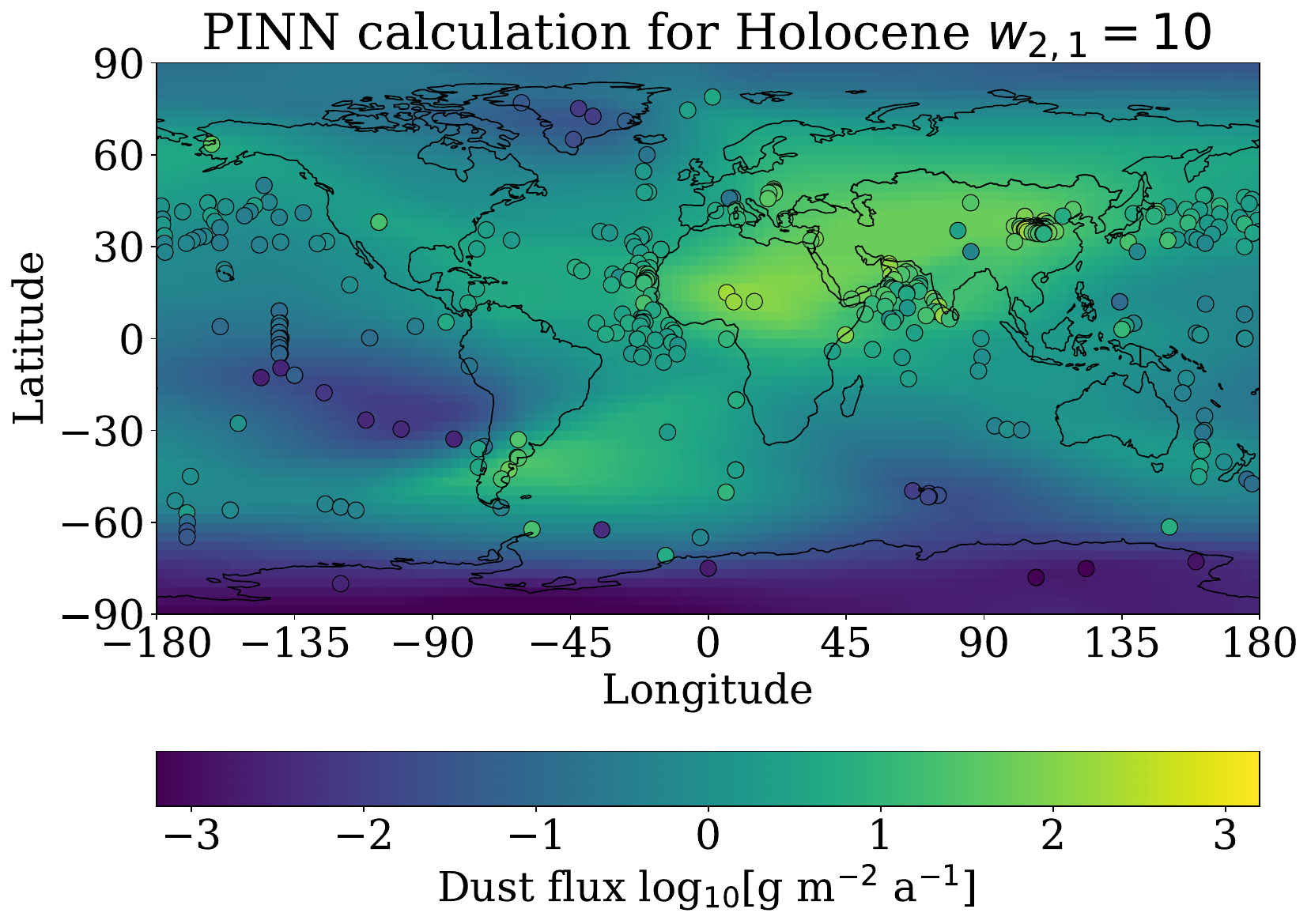}
	(d)
	\includegraphics[width=0.44\textwidth]{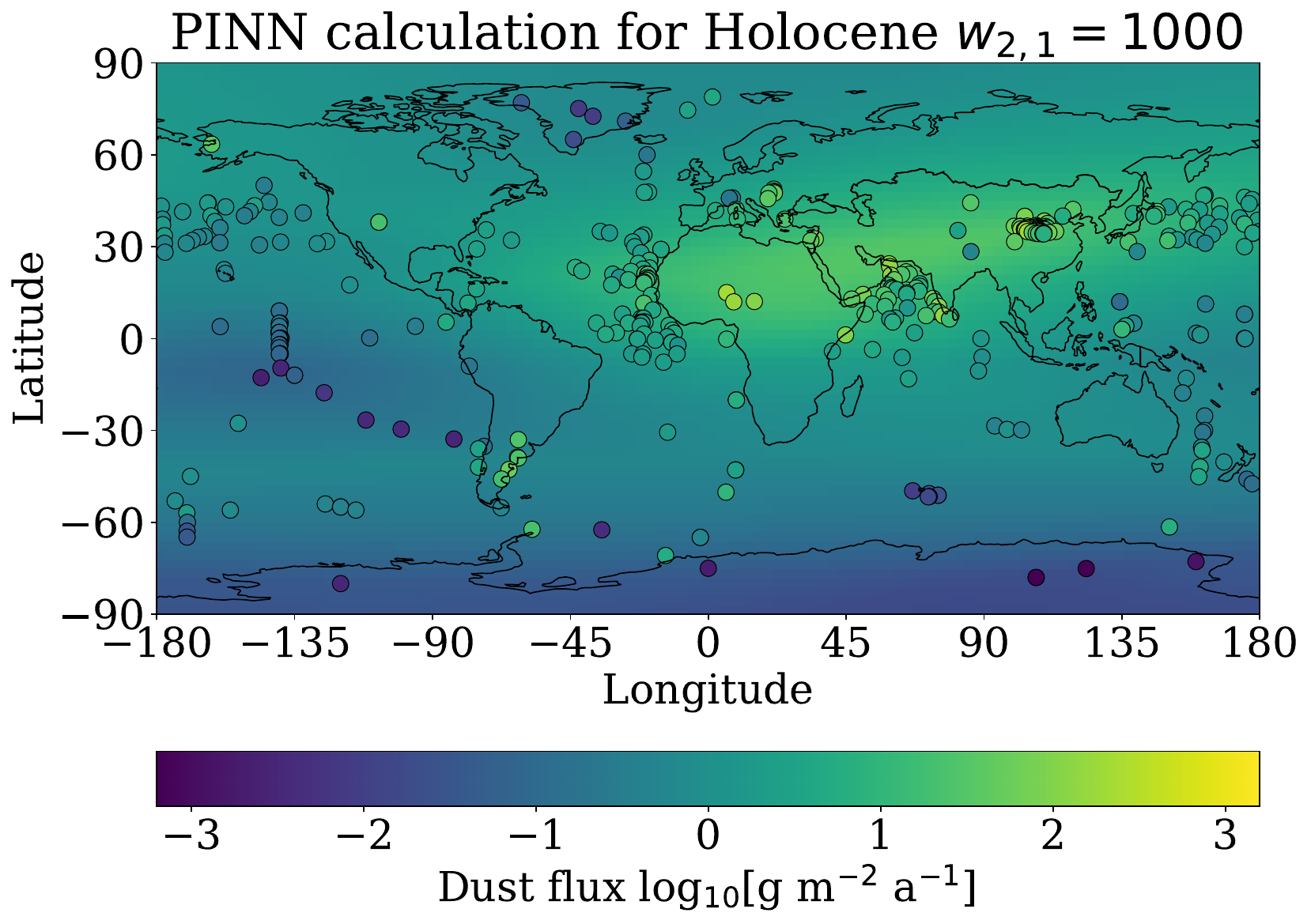}
	\caption{The global dust flux calculated by the PINN from empirical data in the Holocene. Each panel uses a different model weight, with values of 0, 0.1, 10, and 1000, respectively. The main text shows the field for $w_{2,1}=1$ in Figure~\ref{fig: deposition maps}.}
	\label{fig: model weight field holocene}
\end{figure}

\begin{figure}[!ht] 
	\centering
	(a)
	\includegraphics[width=0.44\textwidth]{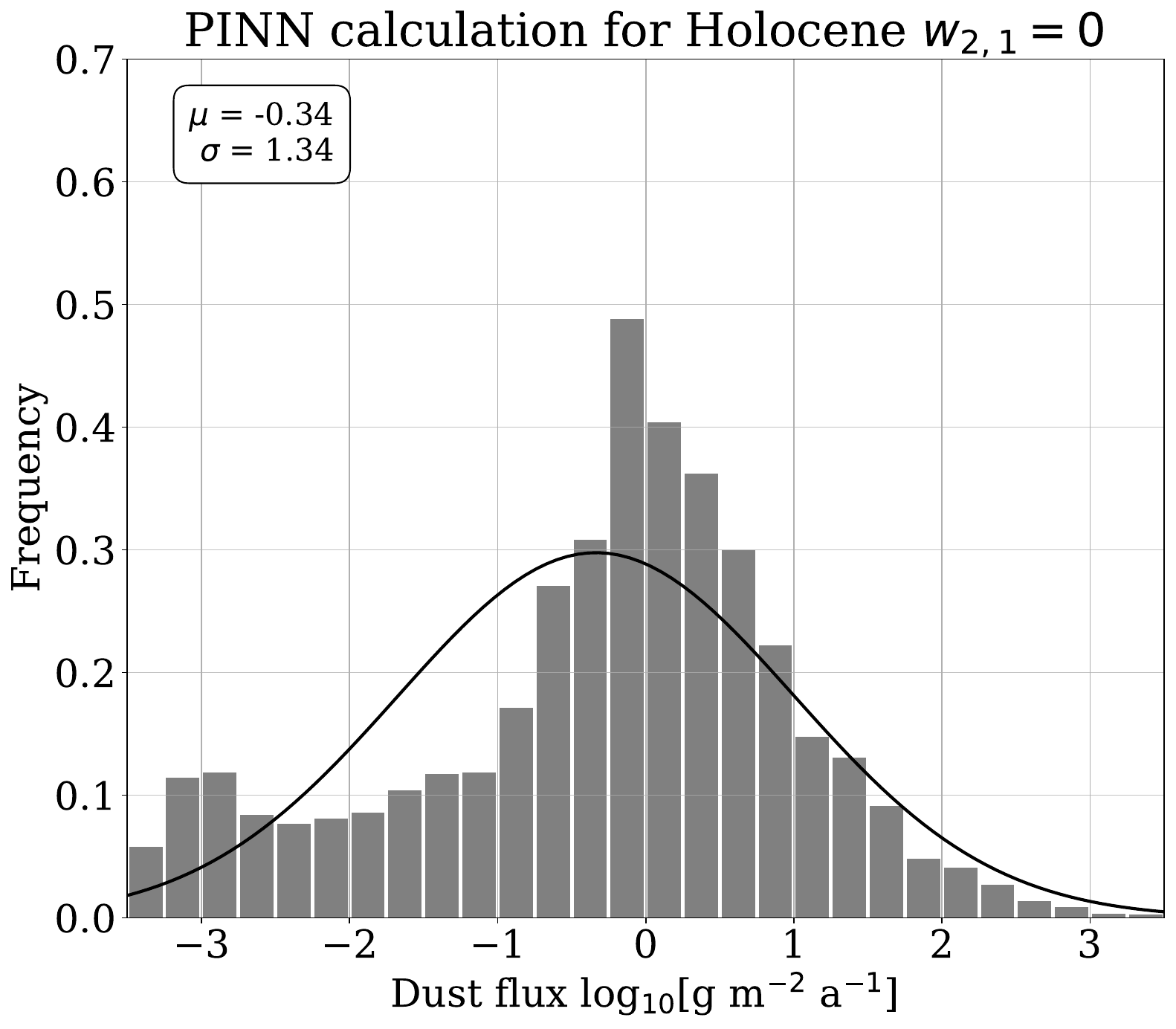}
	(b)
	\includegraphics[width=0.44\textwidth]{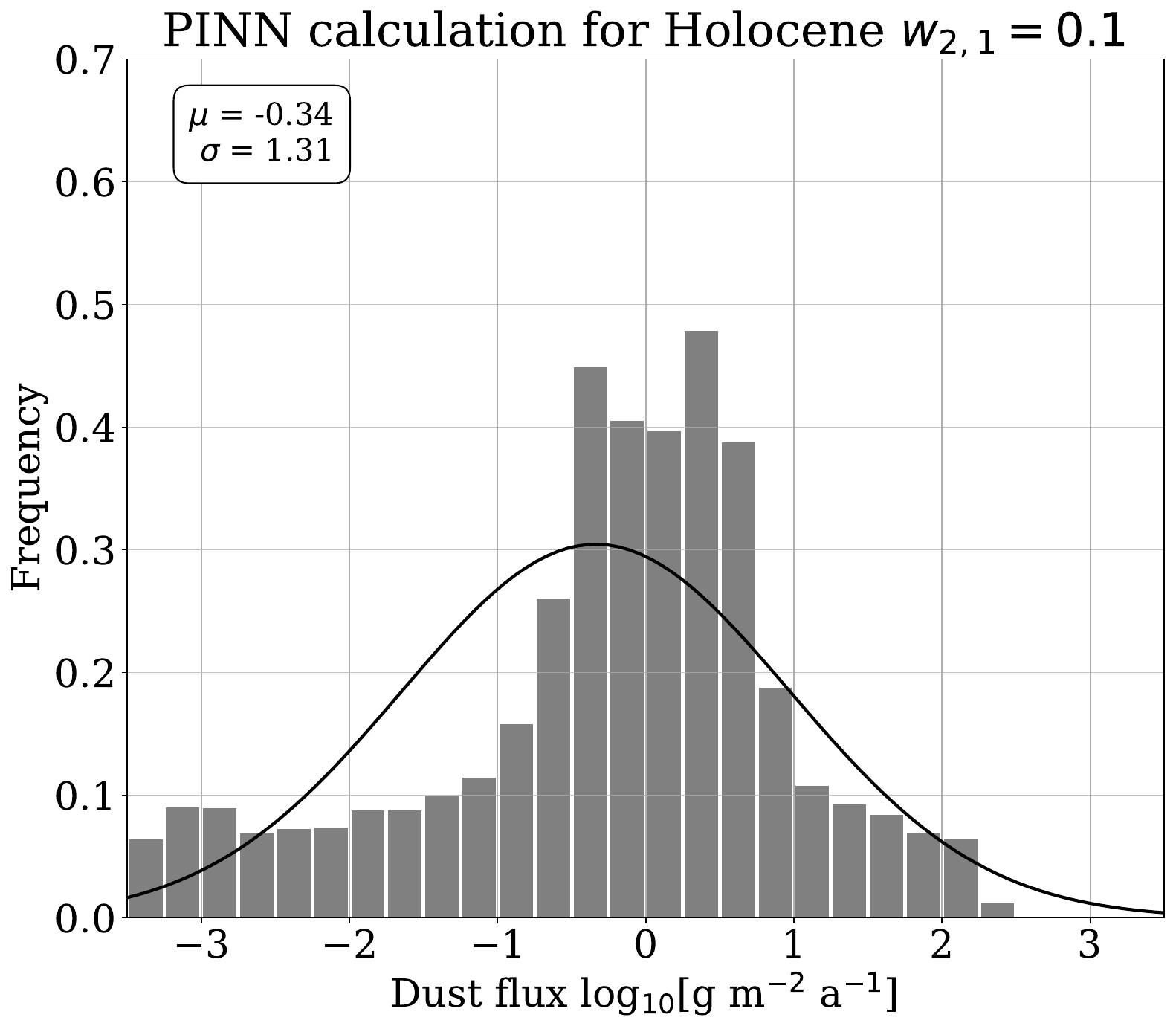}
	\\ (c)
	\includegraphics[width=0.44\textwidth]{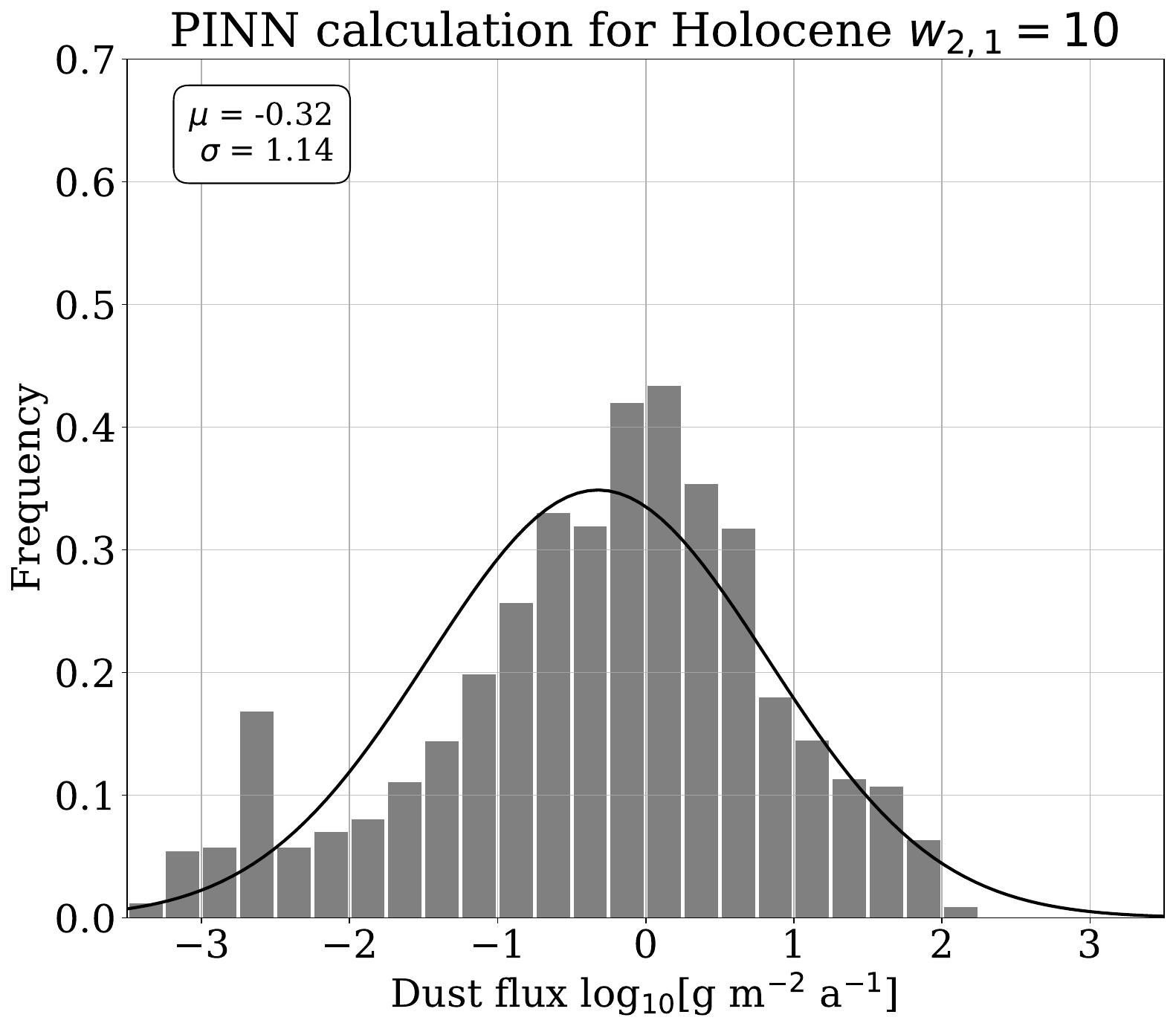}
	(d)
	\includegraphics[width=0.44\textwidth]{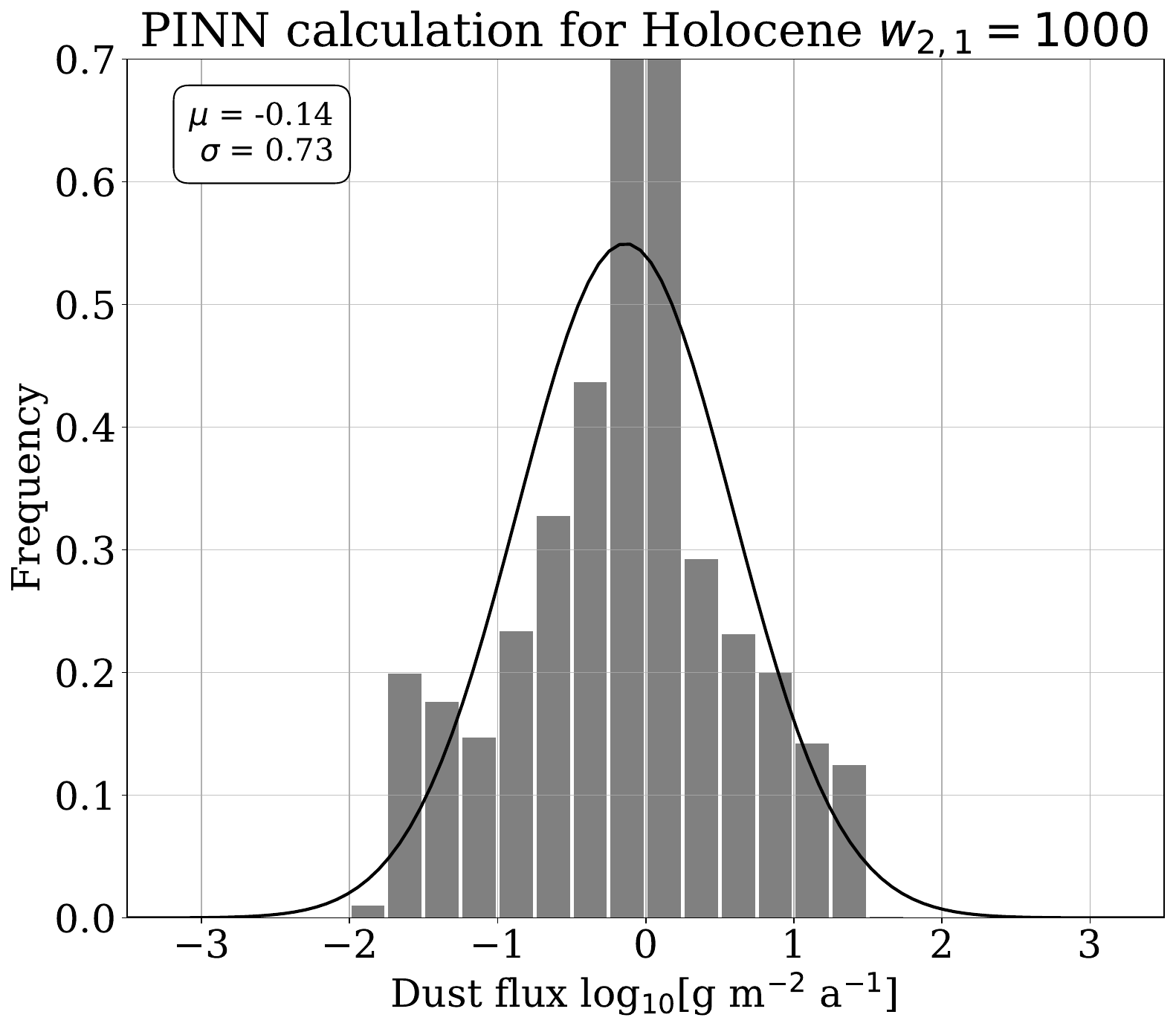}
	\caption{The dust flux distribution calculated by the PINN from empirical data in the Holocene. Each panel uses a different model weight, with values of 0, 0.1, 100, and 1000, respectively. The main text shows the histogram for $w_{2,1}=1$ in Figure~\ref{fig: histograms}.}
	\label{fig: model weight histogram holocene}
\end{figure}

\begin{figure}[!ht] 
	\centering
	(a)
	\includegraphics[width=0.44\textwidth]{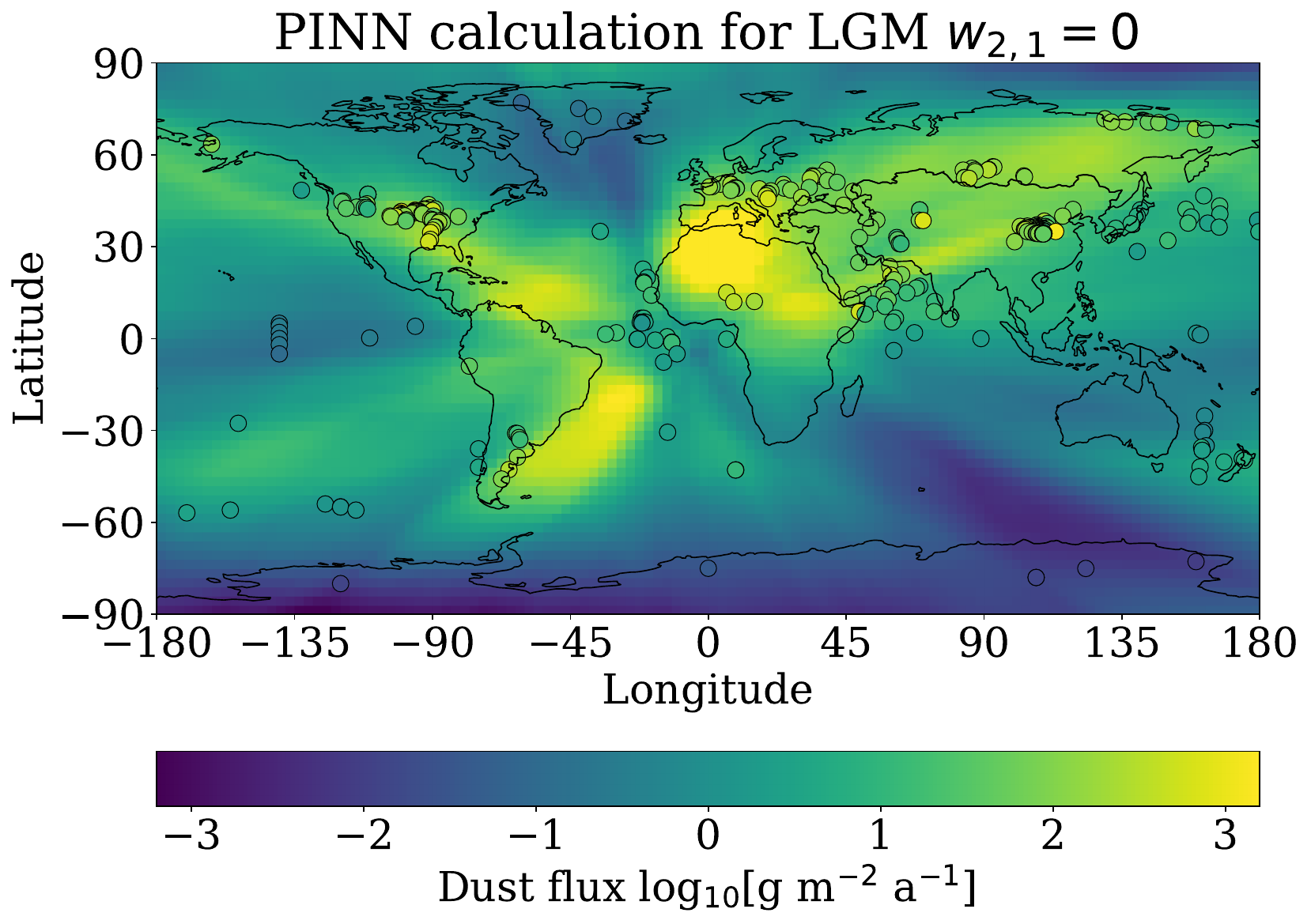}
	(b)
	\includegraphics[width=0.44\textwidth]{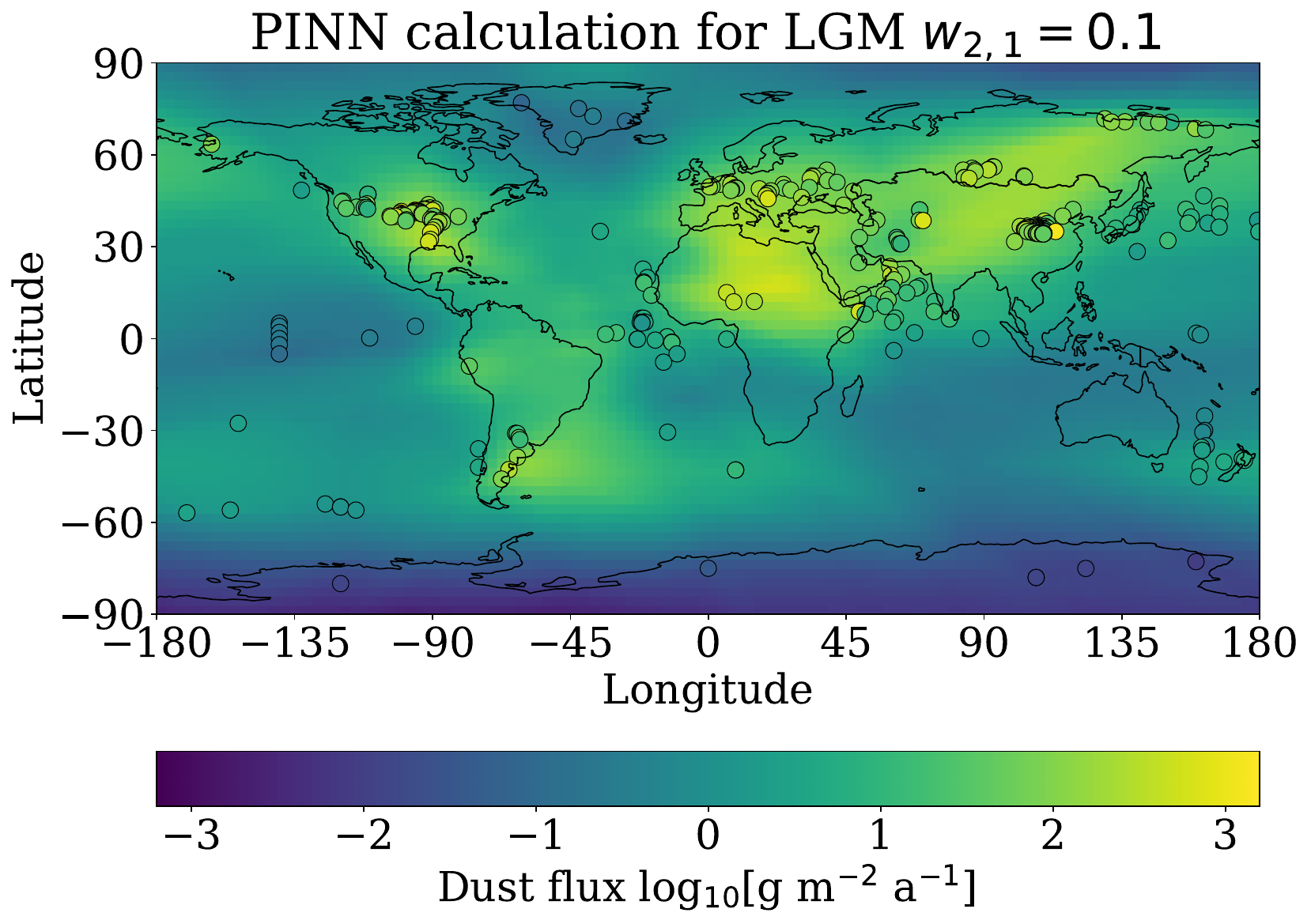}
	\\ (c)
	\includegraphics[width=0.44\textwidth]{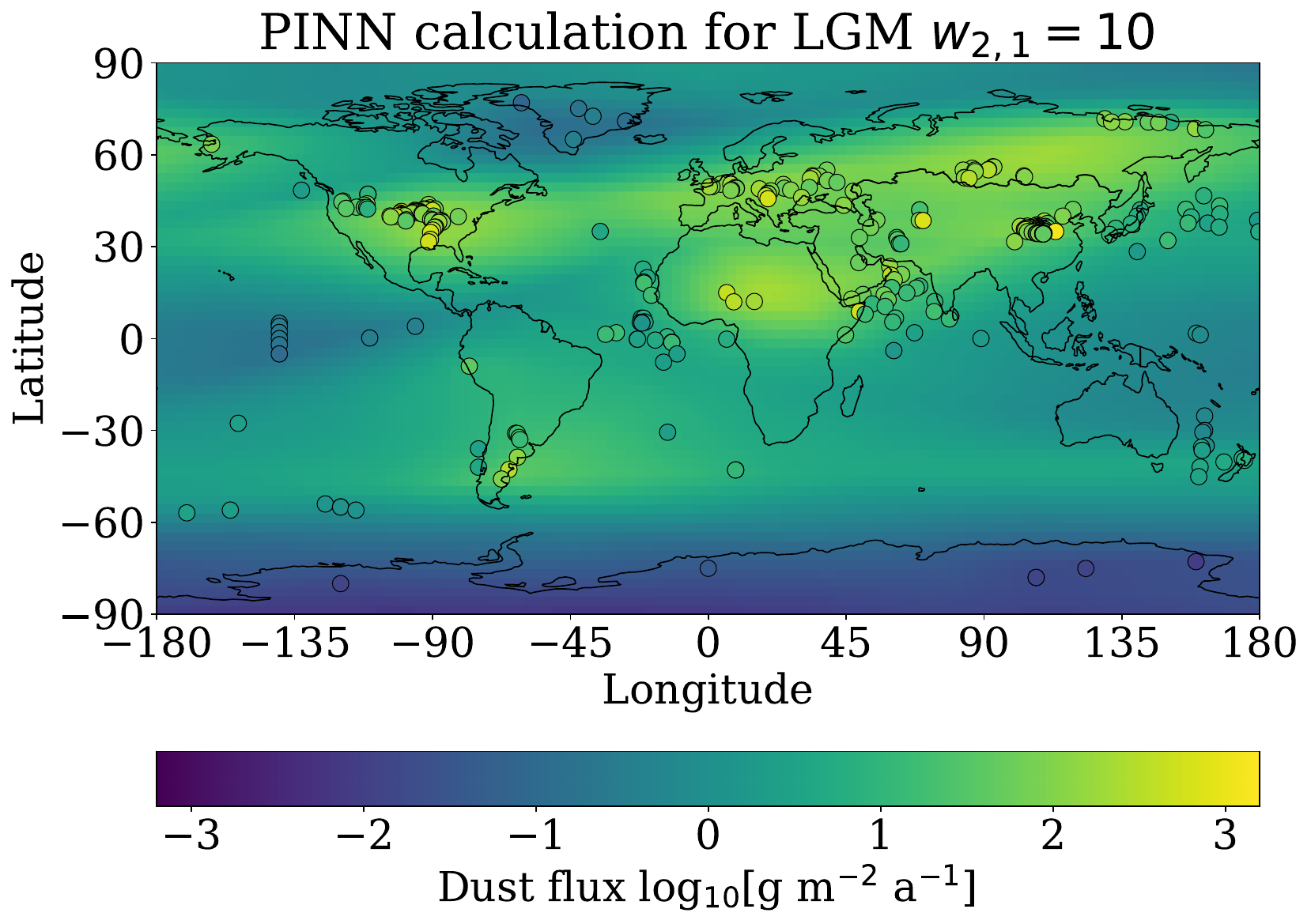}
	(d)
	\includegraphics[width=0.44\textwidth]{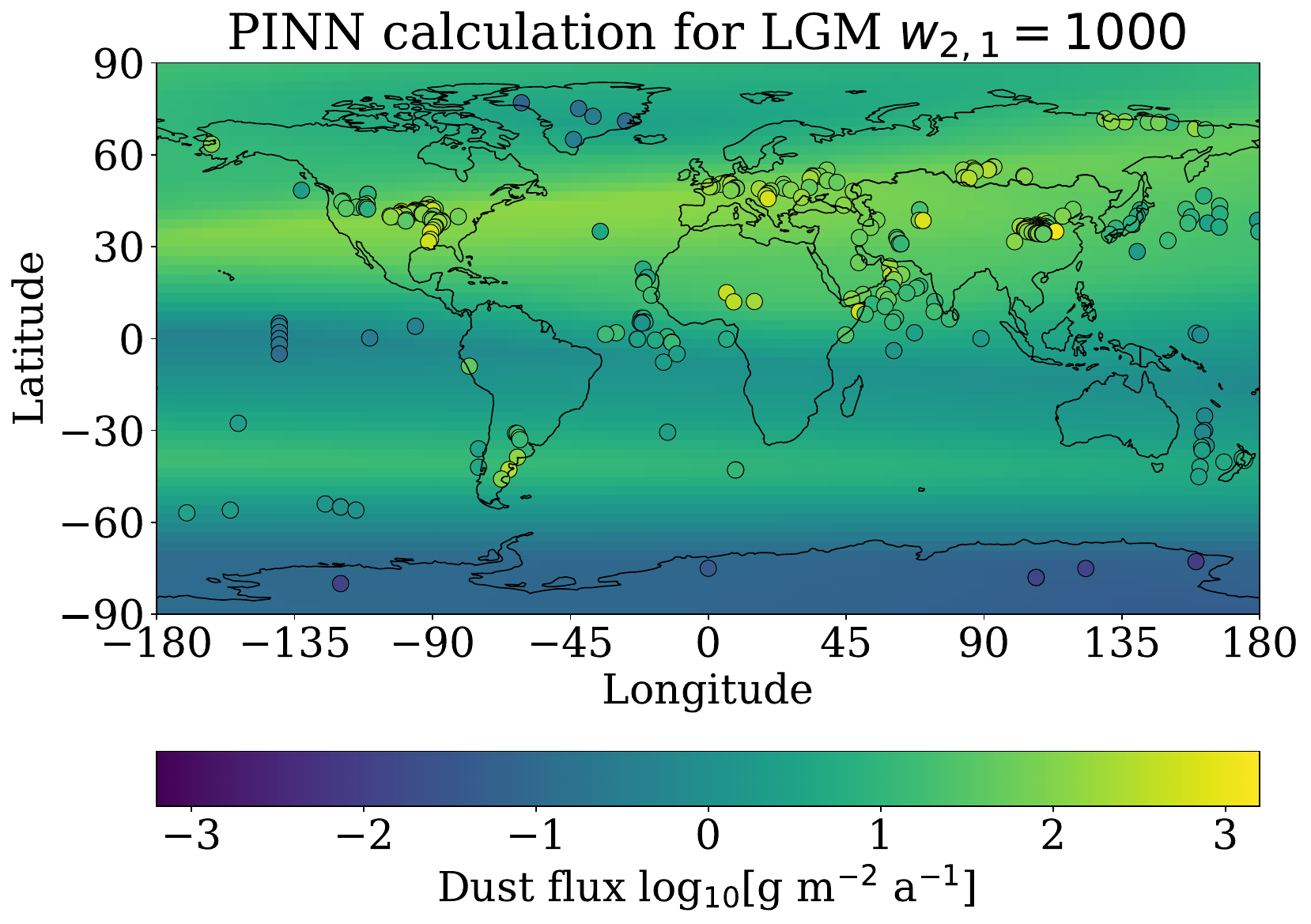}
	\caption{The global dust flux calculated by the PINN from empirical data in the LGM. Each panel uses a different model weight, with values of 0, 0.1, 10, and 1000, respectively. The main text shows the field for $w_{2,1}=1$ in Figure~\ref{fig: deposition maps}.}
	\label{fig: model weight field lgm}
\end{figure}

\begin{figure}[!ht] 
	\centering
	(a)
	\includegraphics[width=0.44\textwidth]{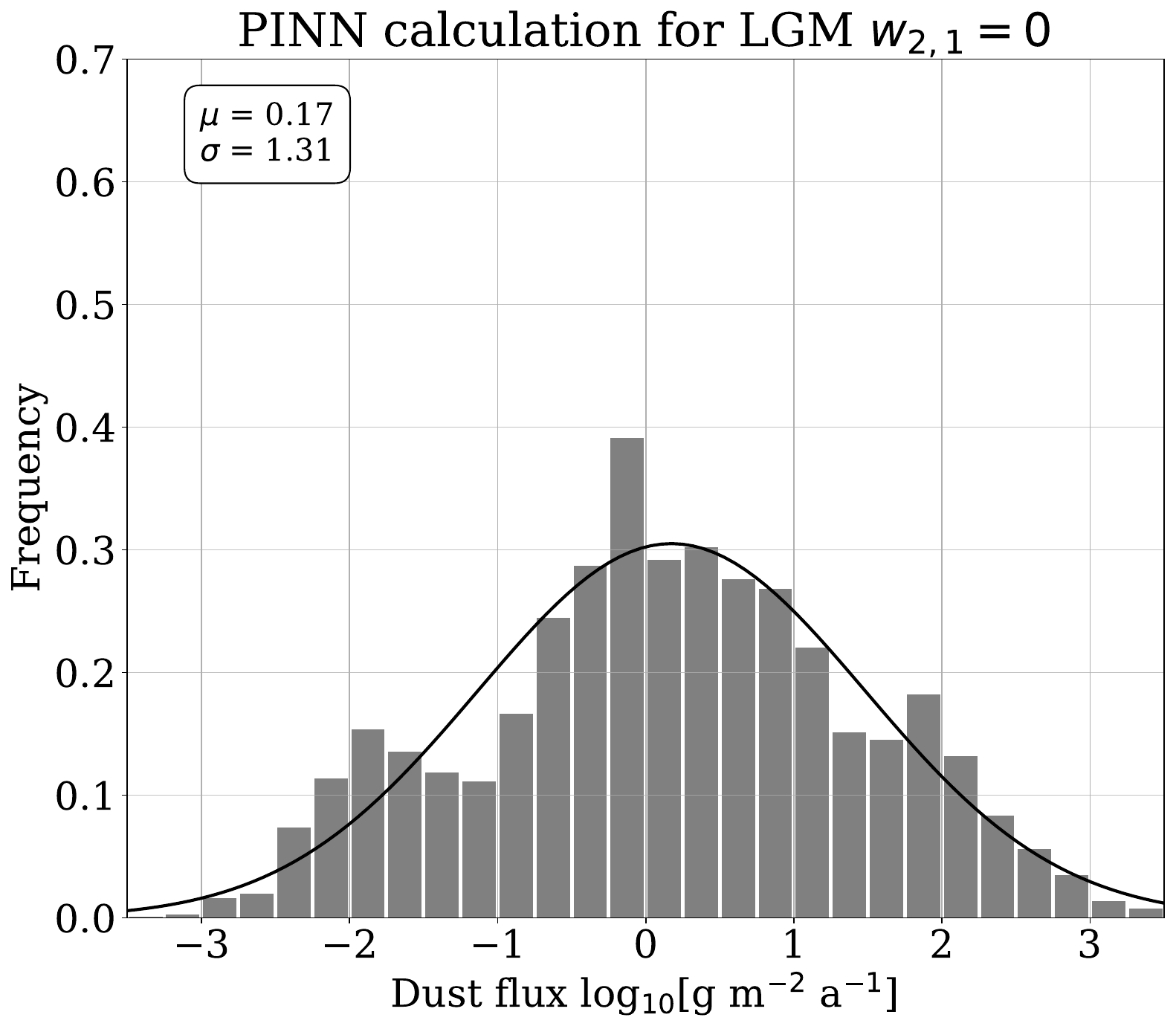}
	(b)
	\includegraphics[width=0.44\textwidth]{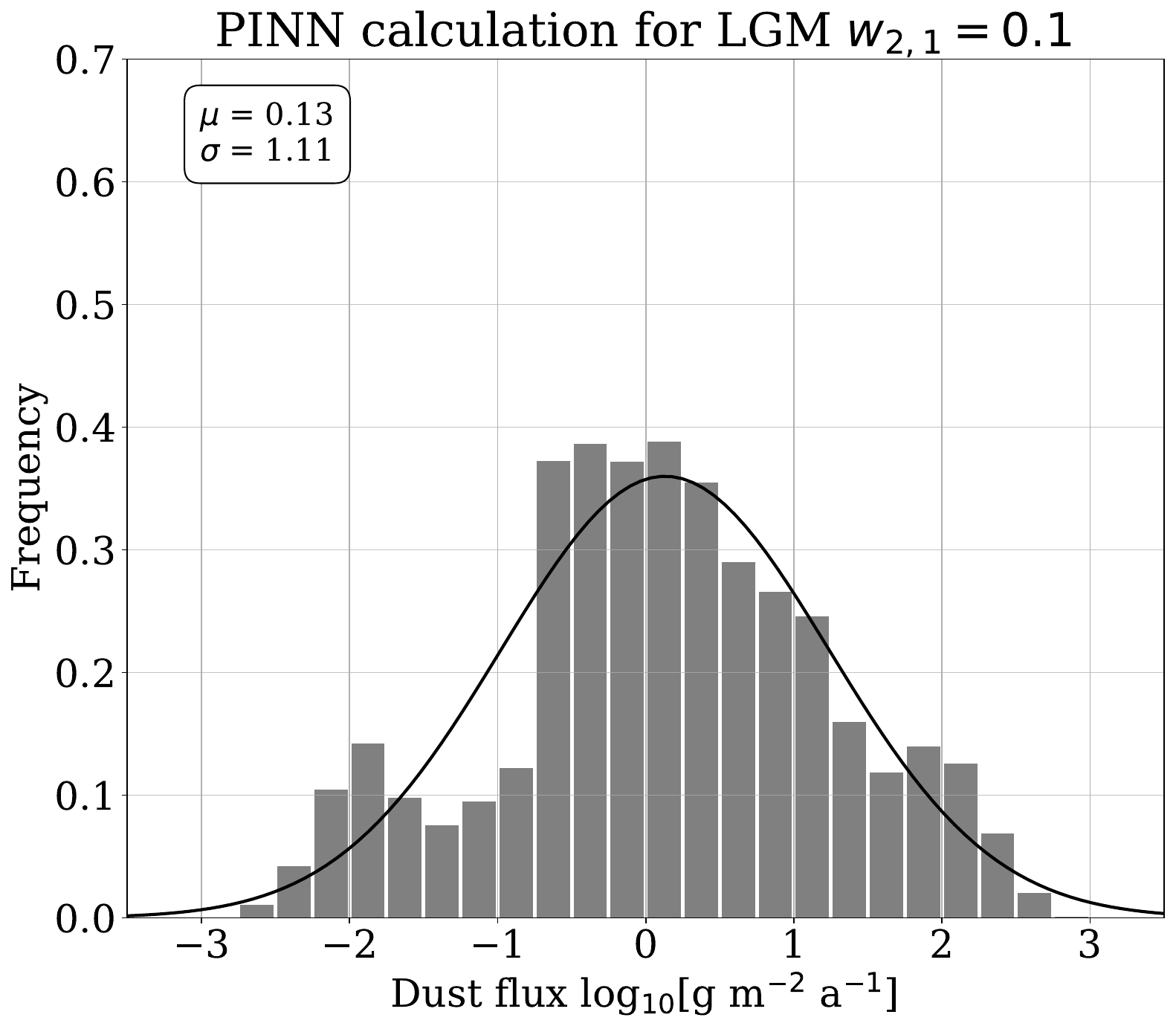}
	\\ (c)
	\includegraphics[width=0.44\textwidth]{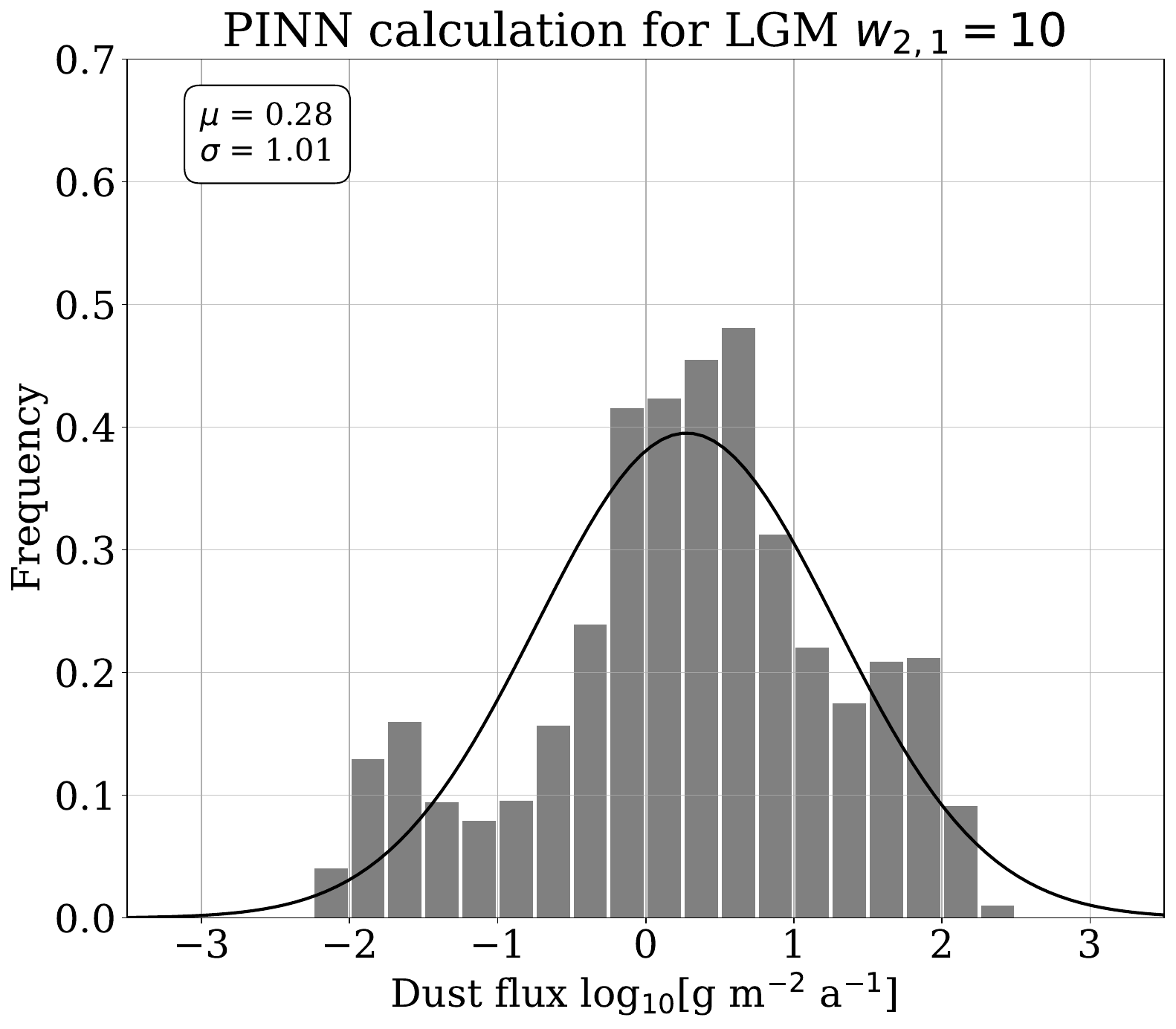}
	(d)
	\includegraphics[width=0.44\textwidth]{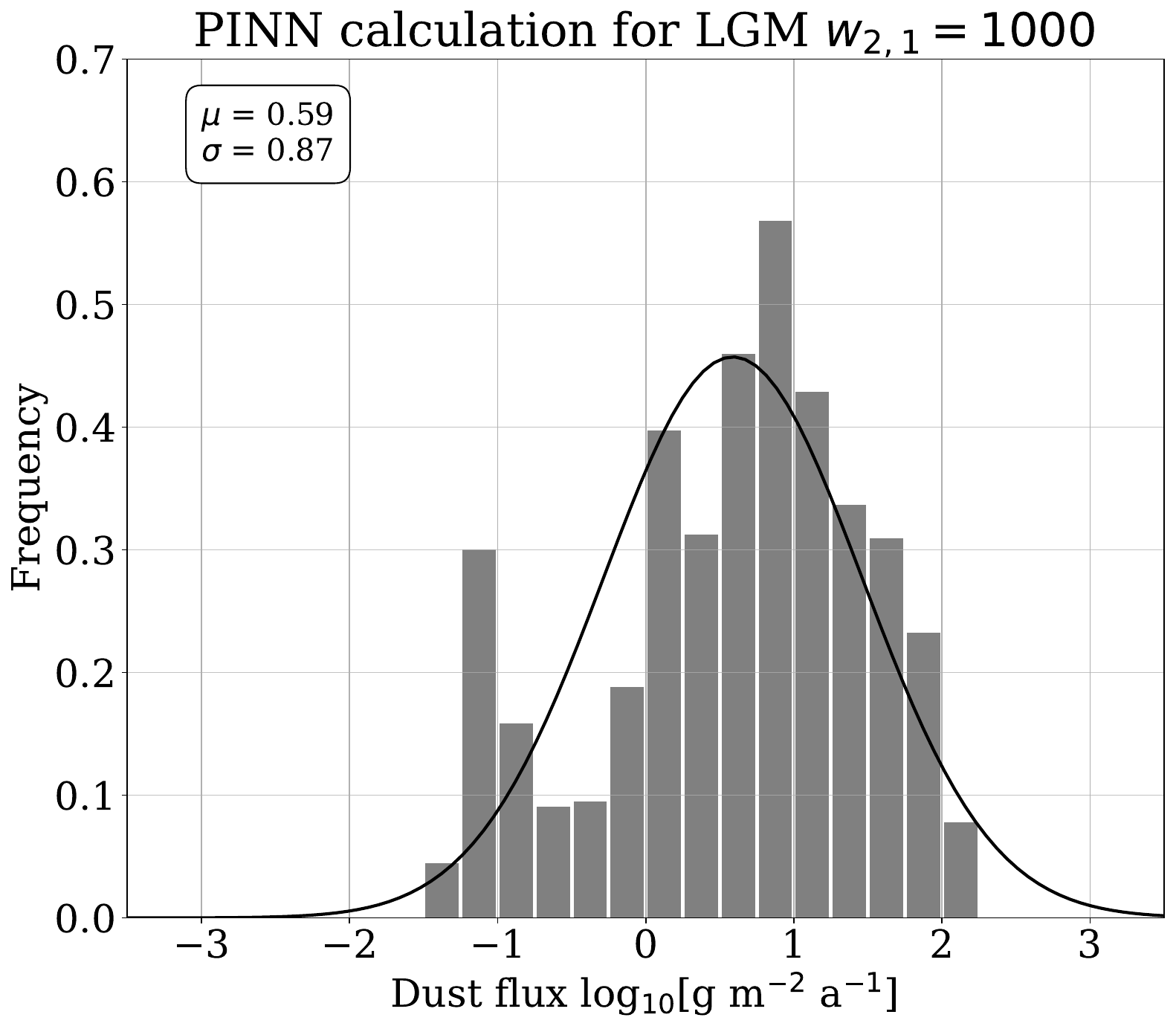}
	\caption{The dust flux distribution calculated by the PINN from empirical data in the LGM. Each panel uses a different model weight, with values of 0, 0.1, 10, and 1000, respectively. The main text shows the histogram for $w_{2,1}=1$ in Figure~\ref{fig: histograms}.}
	\label{fig: model weight histogram lgm}
\end{figure}

As explained in the Methodology section in the main text, the design of a PINN involves setting relative weights for the separate loss terms. This choice is arguably one of the most contentious decisions in the algorithmic design and one for which no clear guidelines are available in the literature. The most common approach is to use domain knowledge to find a suitable range of parameters. Here, we show the influence of the model weight $w_{2,1}$ in Equation~\eqref{loss_function_model}, which is the weight for the model term corresponding to the advection-diffusion equation. Figures~\ref{fig: model weight field holocene} to~\ref{fig: model weight histogram lgm} show the PINN's calculations for a range of values, where $w_{2,1}=1$ was used in the main text as the optimal choice.

Generally speaking, a low value of the model weight overemphasizes the empirical data and leads to unphysical solutions and overfitting issues. For example, when setting $w_{2,1}=0$, the PDE model is not considered, and the neural network only considers the empirical data and boundary conditions. The calculated dust flux indeed suffers from strong overfitting, with reasonable values in the data points, but the dust maps evidently disregard physical principles, displaying sharp peaks and upwind flows. In contrast, setting a very high value of the model weight overemphasizes the model and smooths out the solution, ignoring the spatial variability in the empirical data. Indeed, setting the model weight to $w_{2,1}=1000$ clearly shows an almost constant dust field, with reconstructions displaying global bands of horizontal dust flow along the zonal wind direction. The histograms show distributions that are too sharp and do not resemble the variety of dust deposition rates across the globe.

An expert's evaluation of the results can quickly discard the extreme choices of model weights as unreliable reconstructions. However, finding the optimal setting is difficult in the absence of a ground truth. In any case, the PINN is not overly sensitive to the parameters, with excellent results even for model weights ten times smaller or larger than used as the optimal value in this study.

\end{document}